\documentclass[aps,prd,preprint,nofootinbib,longbibliography,superscriptaddress,eqsecnum,amsmath,amsfonts,amssymb]{revtex4-2}
\usepackage{graphicx} 
\usepackage{physics}
\usepackage{tikz}
\usepackage{hyperref}

\newcommand\nn{\nonumber}

\newcommand{\be}{\begin{equation}}
\newcommand{\ee}{\end{equation}}
\newcommand{\bea}{\setlength\arraycolsep{2pt} \begin{eqnarray}}
\newcommand{\eea}{\end{eqnarray}}

\def\fft#1#2{{\frac{#1}{#2}}}

\def\0{{\sst{(0)}}}
\def\1{{\sst{(1)}}}
\def\2{{\sst{(2)}}}
\def\3{{\sst{(3)}}}
\def\4{{\sst{(4)}}}
\def\5{{\sst{(5)}}}
\def\6{{\sst{(6)}}}
\def\7{{\sst{(7)}}}
\def\8{{\sst{(8)}}}
\def\sst#1{{\scriptscriptstyle #1}}

\allowdisplaybreaks[1]

\begin{document}
\title{Higher-derivative Heterotic Kerr-Sen Black Holes}
\preprint{USTC-ICTS/PCFT-25-23}
\date{\today}

\author{Peng-Ju Hu}\email{pengjuhu@tju.edu.cn}
\affiliation{Center for Joint Quantum Studies and Department of Physics, School of Science, Tianjin University, Tianjin 300350, China}

\author{Liang Ma}\email{maliang0@tju.edu.cn}
\affiliation{Center for Joint Quantum Studies and Department of Physics, School of Science, Tianjin University, Tianjin 300350, China}

\author{Yi Pang}\email{pangyi1@tju.edu.cn}
\affiliation{Center for Joint Quantum Studies and Department of Physics, School of Science, Tianjin University, Tianjin 300350, China}
\affiliation{Peng Huanwu Center for Fundamental Theory, Hefei, Anhui 230026, China}

\author{Robert J. Saskowski}\email{robert\_saskowski@tju.edu.cn}
\affiliation{Center for Joint Quantum Studies and Department of Physics, School of Science, Tianjin University, Tianjin 300350, China}

\begin{abstract}
    We obtain the four-derivative corrections to the Kerr-Sen solution in heterotic supergravity, which includes the Gibbons-Maeda-Garfinkle-Horowitz-Strominger solution as a limiting case. In particular, we first embed the Kerr solution into heterotic supergravity and compute the higher-derivative corrections. We then obtain the corrections to the Kerr-Sen solution by performing an $O(2,1)$ boost of the Kerr solution, which, in contrast to the two-derivative case, requires field redefinitions to make the $O(2,1)$ invariance of the action manifest. Finally, we compute the multipole moments and find that they are distinct from those of the Kerr solution at the four-derivative level. We also find that the multipole moments are distinct from those of the Kerr-Newman solution in Einstein-Maxwell theory at the four-derivative level, even for the most general choice of four-derivative corrections. This gives a way to experimentally distinguish traces of string theory in gravitational wave data.
\end{abstract}

\maketitle
\newpage

\tableofcontents

\section{Introduction and summary}
String theory is the only known UV-complete description of quantum gravity. Despite this, the only observed experimental prediction of string theory is the very existence of gravity itself. On the other hand, it is known that Einstein gravity cannot be the fundamental quantum description of gravity, so it is natural to consider deviations at low energy within an effective field theory framework. In particular, if we are at energies well below the ultraviolet cutoff scale $\Lambda_c$, then the effective action is given by a higher-derivative expansion
\begin{equation}
    S=\frac{1}{16\pi G_N}\int\dd[4]x\sqrt{-g}\qty(\mathcal L_{2\partial}+\Lambda_c^{-2}\mathcal L_{4\partial}+\mathcal O(\Lambda_c^{-4})).
\end{equation}
We expect that the two-derivative Lagrangian $\mathcal L_{2\partial}$ is just general relativity with some choices of matter, including the standard model, dark matter, \emph{etc}. The higher-derivative corrections then subtly encode the influence of both quantum effects and unknown UV physics. One distinguished choice of effective field theory, which we might compare with experimental results, comes from the low-energy limit of strings, where the ultraviolet cutoff is given by the string scale $\Lambda_c^{-2}=\alpha'$.

Within the context of string theory, it is known that reduction on a $d$-dimensional torus $T^d$ leads to the appearance of a hidden $O(d,d;\mathbb Z)$ symmetry arising from T-duality. In the case of heterotic supergravity, if the gauge field configuration lies along a $U(1)^p$ subgroup of the gauge group, this is further enhanced to $O(d+p,d;\mathbb{Z})$, and in the low-energy effective theory this appears as a continuous $O(d+p,d;\mathbb{R})$ symmetry. The scalars then parametrize a $O(d+p,d)/(O(d+p-1,1)\times O(d-1,1))$ coset,\footnote{Note that this is in Lorentzian signature since we will be interested in compactifying time.} the gauge fields transform as a vector under $O(d+p,d)$, and the fermions transform under $O(d+p-1,1)\times O(d-1,1)$. Hence, heterotic supergravity solutions that are independent of $d$ coordinates possess an $O(d+p,d)$ symmetry, of which the ${O(d+p-1,1)\times O(d-1,1)}$ subgroup can be used to generate new, inequivalent solutions (see \emph{e.g.}~\cite{Veneziano:1991ek,Meissner:1991zj,Sen:1991zi,Sen:1991cn,Gasperini:1991qy,Hassan:1991mq,Sen:1992ua,Cvetic:1995sz,Cvetic:1995kv,Cvetic:1996xz}). In particular, this was exploited in~\cite{Sen:1992ua} by embedding the Kerr solution in heterotic supergravity and then performing an $O(2,1)$ boost that mixes the time and gauge field directions. The resulting family of exact solutions describes non-extremal, charged, rotating black holes in heterotic supergravity. This is generally referred to as the Kerr-Sen solution and is the unique stationary axisymmetric solution to the low-energy effective theory~\cite{Rogatko:2010hf}.

From a phenomenological viewpoint, heterotic supergravity reduced on $T^6$ is just Einstein-Maxwell-dilaton-axion gravity. Although massless scalar fields are usually problematic due to the emergence of long-range scalar forces, the weak equivalence principle is protected in scalar-tensor theories~\cite{Fierz:1956zz,Brans:1961sx,Damour:1992kf,Damour:1994zq, Catena:2006bd} by the universal coupling of gravity to the scalar sector~\cite{Damour:1996xx} and can thus provide a candidate for dynamical dark energy~\cite{Bartolo:1999sq,Esposito-Farese:2000pbo,Gannouji:2006jm} and inflation~\cite{Garcia-Bellido:1995him}. Hence, the dilaton and axion that appear in heterotic supergravity have important implications for inflationary cosmology~\cite{Sonner:2006yn,Catena:2007jf}, and models with this scalar content have been seriously considered~\cite{Linde:1991km}. This also potentially leads to modifications of Newton's laws and the Hubble expansion~\cite{Damour:1996xx,Catena:2004ba,Catena:2004pz,Schelke:2006eg}. Kerr-Sen black holes are thus a natural candidate to compare with experimental observations~\cite{Campbell:1992hc,Dastan:2016bfy,Cvetic:2017zde,Guo:2019lur,Narang:2020bgo,Xavier:2020egv,Shahzadi:2022rzq,DellaMonica:2023ydm,Sahoo:2023czj,Feng:2024iqj}. 

One observable of great interest is gravitational multipole moments, which influence the emission of gravitational waves and can serve as a test of the deviation of quantum gravity from general relativity~\cite{Bena:2020see,Bianchi:2020bxa,Bianchi:2020miz,Bah:2021jno}. In particular, extreme mass ratio inspiral (EMRI) experiments~\cite{LISA:2017pwj,Barack:2006pq,Cardoso:2016ryw,Babak:2017tow,Ryan:1995wh,Krishnendu:2018nqa} in the near future may be able to probe the multipole structure of black holes precisely. If we view the heterotic $U(1)$ gauge field as being the dark photon, we can then ask if the Kerr-Sen and Kerr solutions have the same multipole moments, which would distinguish them within heterotic supergravity. Alternatively, if we view the Maxwell field as being the ``real'' photon of the standard model, we can ask how heterotic supergravity compares to the more austere Einstein-Maxwell theory. There, the unique stationary axisymmetric solution is the Kerr-Newman one.  However, as we will see, the Kerr, Kerr-Newman, and Kerr-Sen solutions have the same gravitational multipole moments at the two-derivative level, and the Kerr-Newman and Kerr-Sen solutions have the same electromagnetic multipole moments at the two-derivative level. This differs from what we might expect, as, for example, the Larsen-Rasheed solution~\cite{Rasheed:1995zv,Larsen:1999pp} to Kaluza-Klein gravity has a different multipole structure at the two-derivative level~\cite{Bena:2020uup}. Thus, in order to distinguish the solutions, we must compare with the four-derivative corrected Kerr-Sen solution, which can be generated via the higher-derivative extension of the $O(2,1)$ transformation.

Fortunately, it is expected that the $O(d+p,d)$ symmetry persists to all orders in the $\alpha'$ expansion~\mbox{\cite{Sen:1991zi,Hohm:2014sxa}}, and the very presence of this symmetry is quite constraining for higher-derivative couplings~\mbox{\cite{Godazgar:2013bja,Marques:2015vua,Hohm:2015doa,Garousi:2019wgz,Garousi:2019mca,Garousi:2020gio,Codina:2020kvj,David:2021jqn,David:2022jcl,Ozkan:2024euj}}. One approach to the construction of such $O(d+p,d)$ invariant actions is double field theory~\mbox{\cite{Siegel:1993th,Hull:2009mi,Hohm:2010jy,Hohm:2010pp,Hohm:2011ex}}, and, in particular, the four-derivative extensions thereof~\mbox{\cite{Marques:2015vua,Hohm:2013jaa,Bedoya:2014pma,Hohm:2014xsa,Coimbra:2014qaa,Lee:2015kba,Baron:2017dvb,Lescano:2021guc}}.\footnote{Note that such DFT formulations break down at higher orders in $\alpha'$, namely at $\mathcal O(\alpha'^3)$~\cite{Hronek:2020xxi,Hsia:2024kpi}.} Indeed, some $\alpha'$ corrected dolutions have been obtained directly within the DFT framework~\cite{Lunin:2024vsx}. One reason why the Kerr-Sen procedure works so beautifully at the two-derivative level is that upon dimensional reduction, the action is automatically $O(d+p,d)$ invariant. However, this is no longer true at the four-derivative level. Instead, one must perform field redefinitions to bring the reduced theory into an $O(d+p,d)$ symmetric form. There has been much recent progress on explicitly obtaining $O(d,d)$ invariant actions and the field redefinitions required to reach them from the dimensional reduction of bosonic and heterotic string effective actions~\cite{Eloy:2020dko,Elgood:2020xwu,Ortin:2020xdm,Jayaprakash:2024xlr}.

\subsection{Summary}

We begin by embedding the four-dimensional Kerr solution into heterotic supergravity and obtaining the four-derivative corrections. Following~\cite{Cano:2019ore,Cano:2021rey}, the two-form NS field can be dualized to an axion. The Einstein frame metric receives no corrections, and so the problem reduces to that of solving for the scalar hair of the Kerr black hole. The result cannot be computed in closed form and instead must be expressed as a series in small rotation. This result is equivalent to that of~\cite{Cano:2021rey}, albeit in a different field redefinition frame. We then dualize the axion back to a two-form to identify the embedding into heterotic supergravity.

Next, we perform an $O(2,1)$ boost to obtain the corrected Kerr-Sen solution. In particular, four-derivative heterotic supergravity reduced on $S^1$ does not automatically have $O(2,1)$ invariance and must be field redefined. We choose to use the minimal set of field redefinitions, given by~\cite{Jayaprakash:2024xlr}. However, there is a catch: The authors of~\cite{Jayaprakash:2024xlr} truncate the heterotic gauge fields. In order to get around this, we uplift the solution to five dimensions. Using the result of~\cite{Liu:2023fqq}, there is a particular consistent truncation on $S^1$ that allows us to obtainthe action with a $U(1)$ gauge field. We then reduce the five-dimensional Kerr solution to three dimensions and perform the necessary field redefinitions to make the $O(2,2)$ invariance manifest. We can then pick an $O(2,1)\subset O(2,2)$ transformation, matching that of~\cite{Sen:1992ua}, and use it to generate our new solution. We then field redefine back and uplift to five dimensions before finally reducing to four dimensions to obtain the four-derivative corrected Kerr-Sen solution.

Using this four-derivative corrected Kerr-Sen solution, we then compute the multipole moments. The $O(2,1)$ boost procedure we used results in four-derivative corrections to the charges that can be shifted away by a redefinition of the solution parameters. After performing this redefinition, we are able to compare with the Kerr solution, which has no four-derivative corrections to the multipole moments in heterotic supergravity. We then compare it with the Kerr-Newman solution in Einstein-Maxwell theory. In particular, we consider the most general four-derivative extension of Einstein-Maxwell theory, for which the Kerr-Newman multipole moments were computed in~\cite{Ma:2024ulp}. We find that there is no choice of parameters that allows the Kerr-Newman and Kerr-Sen to have the same multipole moments. In both cases, the Kerr-Sen solution has a distinct multipole structure that makes it distinguishable in gravitational wave experiments. The first case allows us to distinguish solutions within heterotic supergravity, while the second case allows us to distinguish heterotic supergravity from Einstein-Maxwell theory.

The rest of this paper is organized as follows. In Section \ref{sec:Kerr}, we obtain the higher-derivative corrections to the Kerr solution, and in Section \ref{sec:Odd}, we perform an $O(2,1)$ boost to obtain the higher-derivative corrections to the Kerr-Sen solution. We then compute the multipole moments in Section \ref{sec:multipoles}. We conclude and discuss some future directions in Section \ref{sec:disc}. Some technical details are relegated to the appendices. In Appendix \ref{app:gaugeFields}, we demonstrate how $O(d+p,d)$ invariance in heterotic supergravity with gauge fields reduced on $T^d$ arises from the $O(d+p,d+p)$ invariance of heterotic supergravity without gauge fields reduced on $T^{d+p}$. In Appendix \ref{app:JimTranslate}, we review how to convert $O(d)_-\times O(d)_+$ invariant actions to an explicitly $O(d,d)$ invariant form.

\section{Higher-derivative corrections to Kerr}\label{sec:Kerr}
In this section, we review how to reformulate the four-dimensional Bergshoeff-de Roo (BdR) action as a higher-derivative Einstein-scalar-axion theory~\cite{Cano:2021rey} and then derive the higher-derivative corrected Einstein-scalar-axion black hole solution~\cite{Cano:2019ore}. We then map the theory back to the original BdR framework. This yields a rotating black hole solution in the $\alpha'$-corrected heterotic string theory.

\subsection{4d heterotic string at order $\alpha'$}
Heterotic supergravity describes a half-maximal supergravity theory consisting of gravity $(g_{\hat\mu\hat\nu},\psi_{\hat\mu},B_{\hat\mu\hat\nu},\lambda,\phi)$ coupled to vector multiplets $(\mathcal A_{\hat\mu}^{\mathfrak a},\chi^{\mathfrak a})$ transforming under either $E_8\times E_8$ or $SO(32)$. The field content is thus the metric $g$, the gravitini $\psi$, the two-form $B$, the dilatino $\lambda$, the dilaton $\phi$, the gauge fields $\mathcal A^{\mathfrak a}$, and the gaugini $\chi^{\mathfrak a}$. For now, we will not consider the vector multiplets as they will be irrelevant to the Kerr solution, although we will reinstate them when we boost the solution in the next section. We begin with the BdR form of the bosonic action for heterotic supergravity in the string frame~\cite{Bergshoeff:1988nn,Bergshoeff:1989de}
\begin{equation}
    e^{-1}\mathcal{L}=e^{-2\phi}\qty[R(\Omega)+4(\partial\phi)^2-\frac{1}{12}\tilde H^2+\frac{\alpha'}{8}R_{\hat\mu\hat\nu\hat\rho\hat\sigma}(\Omega_-)^2],\label{eq:BdR4d}
\end{equation}
where $R$ is the Ricci scalar and $\Omega$ the Levi-Civita spin connection, and we have defined
\begin{equation}
    \tilde H= H+\frac{\alpha'}{4}\omega_{3L}(\Omega_-),
\label{eq:Htilde}
\end{equation}
where $H=\dd B$ is the three-form NS flux. Here we have introduced the torsionful connection
\begin{equation}
    \Omega_\pm=\Omega\pm\fft12 H.
\end{equation}
The corresponding curvature is given by
\begin{equation}
    R(\Omega_\pm)=\dd\Omega_\pm+\Omega_\pm\wedge\Omega_\pm,
\end{equation}

while the Lorentz-Chern-Simons form is defined by
\begin{equation}
    \omega_{3L}(\Omega_\pm)=\Tr\left(\Omega_\pm\wedge \dd\Omega_\pm+\fft23\Omega_\pm\wedge\Omega_\pm\wedge\Omega_\pm\right).
\end{equation}
The interplay between $\Omega_+$ and $\Omega_-$ plays an important role in supersymmetry~\cite{Bergshoeff:1988nn,Bergshoeff:1989de}, and the presence of the Lorentz-Chern-Simons form is necessary for anomaly cancellation~\cite{Sen:1985tq}. Note that the three-form flux satisfies the Bianchi identity
\begin{equation}
    \dd\tilde H=\frac{\alpha'}{4}\Tr \qty[R(\Omega_-)\land R(\Omega_-)].\label{eq:Bianchi}
\end{equation}
For notational convenience, we define the Hodge dual of the Riemann tensor to be
\begin{equation}
    \tilde{R}^{\hat\mu\hat\nu}{}_{\hat\rho\hat\sigma}(\Omega)=\frac{1}{2}\epsilon^{\hat\mu\hat\nu\hat\lambda\hat\tau}R_{\hat\lambda\hat\tau\hat\rho\hat\sigma}(\Omega).
\end{equation}

Since we are working in four dimensions, the two-form field $B$ can be dualized to an axion~$\varphi$. Given the Bianchi identity \eqref{eq:Bianchi}, we add a Lagrange multiplier 
\begin{equation}
    e^{-1}\Delta\mathcal L=-\varphi\qty[\dd\tilde H-\frac{\alpha'}{4}\Tr R(\Omega_-)\land R(\Omega_-)],
\end{equation}
so that, after integration by parts, the Lagrangian becomes
\begin{align}
    e^{-1}\widetilde{\mathcal L}&=e^{-2\phi}\qty[R(\Omega)+4(\partial\phi)^2-\frac{1}{12}\tilde H^2+\frac{\alpha'}{8}R_{\hat\mu\hat\nu\hat\rho\hat\sigma}(\Omega_-)^2]\nn\\
    &\quad+\frac{1}{3!}\epsilon^{\hat\mu\hat\nu\hat\rho\hat\sigma}\tilde{H}_{\hat\mu\hat\nu\hat\rho}\nabla_{\hat\sigma}\varphi+\frac{\alpha'}{8}\varphi R_{\hat\mu\hat\nu\hat\rho\hat\sigma}(\Omega_-)\tilde{R}^{\hat\mu\hat\nu\hat\rho\hat\sigma}(\Omega_-).
\end{align}
As in~\cite{Cano:2021rey}, we separate $\widetilde{\mathcal{L}}$ into two- and four-derivative parts
\begin{align}
    \widetilde{\mathcal L}&=\widetilde{\mathcal L}^{(2)}+\frac{\alpha'}{8}\widetilde{\mathcal L}^{(4)},\nn\\
    \widetilde{\mathcal{L}}^{(2)}&=e^{-2\phi}\qty[R(\Omega)+4(\partial\phi)^2-\frac{1}{12}\tilde H^2]+\frac{1}{3!}\epsilon^{\hat\mu\hat\nu\hat\rho\hat\sigma}\tilde H_{\hat\mu\hat\nu\hat\rho}\nabla_{\hat\sigma}\varphi,\nn\\
    \widetilde{\mathcal{L}}^{(4)}&=e^{-2\phi}R_{\hat\mu\hat\nu\hat\rho\hat\sigma}(\Omega_-)R^{\hat\mu\hat\nu\hat\rho\hat\sigma}(\Omega_-)+\varphi R_{\hat\mu\hat\nu\hat\rho\hat\sigma}(\Omega_-)\tilde{R}^{\hat\mu\hat\nu\hat\rho\hat\sigma}(\Omega_-).
\end{align}
Note that, with an eye to dualization, we are treating $\tilde H$ as a single object.\footnote{Note that it is equivalent to dualize $H$ using the Bianchi identity $\dd H=0$ or to dualize $\tilde H$ using $\dd\tilde H=\tfrac{\alpha'}{4}\Tr[R(\Omega_-)\land R(\Omega_-)]$. The resulting four-derivative actions will be the same.} Varying $\widetilde{\mathcal L}$ with respect to $\varphi$ yields the Bianchi identity \eqref{eq:Bianchi}, and so integrating out $\varphi$ just yields our original Lagrangian \eqref{eq:BdR4d}. On the other hand, the variation with respect to $\tilde H$ yields the relation
\begin{equation}
    \tilde H_{\hat\mu\hat\nu\hat\rho}=e^{2\phi}\epsilon_{\hat\mu\hat\nu\hat\rho\hat\sigma}\nabla^{\hat\sigma}\varphi+e^{2\phi}\frac{3\alpha'}{4}\frac{\delta\widetilde{\mathcal{L}}^{(4)}}{\delta\tilde{H}^{\hat\mu\hat\nu\hat\rho}}\bigg\vert_{\tilde H=\dd B}.\label{H varphi}
\end{equation}
Substituting this back into the action, we get the dualized Lagrangian
\begin{align}
    \widetilde{\mathcal{L}}_S&=\widetilde{\mathcal{L}}^{(2)}_S+\frac{\alpha'}{8}\widetilde{\mathcal{L}}^{(4)}_S,\nn\\
    e^{-1}\widetilde{\mathcal{L}}^{(2)}_S&=e^{-2\phi}\qty[R+4(\partial\phi)^2]-\frac{1}{2}e^{2\phi}(\partial\varphi)^2,\nn\\
    e^{-1}\widetilde{\mathcal{L}}^{(4)}_S&=e^{-2\phi}\Big[R_{\hat\mu\hat\nu\hat\rho\hat\sigma}(\Omega)R^{\hat\mu\hat\nu\hat\rho\hat\sigma}(\Omega)-6G_{\hat\mu\hat\nu}C^{\hat\mu} C^{\hat\nu}+\frac{7}{4}C^4-2(\nabla_{\hat\mu} C_{\hat\nu})^2-(\nabla_{\hat\mu} C^{\hat\mu})^2\Big]\nn\\
    &\quad+\varphi R_{\hat\mu\hat\nu\hat\rho\hat\sigma}(\Omega)\tilde{R}^{\hat\mu\hat\nu\hat\rho\hat\sigma}(\Omega),\label{dual theory String frame}
\end{align}
where $G_{\hat\mu\hat\nu}=R_{\hat\mu\hat\nu}(\Omega)-\frac{1}{2}g_{\hat\mu\hat\nu}R(\Omega)$ is the Einstein tensor and we have defined $C_{\hat\mu}=e^{2\phi}\nabla_{\hat\mu}\varphi$.

The Lagrangian \eqref{dual theory String frame} is in the string frame, and so we perform a conformal scaling
\begin{equation}
    g_{\hat\mu\hat\nu}\to e^{2\phi}g_{\hat\mu\hat\nu},\label{conformal transformation}
\end{equation}
which yields the Einstein frame Lagrangian
\begin{align}
    \widetilde{\mathcal{L}}_E&=\widetilde{\mathcal{L}}^{(2)}_E+\frac{\alpha'}{8}\widetilde{\mathcal{L}}^{(4)}_E,\nn\\
    e^{-1}\widetilde{\mathcal{L}}^{(2)}_E&=R-2(\partial\phi)^2-\frac{1}{2}e^{4\phi}(\partial\varphi)^2,\nn\\
    e^{-1}\widetilde{\mathcal{L}}^{(4)}_E&=e^{-2\phi}\Bigg\{R_{\hat\mu\hat\nu\hat\rho\hat\sigma}R^{\hat\mu\hat\nu\hat\rho\hat\sigma}-4R^{\hat\mu\hat\nu}\left(4\nabla_{\hat\mu}\phi\nabla_{\hat\nu}\phi+C_{\hat\mu} C_{\hat\nu}\right)-R\left(4\Box\phi-4(\partial\phi)^2-3C^2\right)\nn\\
    &\qquad\qquad+12(\partial\phi)^4+12(\Box\phi)^2+\frac{7}{4}C^4-12(\nabla_{\hat\mu}\phi C^{\hat\mu})^2-2C^2(\partial\phi)^2-8C^2\Box\phi\nn\\
    &\qquad\qquad-16(\nabla_{\hat\mu}\phi C^{\hat\mu})\nabla_{\hat\nu} C^{\hat\nu}-3(\nabla_{\hat\mu} C^{\hat\mu})^2\Bigg\}+\varphi R_{\hat\mu\hat\nu\hat\rho\hat\sigma}\tilde{R}^{\hat\mu\hat\nu\hat\rho\hat\sigma}.
\label{dual theory Einstein frame}
\end{align}
This then reduces the problem of finding corrections to the Kerr solution to a problem of solving for scalar fluctuations.

\subsection{Rotating black hole solution}

We now wish to discuss the corrections to the Kerr solution due to these higher-derivative corrections. The Kerr black hole, expressed in Boyer-Lindquist coordinates as
\begin{align}
    \dd s^2&=-\left(1-\frac{2\mu r}{\Sigma}\right)\dd t^2-\frac{4\mu ar(1-x^2)}{\Sigma}\,\dd t\,\dd y+\Sigma\left(\frac{\dd r^2}{\Delta}+\frac{\dd x^2}{1-x^2}\right)\nn\\
    &\quad+\qty(r^2+a^2+\frac{2\mu ra^2(1-x^2)}{\Sigma})(1-x^2)\,\dd y^2,\nn\\
    \phi&=0,\qquad \varphi=0,\label{Kerr}
\end{align}
where
\begin{equation}
    \Sigma=r^2+a^2x^2,\qquad \Delta=r^2-2\mu r+a^2,
\end{equation}
is a solution to the two-derivative theory given by $\widetilde{\mathcal{L}}^{(2)}_E$ in Eq.~\eqref{dual theory Einstein frame}. This is a stationary, axisymmetric, Ricci flat solution with mass $\mu$ and angular momentum $J=a \mu$.

As noted in~\cite{Cano:2021rey}, unlike the case for charged black holes~\cite{Cano:2018qev,Chimento:2018kop,Cano:2018brq,Cano:2019ycn}, the (Einstein frame) Kerr black hole metric \eqref{Kerr} remains unchanged by the four-derivative corrections, although the scalar fields $\phi$, $\varphi$ are modified. Unfortunately, it is not known how to obtain a closed-form analytic expression for these corrections; instead, the corrections must be expanded as a series in the spin~\cite{Pani:2009wy,Yunes:2009hc,Konno:2009kg,Yagi:2012ya,Mignemi:1992pm,Cano:2019ore}. Following~\cite{Cano:2019ore}, we introduce the dimensionless spin parameter
\begin{equation}
    \chi=\frac{a}{\mu},
\end{equation}
and expand the scalar fields in powers of $\chi$ as
\begin{equation}
    \phi=\alpha'\sum_{n=0}^\infty\delta\phi_i^{(n)}(r,x)\chi^n,\qquad\varphi=\alpha'\sum_{n=0}^\infty\delta\varphi_i^{(n)}(r,x)\chi^n.
\end{equation}
Here, $\delta\phi_i^{(n)}$ and $\delta\varphi_i^{(n)}$ can always be expressed as polynomials in $x$ and $r^{-1}$
\bea
\delta\phi_i^{(n)}(r,x)=\sum_{p=0}^n\sum_{k=0}^{k_{max}}\delta\phi_i^{(n,p,k)}\frac{x^p}{r^k},\qquad
\delta\varphi_i^{(n)}(r,x)=\sum_{p=0}^n\sum_{k=0}^{k_{max}}\delta\varphi_i^{(n,p,k)}\frac{x^p}{r^k}.
\eea
Although we have calculated up to $\mathcal{O}(\chi^{20})$, for the sake of space, we only present the first few terms
\begin{align}
    \phi&=\alpha'\left[-\frac{\mu }{6 r^3}-\frac{1}{8 r^2}-\frac{1}{8 \mu  r}+ \chi^2\left(\frac{\mu ^2 }{80 r^4}+\frac{\mu }{40 r^3}+\frac{1}{32 r^2}+\frac{1}{32 \mu  r}\right)\right.\nn\\
    &\qquad+\chi ^2 x^2 \left(\frac{6 \mu ^3 }{5 r^5}+\frac{21 \mu ^2 }{40 r^4}+\frac{7 \mu }{40 r^3}\right)+\chi ^4 \left(\frac{\mu ^3 }{280 r^5}+\frac{\mu ^2 }{112 r^4}+\frac{3 \mu  }{224 r^3}+\frac{1}{64 r^2}+\frac{1}{64 \mu  r}\right)\nn\\
    &\qquad\left.+\chi ^4 x^4 \left(-\frac{45 \mu ^5}{14 r^7}-\frac{55 \mu ^4 }{56 r^6}-\frac{11 \mu ^3 }{56 r^5}\right)+\chi ^4 x^2 \left(-\frac{\mu ^4}{28 r^6}-\frac{3 \mu ^3}{70 r^5}-\frac{3 \mu ^2 }{112 r^4}-\frac{\mu }{112 r^3}\right)+\mathcal O(\chi^6)\right],\nn\\
    \varphi&=\alpha'\left[ \chi  x \left(\frac{9 \mu ^2 }{8 r^4}+\frac{5 \mu  }{8 r^3}+\frac{5 }{16 r^2}\right)+\chi ^3 x \left(-\frac{\mu ^3 }{20 r^5}-\frac{3 \mu ^2 }{40 r^4}-\frac{\mu  }{16 r^3}-\frac{1}{32 r^2}\right)\right.\nn\\
    &\qquad+\chi ^3 x^3 \left(-\frac{25 \mu ^4 }{6 r^6}-\frac{3 \mu ^3 }{2 r^5}-\frac{3 \mu ^2 }{8 r^4}\right)+\chi ^5 x \left(-\frac{3 \mu ^4 }{224 r^6}-\frac{3 \mu ^3 }{112 r^5}-\frac{27 \mu ^2 }{896 r^4}-\frac{3 \mu  }{128 r^3}-\frac{3 }{256 r^2}\right)\nn\\
    &\qquad\left.+\chi ^5 x^5 \left(\frac{147 \mu ^6 }{16 r^8}+\frac{39 \mu ^5 }{16 r^7}+\frac{13 \mu ^4 }{32 r^6}\right)+\chi ^5 x^3 \left(\frac{5 \mu ^5 }{56 r^7}+\frac{5 \mu ^4 }{56 r^6}+\frac{5 \mu ^3 }{112 r^5}+\frac{5 \mu ^2 }{448 r^4}\right)+\mathcal O(\chi^7)\right].\label{eq:ogScalars}
\end{align}
This solution should be interpreted as imbuing the Kerr solution with axidilatonic hair~\mbox{\cite{Campbell:1991kz,Campbell:1992hc,Mignemi:1992pm}}. 

Given $\varphi$, we can use \eqref{H varphi} to obtain the three-form flux $\tilde H$,\footnote{Note that the second term in \eqref{H varphi} does not contribute because the value of $\frac{\delta\widetilde{\mathcal{L}}^{(2)}}{\delta\tilde{H}^{\hat\mu\hat\nu\hat\rho}}$ is zero to leading order.} from which we obtain the solution for $B$ using equation \eqref{eq:Htilde},
\bea
B=\alpha'\Lambda(r,x)\,\dd t\wedge \dd y.
\eea
Here, the function $\Lambda$ is again expressed as a series in $\chi$
\bea
\Lambda(r,x)=\sum_{n=0}^\infty \Lambda_\chi^{(n)}(r,x)\,\chi^n,
\eea
with the first several terms given by
\bea
\Lambda_\chi^{(0)}&=&0,\qquad \Lambda_\chi^{(2)}=0,\qquad \Lambda_\chi^{(4)}=0,\cr
\Lambda_\chi^{(1)}&=&\frac{x^2   \left(-18 \mu ^2+5 r^2+5 \mu  r\right)}{16 r^3}-\frac{ \left(-2 \mu ^2+5 r^2+5 \mu  r\right)}{16 r^3},\cr
\Lambda_\chi^{(3)}&=&-\frac{x^2  }{160 r^5}\Big[5 r^4+5 \mu  r^3+4 \mu ^2 r^2 \left(15 x^2-14\right)+2 \mu ^3 r \left(90 x^2-89\right)\cr
&&+80 \mu ^4 \left(1-5 x^2\right)\Big]+\frac{ \left(2 \mu ^3+5 r^3+5 \mu  r^2+4 \mu ^2 r\right)}{160 r^4},\cr
\Lambda_\chi^{(5)}&=&\frac{x^2  }{8960 r^7}\Big[-105 r^6-105 \mu  r^5+10 \mu ^2 r^4 \left(10 x^2-19\right)+60 \mu ^3 r^3 \left(5 x^2-6\right)\cr
&&+8 \mu ^4 r^2 \left(455 x^4-395 x^2-63\right)+200 \mu ^5 r \left(91 x^4-89 x^2-2\right)+1680 \mu ^6 x^2 \left(5-21 x^2\right)\Big]\cr
&&+\frac{3  \left(8 \mu ^4+35 r^4+35 \mu  r^3+30 \mu ^2 r^2+20 \mu ^3 r\right)}{8960 r^5}.
\eea

The metric in the string frame can be obtained via a conformal transformation, the inverse of \eqref{conformal transformation}.

\section{Boosting the Kerr solution}\label{sec:Odd}
In this section, we will boost the Kerr solution found in Section \ref{sec:Kerr} using an $O(2,1)$ transformation. This is the four-derivative generalization of the two-derivative boost performed in~\cite{Sen:1992ua}. The key insight of~\cite{Sen:1992ua} is that a $U(1)^d$ symmetry $\partial_i=0$ is (mathematically) equivalent to compactifying the $y^i$ directions on a torus. Thus, given a theory with a $U(1)^p$ gauge symmetry, any solution with a $U(1)^d$ isometry has a hidden $O(d+p,d)$ that rotates solutions into inequivalent solutions.\footnote{Said another way, this $O(d+p,d)$ is a symmetry of the action but not the solution.} At the two-derivative level, the heterotic action has a manifest $O(d+p,d)$ symmetry; however, when higher-derivative corrections are included, the action in the na\"ive field redefinition frame is not automatically $O(d+p,d)$ invariant, but rather must be field redefined to make this symmetry manifest~\cite{Eloy:2020dko,Elgood:2020xwu,Ortin:2020xdm,Jayaprakash:2024xlr}. In many cases, these authors put the action into the simplest form, without any derivatives on field strengths, which is generally at the cost of requiring more complicated field redefinitions. The minimal set of field redefinitions required to have an $O(d,d)$ invariant action is given in~\cite{Jayaprakash:2024xlr}, which we will make use of. One issue is that the authors of~\cite{Jayaprakash:2024xlr} truncate the heterotic gauge fields, and so only elucidate the $O(d,d)$ symmetry. Hence, we need to couple a $U(1)$ gauge field to the action in a way that enhances the $O(d,d)$ symmetry to an $O(d+p,d)$ symmetry. To do this, we can uplift to five dimensions as a trick to recover the gauge fields from an $O(2,1)$ subgroup of $O(2,2)$. More generally, as we show in Appendix \ref{app:gaugeFields}, uplifting on $T^p$ allows one to recover $O(d+p,d)$ as a diagonal subgroup of $O(d+p,d+p)$.

\begin{figure}
    \centering

\tikzset{every picture/.style={line width=0.75pt}} 

\begin{tikzpicture}[x=0.75pt,y=0.75pt,yscale=-1,xscale=1]

\draw   (40,118) .. controls (40,113.58) and (43.58,110) .. (48,110) -- (102,110) .. controls (106.42,110) and (110,113.58) .. (110,118) -- (110,142) .. controls (110,146.42) and (106.42,150) .. (102,150) -- (48,150) .. controls (43.58,150) and (40,146.42) .. (40,142) -- cycle ;
\draw   (140,58) .. controls (140,53.58) and (143.58,50) .. (148,50) -- (202,50) .. controls (206.42,50) and (210,53.58) .. (210,58) -- (210,82) .. controls (210,86.42) and (206.42,90) .. (202,90) -- (148,90) .. controls (143.58,90) and (140,86.42) .. (140,82) -- cycle ;
\draw   (140,178) .. controls (140,173.58) and (143.58,170) .. (148,170) -- (202,170) .. controls (206.42,170) and (210,173.58) .. (210,178) -- (210,202) .. controls (210,206.42) and (206.42,210) .. (202,210) -- (148,210) .. controls (143.58,210) and (140,206.42) .. (140,202) -- cycle ;
\draw   (300,58) .. controls (300,53.58) and (303.58,50) .. (308,50) -- (362,50) .. controls (366.42,50) and (370,53.58) .. (370,58) -- (370,82) .. controls (370,86.42) and (366.42,90) .. (362,90) -- (308,90) .. controls (303.58,90) and (300,86.42) .. (300,82) -- cycle ;
\draw   (300,178) .. controls (300,173.58) and (303.58,170) .. (308,170) -- (362,170) .. controls (366.42,170) and (370,173.58) .. (370,178) -- (370,202) .. controls (370,206.42) and (366.42,210) .. (362,210) -- (308,210) .. controls (303.58,210) and (300,206.42) .. (300,202) -- cycle ;
\draw   (400,118) .. controls (400,113.58) and (403.58,110) .. (408,110) -- (472,110) .. controls (476.42,110) and (480,113.58) .. (480,118) -- (480,142) .. controls (480,146.42) and (476.42,150) .. (472,150) -- (408,150) .. controls (403.58,150) and (400,146.42) .. (400,142) -- cycle ;
\draw [color={rgb, 255:red, 0; green, 0; blue, 0 }  ,draw opacity=1 ]   (80,110) -- (137.5,71.66) ;
\draw [shift={(140,70)}, rotate = 146.31] [fill={rgb, 255:red, 0; green, 0; blue, 0 }  ,fill opacity=1 ][line width=0.08]  [draw opacity=0] (10.72,-5.15) -- (0,0) -- (10.72,5.15) -- (7.12,0) -- cycle    ;
\draw [color={rgb, 255:red, 0; green, 0; blue, 0 }  ,draw opacity=1 ]   (180,90) -- (180,167) ;
\draw [shift={(180,170)}, rotate = 270] [fill={rgb, 255:red, 0; green, 0; blue, 0 }  ,fill opacity=1 ][line width=0.08]  [draw opacity=0] (10.72,-5.15) -- (0,0) -- (10.72,5.15) -- (7.12,0) -- cycle    ;
\draw [color={rgb, 255:red, 0; green, 0; blue, 0 }  ,draw opacity=1 ]   (210,190) -- (297,190) ;
\draw [shift={(300,190)}, rotate = 180] [fill={rgb, 255:red, 0; green, 0; blue, 0 }  ,fill opacity=1 ][line width=0.08]  [draw opacity=0] (10.72,-5.15) -- (0,0) -- (10.72,5.15) -- (7.12,0) -- cycle    ;
\draw [color={rgb, 255:red, 0; green, 0; blue, 0 }  ,draw opacity=1 ]   (330,170) -- (330,93) ;
\draw [shift={(330,90)}, rotate = 90] [fill={rgb, 255:red, 0; green, 0; blue, 0 }  ,fill opacity=1 ][line width=0.08]  [draw opacity=0] (10.72,-5.15) -- (0,0) -- (10.72,5.15) -- (7.12,0) -- cycle    ;
\draw [color={rgb, 255:red, 0; green, 0; blue, 0 }  ,draw opacity=1 ]   (370,70) -- (437.4,108.51) ;
\draw [shift={(440,110)}, rotate = 209.74] [fill={rgb, 255:red, 0; green, 0; blue, 0 }  ,fill opacity=1 ][line width=0.08]  [draw opacity=0] (10.72,-5.15) -- (0,0) -- (10.72,5.15) -- (7.12,0) -- cycle    ;

\draw (61,131) node [anchor=north west][inner sep=0.75pt]   [align=left] {Kerr};
\draw (54,112.4) node [anchor=north west][inner sep=0.75pt]    {$D=4$};
\draw (154,62.4) node [anchor=north west][inner sep=0.75pt]    {$D=5$};
\draw (154,182.4) node [anchor=north west][inner sep=0.75pt]    {$D=3$};
\draw (314,62.4) node [anchor=north west][inner sep=0.75pt]    {$D=5$};
\draw (314,182.4) node [anchor=north west][inner sep=0.75pt]    {$D=3$};
\draw (414,112.4) node [anchor=north west][inner sep=0.75pt]    {$D=4$};
\draw (407,132) node [anchor=north west][inner sep=0.75pt]   [align=left] {Kerr-Sen};
\draw (228,162.4) node [anchor=north west][inner sep=0.75pt]    {$O( 2,2)$};

\end{tikzpicture}
    \caption{A schematic depiction of the series of uplifts, field redefinitions, and transformations we perform to obtain the Kerr-Sen solution.}
    \label{fig:OddProcedure}
\end{figure}
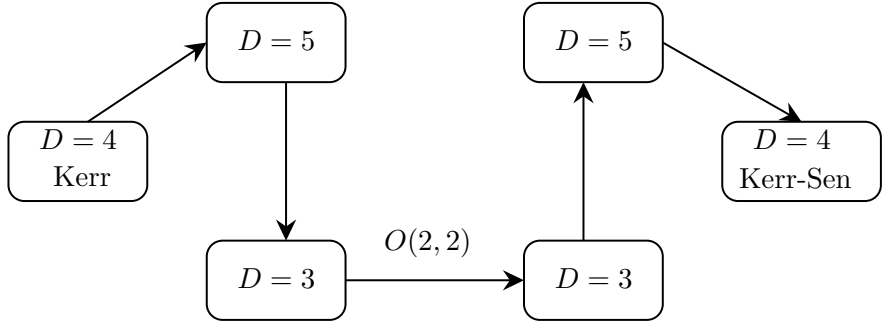

Since this whole process becomes rather technically involved, we give a brief outline of our procedure here:
\begin{enumerate}
    \item Start with the 4d string frame Kerr solution and embed it into 5d. Because $\mathcal A=0$, there are no field redefinitions required.
    \item Reduce to 3d and do the appropriate field redefinitions to make the $O(2,2)$ symmetry manifest.
    \item Do an $O(2,1)\subset O(2,2)$ transformation to get the boosted solution in 3d.
    \item Field redefine back to the frame without $O(2,2)$ covariance and uplift to 5d.
    \item Reduce back down to 4d.
    \item Field redefine to the BdR frame.
\end{enumerate}
This is represented schematically in Figure \ref{fig:OddProcedure}.

\let\oldaddcontentsline\addcontentsline
\renewcommand{\addcontentsline}[3]{}
\subsection*{Notation}
To summarize our notation, we use $\Phi$, $\phi$, and $\bar\phi$ for the five-, four-, and three-dimensional dilaton, respectively. $\hat B$, $B$, and $b$ denote the two-form field in five, four, and three dimensions, respectively, with corresponding field strengths $\hat H$, $H$, and $h$. The spin connection is denoted by $\hat\Omega$, $\Omega$, and $\omega$ in five, four, and three dimensions, respectively, and likewise for the torsionful spin connections $\hat\Omega_\pm$, $\Omega_\pm$, and $\omega_\pm$. We use capital Latin letters $M,N,P,...$ for 5d curved indices, hatted Greek letters $\hat\mu,\hat\nu,\hat\rho,...$ for 4d curved indices, and unhatted Greek letters $\mu,\nu,\rho,...$ for 3d curved indices. The 3d coordinates $x^\mu$ are taken to be $\{r,x,y\}$, while in 4d $\hat x^{\hat\mu}$ corresponds to $\{t,r,x,y\}$ and the 5d coordinates are $X^M=\{t,r,x,y,z\}$. These are related in reductions from 5 to 3 dimensions as $X^M=\{x^\mu,y^i\}$ with $y^i=\{t,z\}$, from 5 to 4 dimensions as $X^M=\{\hat x^{\hat\mu},z\}$, and from 4 to 3 dimensions as $\hat x^{\hat\mu}=\{x^\mu,t\}$.
\let\addcontentsline\oldaddcontentsline

\subsection{3d effective action}
The four-dimensional four-derivative Kerr solution of Section \ref{sec:Kerr} can be embedded trivially as a solution to the 5d heterotic action (without gauge fields)
\begin{align}
    e^{-1}\mathcal L=e^{-2\Phi}\qty[R(\hat\Omega)+4(\partial\Phi)^2-\frac{1}{12}\tilde H_{MNP}^2+\frac{\alpha'}{8}R(\hat\Omega_-)_{MNAB}^2+\mathcal O(\alpha'^3)],
\end{align}
by setting~\cite{Liu:2023fqq}
\begin{equation}
    \dd s_5^2=\dd s_4^2+\dd z^2,\qquad \hat B=\frac{1}{2}B_{\hat\mu\hat \nu}\dd\hat x^{\hat \mu}\land\dd\hat x^{\hat \nu},\qquad \Phi=\phi.
\end{equation}

Since the Kerr solution is time-independent, we are free to reduce along the time direction, as well as the extra $z$-direction that we have added. Following~\cite{Jayaprakash:2024xlr}, we ansatz
\begin{align}
    \dd s_5^2&=g_{\mu\nu}\dd x^\mu\dd x^\nu+g_{ij}\eta^i\eta^j,\qquad \eta^i=\dd y^i+A^i_\mu\dd x^\mu,\nn\\
    \hat B&=\frac{1}{2}b_{\mu\nu}\dd x^\mu\land\dd x^\nu+B_{\mu i}\dd x^\mu\land\eta^i+\frac{1}{2}b_{ij}\eta^i\land\eta^j,\nn\\
    \Phi&=\bar\phi+\frac{1}{4}\log|\det g_{ij}|,\label{eq:5to3ansatz}
\end{align}
where $y^i=\{t,z\}$. $A_\mu^i$ should be thought of as a principal $U(1)^2$ connection with curvature $F^i=\dd A^i$, $g_{\mu\nu}$ the effective three-dimensional metric, and $g_{ij}$ a symmetric matrix of scalars. It then follows that
\begin{equation}
    H=\dd B=\frac{1}{6}h_{\mu\nu\rho}\dd x^\mu\land\dd x^\nu\land\dd x^\rho+\frac{1}{2}\tilde G_{\mu\nu i}\dd x^\mu\land\dd x^\nu\land\eta^i+\frac{1}{2}\dd b_{ij}\eta^i\land\eta^j,
\end{equation}
where
\begin{align}
    h=\dd b-B_i\land F^i,\qquad \tilde G_i=G_i-b_{ij}F^j,\qquad G_i=\dd B_i\,,
\end{align}
The scalar kinetic terms are given by
\begin{align}
    P^{(-+)}_{ij}=P^{(+-)}_{ji}=\frac{1}{2}\dd(g_{ij}+b_{ij}).
\end{align}
Choosing a dreibein $g_{\mu\nu}=e_\mu^\alpha e_\nu^\beta\delta_{\alpha\beta}$ and a scalar ``zweibein'' $g_{ij}=e_i^a e_j^b\eta_{ab}$, there are also composite $O(1,1)\times O(1,1)$ connections
\begin{align}
    Q^{(\pm\pm)}_{ij}&=\frac{1}{2}\qty(e^T\dd e-\dd e^T\,e\mp\dd b)_{ij}.
\end{align}
Notably, the form of $\tilde G_i$ tells us that $B_{\mu i}$ does not transform merely as a gauge field but rather has extra structure arising from the two-group gauge symmetry of the $B$ field. 

This yields an effective three-dimensional action describing the gravity multiplet coupled to two vector multiplets. From the supersymmetry variations, one finds that $F^{(+)\,i}$ is part of the graviton multiplet and $(F^{(-)\,i},P^{(-+)}_{ij})$ are part of the vector multiplets~\cite{Liu:2023fqq}, where
\begin{equation}
    F^{(\pm)i}=F^i\pm g^{ij}\tilde G_j.
\end{equation}
Note that the $(\pm)$ indices on $P^{(\pm\mp)}$, $Q^{(\pm\pm)}$, and $F^{(\pm)}$ indicate that the (rigid) indices transform under $O(1,1)_{\pm}$ (see~\cite{Jayaprakash:2024xlr} for more details).

After performing the field redefinitions of the form $\Phi\to\Phi+\alpha'\delta\Phi$~\cite{Jayaprakash:2024xlr},\footnote{Note that the $B_{\mu i}$ field redefinition is non-covariant.}\footnote{Here we denote $F_i\equiv g_{ij}F^j$.}
\begin{align}
    \delta g_{ij}&=\frac{1}{16}F^{(+)}_{\mu\nu i}F^{(+)}_{\mu\nu j}+\frac{1}{2}P_{\mu ki}^{(-+)}P_{\mu lj}^{(-+)}g^{kl},\nn\\
    \delta B_{\mu i}&=\frac{1}{4}\qty(\frac{1}{2}\omega_{-\mu}^{\alpha\beta}F^{(+)}_{\alpha\beta i}+F^{(-)\,j}_{\mu\nu}P^{(-+)\,\nu}{}_{ji})\,,
\end{align}
we are left with the $O(2,2)$ covariant action,
\begin{align}
    \mathcal L_{O(2,2)}&=\mathcal L_{2\partial}+\frac{\alpha'}{8}\qty(\mathcal L_1+\mathcal L_2+\mathcal L_3),\nn\\
     e^{-1}\mathcal L_{2\partial}&=e^{-2\bar\phi}\left[R(\omega)+4(\partial\bar\phi)^2-\frac{1}{12}h^2-\frac{1}{4}\qty(g_{ij}F^i_{\mu\nu}F^{\mu\nu j}+g^{ij}\tilde G_{\mu\nu i}\tilde G^{\mu\nu}_ j)\right.\nn\\
    &\qquad\qquad\left.-\frac{1}{4}\Tr\qty(g^{-1}\partial_\mu g g^{-1}\partial^\mu g-g^{-1}\partial_\mu b g^{-1}\partial^\mu b)\right],\label{eq:OddActionEff3d}
\end{align}
where the gravity multiplet couplings are given by
\begin{align}
e^{-1} \mathcal{L}_1&=e^{-2 \bar\phi}\left[\left(R_{\alpha \beta \gamma \delta}\left(\omega_{-}\right)\right)^2-\frac{1}{3} h_{\alpha \beta \gamma} \omega_{3 L}\left(\omega_{-}\right)_{\alpha \beta \gamma}-R_{\alpha \beta \gamma \delta}\left(\omega_{-}\right) F_{\alpha \beta}^{(+) a} F_{\gamma \delta}^{(+) a}\right. \nn\\
& \qquad\qquad+\frac{1}{2}\left(\mathcal{D}_\gamma^{(-)} F_{\alpha \beta}^{(+) a}\right)^2+\frac{1}{8} F_{\alpha \beta}^{(+) a} F_{\beta \gamma}^{(+) a} F_{\gamma \delta}^{(+) b} F_{\delta \alpha}^{(+) b}-\frac{1}{8} F_{\alpha \beta}^{(+) a} F_{\beta \gamma}^{(+) b} F_{\gamma \delta}^{(+) a} F_{\delta \alpha}^{(+) b} \nn\\
& \left.\qquad\qquad+\frac{1}{8} F_{\alpha \beta}^{(+) a} F_{\alpha \beta}^{(+) b} F_{\gamma \delta}^{(+) a} F_{\gamma \delta}^{(+) b}\right],
\end{align}
the vector multiplet couplings are given by
\begin{align}
e^{-1} \mathcal{L}_2&=e^{-2 \bar\phi}\left[\frac{1}{8} F_{\alpha \beta}^{(-) a} F_{\alpha \beta}^{(-) b} F_{\gamma \delta}^{(-) a} F_{\gamma \delta}^{(-) b}-\frac{1}{4} F_{\alpha \beta}^{(-) a} F_{\beta \gamma}^{(-) b} F_{\gamma \delta}^{(-) a} F_{\delta \alpha}^{(-) b}\right.\nn \\
& \qquad\qquad+\frac{1}{8} F_{\alpha \beta}^{(-) a} F_{\beta \gamma}^{(-) a} F_{\gamma \delta}^{(-) b} F_{\delta \alpha}^{(-) b}+\left(\mathcal{D}_\mu^{\prime(-)} F_{\nu \gamma}^{(-) a}\right)^2-\mathcal{D}_\mu^{\prime(-)} F_{\nu \gamma}^{(-) a} \mathcal{D}_\nu^{\prime(-)} F_{\mu \gamma}^{(-) a}\nn \\
& \qquad\qquad+P_\gamma^{(-+) a c} P_\gamma^{(-+) b c} F_{\alpha \beta}^{(-) a} F_{\alpha \beta}^{(-) b}-3 P_\alpha^{(-+) a c} P_\beta^{(-+) b c} F_{\alpha \gamma}^{(-) b} F_{\beta \gamma}^{(-) a} \nn\\
& \qquad\qquad+P_\alpha^{(-+) a c} P_\beta^{(-+) b c} F_{\alpha \gamma}^{(-) a} F_{\beta \gamma}^{(-) b}+P_\alpha^{(-+) a b} P_\beta^{(-+) a b} F_{\alpha \gamma}^{(-) c} F_{\beta \gamma}^{(-) c} \nn\\
& \qquad\qquad+2 P_\alpha^{(-+) a b} P_\alpha^{(-+) c b} P_\beta^{(-+) c d} P_\beta^{(-+) a d}+6 P_\alpha^{(-+) a b} P_\beta^{(-+) c b} P_\beta^{(-+) c d} P_\alpha^{(-+) a d} \nn\\
& \qquad\qquad+2 P_\alpha^{(-+) a b} P_\beta^{(-+) a b} P_\alpha^{(-+) c d} P_\beta^{(-+) c d}-6 P_\alpha^{(-+) a b} P_\beta^{(-+) c b} P_\alpha^{(-+) c d} P_\beta^{(-+) a d} \nn\\
& \qquad\qquad\left.+4\left(\mathcal{D}_\gamma^{(-)} P_\alpha^{(-+) b d}\right)^2\right],
\end{align}
and the mixed gravity-vector couplings are given by
\begin{align}
e^{-1} \mathcal{L}_3&=e^{-2 \bar\phi}\left[-\frac{1}{3} h_{\alpha \mu \nu}\left(\omega_{3}\left(-Q^{(--)}\right)_{\alpha \mu \nu}-3 F_{\alpha \gamma}^{(-) a} \mathcal{D}_\mu^{\prime(-)} F_{\nu \gamma}^{(-) a}\right)-R_{\alpha \beta \gamma \delta}\left(\omega_{-}\right) F_{\alpha \gamma}^{(-) a} F_{\beta \delta}^{(-) a}\right.\nn \\
& \qquad\qquad+\frac{1}{2} F_{\alpha \beta}^{(+) a} F_{\beta \gamma}^{(-) b} F_{\gamma \delta}^{(+) a} F_{\delta \alpha}^{(-) b}+\frac{1}{4} F_{\alpha \beta}^{(+) a} F_{\beta \gamma}^{(+) a} F_{\gamma \delta}^{(-) b} F_{\delta \alpha}^{(-) b} \nn\\
& \qquad\qquad+\frac{3}{2} P_\gamma^{(-+) c a} P_\gamma^{(-+) c b} F_{\alpha \beta}^{(+) a} F_{\alpha \beta}^{(+) b}+P_\alpha^{(-+) a b} P_\beta^{(-+) a b} F_{\alpha \gamma}^{(+) c} F_{\beta \gamma}^{(+) c} \nn\\
& \qquad\qquad+P_\alpha^{(-+) c a} P_\beta^{(-+) c b} F_{\alpha \gamma}^{(+) a} F_{\beta \gamma}^{(+) b}-2 P_\alpha^{(-+) c a} P_\beta^{(-+) c b} F_{\alpha \gamma}^{(+) b} F_{\beta \gamma}^{(+) a} \nn\\
& \qquad\qquad-4 P_\gamma^{(-+) a b} F_{\mu \nu}^{(+) b} \mathcal{D}_\mu^{\prime(-)} F_{\nu \gamma}^{(-) a}-2 P_\beta^{(-+) a d} F_{\gamma \alpha}^{(-) a} \mathcal{D}_\gamma^{(-)} F_{\alpha \beta}^{(+) d} \nn\\
& \left.\qquad\qquad-2 F_{\alpha \beta}^{(+) d} F_{\gamma \beta}^{(-) b} \mathcal{D}_\gamma^{(-)} P_\alpha^{(-+) b d}\right].
\end{align}
Here, $\mathcal D^{(-)}$ is the covariant derivative with respect to both the torsionful spin connection $\omega_-$ and the composite connection $Q$, while $\mathcal D^{\prime(-)}$ acts as
\begin{equation}
    \mathcal D^{\prime(-)}_\mu F^{(-)\,a}_{\nu\gamma}=\partial_\mu F^{(-)\,a}_{\nu\gamma}-\Gamma_{\mu\nu}^\rho F^{(-)\,a}_{\rho\gamma}+\omega_{-\mu\gamma}{}^{\delta}F^{(-)\,a}_{\nu\delta}-Q_{\mu}^{(--)\,a}{}_bF^{(-)\,b}_{\nu\gamma}.
\end{equation}
Note also that we have defined
\begin{equation}
    F^{(\pm)\,a}\equiv e^a_i F^{(\pm)\,i},\qquad Q_{\mu ab}^{(\pm\pm)}=e_a^iQ_{\mu ij}^{(\pm\pm)}e_b^j,\qquad P_{\mu ab}^{(\mp\pm)}=e_a^iP_{\mu ij}^{(\mp\pm)}e_b^j.
\end{equation}
Notice that $\mathcal L_{O(2,2)}$ is not invariant under the interchange of $O(1,1)_-$ and $O(1,1)_+$. This should come as no surprise since the heterotic string is not invariant under worldsheet parity transformations, $B\to -B$.

\subsection{$O(2,2)$ and boosting the solution}
To make the $O(2,2)$ invariance of the effective 3d action more explicit, we define the $O(2,2)$ covariant objects
\begin{align}
    \mathcal H&=\begin{pmatrix}
        g_{ij}-b_{ik}g^{kl}b_{lj} & \quad b_{ik}g^{kj}\\
        -g^{ik}b_{kj} & \quad g^{ij}
    \end{pmatrix},\qquad
    \mathbb A_\mu=\begin{pmatrix}
        A^i_\mu\\B_{\mu i}
    \end{pmatrix},\qquad\mathbb F=\dd\mathbb A,\qquad \eta = \begin{pmatrix}
        0&\quad\openone\\
        \openone&\quad0
    \end{pmatrix},
\end{align}
which transform as
\begin{equation}
    \mathcal H\to(\Omega^{-1})^T\mathcal H\Omega^{-1},\qquad \mathbb{A}\to \Omega\mathbb A,\qquad \Omega^T\eta\Omega=\eta.
\end{equation}
The last equation is simply the statement that $\Omega\in O(2,2)$. Note that $\mathcal H$ satisfies
\begin{equation}
    \eta\mathcal H\eta = \mathcal H^{-1},
\end{equation}
which implies that $\mathcal H\in O(2,2)$ parametrizes the coset $O(2,2)/(O(1,1)\times O(1,1))$. 

The two-derivative action can straightforwardly be put into an $O(2,2)$ covariant form
\begin{align}
    e^{-1}\mathcal L_{2\partial}&=e^{-2\bar\phi}\left[R+4(\partial\bar\phi)^2-\frac{1}{12} h^2-\frac{1}{4}\mathbb F^T_{\mu\nu}\mathcal H\mathbb F^{\mu\nu}+\frac{1}{8}\Tr\qty(\partial_\mu\mathcal H\eta\partial^\mu\mathcal H\eta)\right],
\end{align}
such that
\begin{equation}
    h=\dd b-\frac{1}{2}\mathbb A^T\land\eta\mathbb F.
\end{equation}
This form of $h$ implicitly requires a shift of $b$ by $A^i\land B_i$, although this shift will vanish for our setup. The four-derivative action is written in a manifestly $O(1,1)_-\times O(1,1)_+$ covariant form, but can be rephrased into an explicitly $O(2,2)$ covariant form using the fact that
\begin{equation}
    F^{(\pm)\,a}_{\mu\nu}F^{(\pm)\,a}_{\rho\sigma}=\mathbb F^T_{\mu\nu}(\mathcal H\pm\eta)\mathbb F_{\rho\sigma},
\end{equation}
as well as other related identities presented in Appendix~\ref{app:JimTranslate}. We do not present this form of the action explicitly, as its existence is sufficient for our purposes.

Note that although the three-dimensional dilaton is left invariant under an $O(2,2)$ transformation, the five-dimensional (and hence, also the four-dimensional) dilaton transforms due to the shift
\begin{equation}
    \Phi=\bar\phi+\frac{1}{4}\log|\det g_{ij}|.
\end{equation}

Following Ref.~\cite{Sen:1992ua}, we can choose a particular $O(1,1)\subset O(2,1)\subset O(2,2)$ boost
\begin{equation}
    \Omega=V\,\Omega'\,V^T,\label{eq:boost}
\end{equation}
where
\begin{equation}
    \Omega'=\begin{pmatrix}
        \cosh\beta &\quad \sinh\beta &\quad 0 &\quad 0 \\
 \sinh\beta &\quad \cosh\beta &\quad 0 &\quad 0 \\
 0 &\quad 0 &\quad 1 &\quad 0 \\
 0  &\quad 0 &\quad 0 &\quad 1
    \end{pmatrix},
\end{equation}
is a boost that mixes the time direction with the gauge field torus direction, and the matrix $V$ is just a change of basis matrix
\begin{equation}
    V=\frac{1}{\sqrt{2}}\begin{pmatrix}
        \eta &\quad -\eta \\
 \openone &\quad \openone
    \end{pmatrix},
\end{equation}
that transforms
\begin{equation}
    V:\eta\to\begin{pmatrix}
        \eta_{ij}&\quad 0\\0&\quad-\eta_{ij}
    \end{pmatrix}.
\end{equation}

\subsection{Uplifting back to 4d}
Upon doing this $O(2,1)$ boost, we can then field redefine back
\begin{align}
    \delta g_{ij}&=-\frac{1}{16}F^{(+)}_{\mu\nu i}F^{(+)}_{\mu\nu j}-\frac{1}{2}P_{\mu ki}^{(-+)}P_{\mu lj}^{(-+)}g^{kl},\nn\\
    \delta B_{\mu i}&=-\frac{1}{4}\qty(\frac{1}{2}\omega_{-\mu}^{\alpha\beta}F^{(+)}_{\alpha\beta i}+F^{(-)\,j}_{\mu\nu}P^{(-+)\,\nu}{}_{ji})\,,
\end{align}
and uplift to 5d via the opposite transformation of \eqref{eq:5to3ansatz}. Note that these are now the fields of the $O(2,1)$-boosted solution.

Once this is done, we can reduce to 4d 
\begin{align}
    \dd s^2_5&=g_{\hat \mu\hat \nu}\dd\hat x^{\hat\mu}\dd\hat x^{\hat\nu}+e^{2\sigma}\qty(\dd z+\frac{1}{\sqrt{2}}\mathcal A_{\hat \mu}\dd\hat x^{\hat \mu})^2,\nn\\
    \hat B&=\frac{1}{2}B_{\hat\mu\hat\nu}\dd\hat x^{\hat \mu}\land\dd\hat x^{\hat \nu}+\frac{1}{\sqrt{2}}\mathcal B_{\hat \mu}\dd\hat x^{\hat \mu}\land \qty(\dd z+\frac{1}{\sqrt{2}}\mathcal A_{\hat \mu}\dd\hat x^{\hat \mu}),\nn\\
    \Phi&=\phi+\frac{1}{2}\sigma,
\end{align}
and field redefine~\cite{Liu:2023fqq}
\begin{align}
    \delta \sigma &=\frac{1}{32}\mathcal F^{(+)}_{\hat\mu\hat\nu}\mathcal F^{(+)}_{\hat\mu\hat\nu}+\frac{1}{4}\partial_{\hat\mu}\sigma\partial^{\hat\mu}\sigma,\nn\\
    \delta\mathcal B_{\hat\mu}&=\frac{1}{4}e^{2\sigma}\qty(\frac{1}{2}\Omega_{-\hat\mu}^{\hat\alpha\hat\beta}\mathcal F^{(+)}_{\hat\alpha\beta}+\mathcal F^{(-)}_{\mu\nu}\partial^\nu\sigma),
\end{align}
where
\begin{equation}
    \mathcal F^{(\pm)}=\mathcal F\pm e^{-2\sigma}\dd\mathcal B.
\end{equation}
This then admits a consistent truncation
\begin{equation}
    \sigma=0,\qquad \mathcal B=\mathcal A.
\end{equation}
In principle, we could stop here and have a solution to the 4d action
\begin{align}
    e^{-1}\mathcal L&=e^{-2\phi}\left\{R(\Omega)+4(\partial\phi)^2-\frac{1}{12}\tilde H^2-\frac{1}{4}\mathcal F^2\right.\nn\\
    &\qquad\qquad\left.+\frac{\alpha'}{8}\qty[R(\Omega_-)_{\hat\mu\hat\nu\hat\rho\hat\sigma}^2-2R(\Omega_-)_{\hat\mu\hat\nu\hat\rho\hat\sigma}\mathcal F^{\hat\mu\hat\nu}\mathcal F^{\hat\rho\hat\sigma}+(\nabla^{(-)} \mathcal F)^2+\frac{1}{2}(\mathcal F^2)^2]+\mathcal O(\alpha'^2)\right\},
\end{align}
where $\nabla^{(-)}$ is covariant with respect to the torsionful connection $\Omega_-$ and
\begin{equation}
    \tilde H=\dd B-\frac{1}{2}\mathcal A\land\mathcal F+\frac{\alpha'}{4}\omega_{3L}(\Omega_-).
\end{equation}
However, we do one final field redefinition~\cite{Liu:2023fqq,Cai:2025yyv}
\begin{align}
    \delta\mathcal A_{\hat\nu}=-\frac{1}{4}e^{2\phi}\nabla^{\hat\mu}\qty(e^{-2\phi}\mathcal F_{\hat\mu\hat\nu}),\qquad \delta g_{\hat\mu\hat\nu}=-\frac{1}{4}\mathcal F^2_{\hat\mu\hat\nu},\qquad \delta\phi&=-\frac{1}{16}\mathcal F^2,\qquad \delta B_{\mu\nu}=\frac{1}{2}\mathcal A_{[\hat\mu}\delta\mathcal A_{\hat\nu]},
\end{align}
to transform to a solution of the action
\begin{align}
    e^{-1}\mathcal L&=e^{-2\phi}\left\{R(\Omega)+4(\partial\phi)^2-\frac{1}{12}\tilde H^2-\frac{1}{4}\mathcal F^2+\frac{\alpha'}{8}\qty[R(\Omega_-)_{\hat\mu\hat\nu\hat\rho\hat\sigma}^2-R(\Omega_-)_{\hat\mu\hat\nu\hat\rho\hat\sigma}\mathcal F^{\hat\mu\hat\nu}\mathcal F^{\hat\rho\hat\sigma}]+\mathcal O(\alpha'^2)\right\},\label{eq:BdR4dGaugeFields}
\end{align}
which is the canonical BdR field redefinition frame for heterotic supergravity.

\subsection{The four-derivative Kerr-Sen solution}
We now apply the procedure outlined in the previous subsections to the four-derivative corrected Kerr solution. Due to the computational complexity, we omit the intermediate results and present only the final solution in the string frame:
\begin{align}
    g'_{tt}&=-\frac{\Sigma\tilde\Delta}{\Upsilon^2}\qty(1+2\frac{\Xi}{\Upsilon}\phi)+\frac{\alpha' \mu^2}{2 \Sigma^2 \Upsilon^4}\bigg[-2 \varpi \cosh\beta \sinh ^2\frac{\beta }{2}\left(\Upsilon ^2-\mu ^2 r^2 \sinh ^2\beta \right)\nn\\
    &\quad+\left(4 a^2 r^2 x^2 (2
   \Delta -\tilde \Delta) \Sigma +2 \mu  r \Xi  \varpi +(\Delta -\mu  r) \Sigma  \varpi \right) \sinh ^2\beta \bigg],\nn\\
    g'_{rr}&=\frac{\Sigma}{\Delta}\qty(1+2\phi)-\alpha'\frac{\mu^2 \sinh ^2\beta \left(r^2-a^2 x^2\right)^2}{2 \Delta \Sigma\Upsilon^2},\nn\\
   g'_{xx}&=\frac{\Sigma}{1-x^2}\qty(1+2\phi)+\alpha'\frac{2 a^2 \mu^2 r^2 x^2 \sinh ^2\beta}{\left(1-x^2\right) \Sigma \Upsilon^2},\nn\\
   g_{ty}'&=-\frac{2 a \mu r \left(1-x^2\right) \cosh ^2\frac{\beta }{2} \Sigma}{\Upsilon^2}\qty(1+2\frac{\Xi}{\Upsilon}\phi)-\frac{\alpha'\tilde\Delta\Sigma\Lambda\sinh^2\frac{\beta}{2}}{\Upsilon^2}\nn\\
   &\quad-\frac{\alpha'a\mu\sinh^2\frac{\beta}{2}}{2\Sigma^2\Upsilon^4}\bigg[\Sigma  \tilde\Delta^2 \left(2 r \Delta +\left(1-x^2\right) \left(r^2 (5 \mu -3 r)-a^2 \left(r \left(2-x^2\right)+\mu 
   x^2\right)\right)\right)\nn\\
   &\qquad\qquad+4\mu\cosh^2\frac{\beta}{2}\bigg(\Sigma ^2 \left(2 x^2 \Sigma ^2+a^2 \left(1+3 x^2-6 x^4\right) \Sigma -4 a^4 x^4 \left(1-x^2\right)\right)\nn\\
   &\qquad\qquad\quad-4 \mu  r x^2 \Sigma ^2 \left(2
   r^2-a^2 \left(-1+x^2\right)\right)+4 \mu ^2 r^2 x^2 \left(2 r^4-a^2 r^2 \left(3-5 x^2\right)+a^4 x^2 \left(1-x^2\right)\right)\bigg)\nn\\
   &\qquad\qquad + 4\mu^2r\cosh^4\frac{\beta}{2}\bigg(2 r^2 \left(r^2+a^2\right) \Delta -\left(1-x^2\right) \left(-2 a^6 x^6+r^5 (r-3 \mu )\right.\nn\\
   &\qquad\qquad\quad\left.+4 a^2 r^3 \left(r+r x^2-\mu -4 \mu  x^2\right)+a^4 r
   \left(r \left(2+2 x^2+x^4\right)+3 x^4 \mu \right)\right)\bigg)\bigg],\nn\\
   g_{yy}'&=\frac{(1-x^2)\Sigma}{\Upsilon^2}\qty(\left(r^2+a^2\right) \Sigma +2 \mu r a^2 \left(1-x^2\right)+4 \mu r \left(r^2+a^2\right) \sinh ^2\frac{\beta }{2}+4 \mu^2 r^2 \sinh
   ^4\frac{\beta }{2})\nn\\
   &\quad +\frac{2\left(1-x^2\right) \Sigma  \phi}{\Upsilon ^3} \left[\Delta \tilde\Delta \Upsilon +4\mu r \Delta    \Upsilon  \cosh ^2\frac{\beta }{2}+4 \mu ^2 r^2 \left(\Upsilon +2 a^2
   \left(1-x^2\right)\right) \cosh ^4\frac{\beta }{2}\right.\nn\\
   &\qquad\qquad\left.-8 \mu ^2 r^2 a^2 \left(1-x^2\right) \cosh ^6\frac{\beta }{2}\right]-\frac{\alpha' a \mu r \left(1-x^2\right) \Sigma  \Lambda  \sinh ^2\beta}{\Upsilon ^2}\nn\\
   &\qquad+\frac{\alpha'a^2\mu^2(1-x^2)\sinh^2\beta}{2\Sigma^2\Upsilon^4}\bigg[\Sigma ^3 \left(2 r^2+\left(1-x^2\right) \left(a^2-\mu r\right)\right)\nn\\
   &\qquad\qquad+2 \mu  r^2 \left(2 r \left(1+x^2\right) \Sigma ^2-\left(1-x^2\right)
   \varpi  \mu \right) \sinh ^2\frac{\beta }{2}\nn\\
   &\qquad\qquad+4 \mu ^2 r^2 x^2 \left(2 r^2 \left(r^2+a^2\right)+a^2 \left(1-x^2\right)
   \left(r^2-a^2 x^2\right)\right) \sinh ^4\frac{\beta }{2}\bigg],\nn\\
   \phi'&=-\frac{1}{2}\ln\frac{\Upsilon}{\Sigma}+\frac{\Sigma\phi}{\Upsilon}\cosh^2\frac{\beta}{2}+\alpha'\frac{\mu^2  \varpi \qty(\Xi\sinh ^2\frac{\beta }{2}-\Sigma\sinh^2\beta)}{4 \Sigma ^3 \Upsilon^2},\nn\\
   \mathcal A_t'&=\frac{\sqrt{2}\mu r\sinh\beta}{\Upsilon}-\frac{\sqrt{2}\tilde\Delta\Sigma\sinh\beta}{\Upsilon^2}\phi\nn\\
   &\quad+\alpha'\frac{\mu^2 \sinh\beta\qty[\varpi \left(2 \mu r \sinh ^2\frac{\beta }{2}-2 \Xi +\Sigma \right)-2a^2\qty(4r^2x^2\tilde\Delta+(1-x^2)\Sigma^2)\sinh^2\frac{\beta}{2}]}{2
   \sqrt{2} \Sigma ^2 \Upsilon ^3},\nn\\
   \mathcal A_y'&=-\frac{\sqrt{2}a \mu r (1-x^2)\sinh\beta}{\Upsilon}\qty(1+\frac{2\Sigma\cosh^2\frac{\beta}{2}}{\Upsilon}\phi)-\frac{\alpha'\Sigma\Lambda\sinh\beta}{\sqrt{2}\Upsilon}\nn\\
   &\quad+\frac{\alpha'a \mu \sinh\beta}{2\sqrt{2}\Sigma^2\Upsilon^3}\bigg[\Upsilon  \bigg(-\Sigma  \left(2 r \Delta +\left(1-x^2\right) \left(r^2 (5 \mu -3 r)-a^2 \left(r \left(2-x^2\right)+\mu 
   x^2\right)\right)\right)\nn\\
   &\qquad\quad\ \qquad\qquad\qquad+2 \mu  r^2 \left(r^2 \left(1-3 x^2\right)-a^2 x^2 \left(3-x^2\right)\right) \cosh ^2\frac{\beta
   }{2}\bigg)\nn\\
   &\qquad\qquad+2 \left(1-x^2\right) \mu  \sinh ^2\frac{\beta }{2} \left(\Delta  \Sigma ^2+2 \mu  r \left(r^2-a^2
   x^2\right)^2 \cosh ^2\frac{\beta }{2}\right)-2 \mu ^2 r \left(1-x^2\right) \varpi  \sinh ^2\beta \bigg],\nn\\
   B_{ty}'&=\frac{2\mu a  r}{\Upsilon}(1-x^2)\sinh^2\frac{\beta}{2}\qty(1+\frac{2\Sigma}{\Upsilon}\cosh^2\frac{\beta}{2}\phi)+\frac{\alpha'\Sigma\Lambda}{\Upsilon }\cosh ^2\frac{\beta }{2}\nn\\
   &\quad+\frac{\alpha' a \mu \sinh^2\frac{\beta}{2}}{4\Sigma^2\Upsilon^3}\bigg[-2 \tilde\Delta \Sigma  \left(a^2 x^2 r \left(3-x^2\right)-r^3 \left(1-3 x^2\right)-\mu  a^2 x^2 \left(1-x^2\right)+\mu  r^2 \left(1-5
   x^2\right)\right)\nn\\
   &\qquad\qquad\qquad\qquad-8 \mu ^2 r \left(a^2 r^2 x^2 \left(1-15
   x^2\right)+2a^4 x^6+r^4(1-x^2)-\left(1+x^2\right) \varpi \right) \cosh ^4\frac{\beta }{2}\nn\\
   &\qquad\qquad\qquad\qquad+4 \mu ^2 r \left(1-x^2\right) \qty(2(r^2-a^2x^2)^2-3\varpi)  \cosh ^2\frac{\beta }{2}\bigg],\label{eq:KerrSen4der}
\end{align}
where we have defined
\begin{align}
    \Upsilon&=r^2+a^2 x^2+2\mu r \sinh^2\frac{\beta}{2},\nn\\
    \Xi&=  \left(r^2+a^2
   x^2-\mu r\right)\cosh \beta+\mu r,\nn\\
   \varpi&=r^4-6r^2a^2x^2+a^4x^4,\nn\\
   \tilde\Delta&=r^2+a^2x^2-2\mu r.
   \end{align}
This is a solution to the action~\eqref{eq:BdR4dGaugeFields}. For ease of use, the solution is included in the companion Mathematica file. Note that $\phi$ in this expression refers to the original dilaton~\eqref{eq:ogScalars} in the corrected Kerr solution. Recall also that $\phi=\mathcal O(\alpha')$. Although $\phi$ and $\Lambda$ themselves are known only as a series in $\chi$, the solution~\eqref{eq:KerrSen4der} is a closed form expression in terms of $\phi$ and $\Lambda$. Note also that $\beta\to-\beta$ transforms $\mathcal A\to-\mathcal A$ and leaves the other fields unchanged, which reflects the fact that the action \eqref{eq:BdR4dGaugeFields} is invariant under $\mathcal A\to-\mathcal A$. We see that this solution indeed reduces to the four-derivative Kerr solution when $\beta\to0$ and to the two-derivative Kerr-Sen solution~\cite{Sen:1992ua} when $\alpha'\to0$.

It is interesting to note that most of the complexity of the solution \eqref{eq:KerrSen4der} arises not from the $O(2,1)$ boost, but rather from the many field redefinitions we had to perform before and after boosting. Only the four-derivative terms proportional to $\phi$ and $\Lambda$ come from the $O(2,1)$ boost itself, whereas all other terms come from field redefinitions. In some sense, this complexity is not so surprising. After all, these are four-derivative corrections to a charged, rotating, non-extremal black hole. 

Notice also that since $g_{rr}\propto\Delta^{-1}$, the horizon position remains unchanged
\begin{equation}
    r_{\pm}=\mu\pm\sqrt{\mu^2-a^2},
\end{equation}
despite the four-derivative corrections.

We also see that the limit $a\to 0$ corresponds to a static charged black hole that corresponds to the $O(2,1)$ boost of the Schwarzschild solution. This yields the exact $\alpha'$-corrected Gibbons-Maeda-Garfinkle-Horowitz-Strominger (GMGHS) solution~\cite{Gibbons:1982ih,Gibbons:1987ps,Garfinkle:1990qj}, which reads
\begin{align}
    g'_{tt}&=-\frac{r(r-2\mu)}{\qty(r+2\mu \sinh^2\frac{\beta}{2})^2}\qty[1-\alpha'\frac{\left(-2 \mu ^2+3 r^2+3 \mu  r\right)\mu+\left(2 \mu ^3+3 r^3+\mu ^2 r\right)\cosh\beta}{12\mu r^3\qty(r+2\mu\sinh^2\frac{\beta}{2})}],\nn\\
    g'_{rr}&=\frac{1}{1-\frac{2\mu}{r}}\qty[1-\frac{\alpha'}{12 r}\qty(\frac{3}{\mu }+\frac{3r+4 \mu}{r^2}+\frac{6 \mu ^2 \sinh ^2\beta}{r \qty(r+2\mu \sinh^2\frac{\beta}{2})^2})],\nn\\
    g'_{xx}&=\frac{r^2}{1-x^2}\qty[1-\frac{\alpha'}{12\mu r^3}\qty(3 r^2 + 3 r\mu + 4\mu^2)],\nn\\
    g'_{yy}&=r^2(1-x^2)\qty[1-\frac{\alpha'}{12\mu r^3}\qty(3 r^2 + 3 r\mu + 4\mu^2)],\nn\\
    \phi'&=-\frac{1}{2}\log\qty(1+\frac{2\mu\sinh^2\frac{\beta}{2}}{r})\nn\\
    &\quad-\frac{\alpha'}{48r^3}\qty(6 \mu +\frac{3 r^3}{\mu ^2}+\frac{3 r^2}{\mu }+10 r+\frac{6 r^2 (r-2 \mu )}{\qty(r+2\mu \sinh^2\frac{\beta}{2})^2}+\frac{r \left(14 \mu ^3-3 r^3+3 \mu  r^2-10
   \mu ^2 r\right)}{\mu ^2 \qty(r+2\mu \sinh^2\frac{\beta}{2})}),\nn\\
    \mathcal A_t'&=\frac{\sqrt{2}\mu\sinh\beta}{r+2\mu \sinh^2\frac{\beta}{2}}+\frac{\alpha'\qty(3 r^4 - 6 r^3 \mu + r^2 \mu^2 - 10 \mu^4 + \mu (3 r^3 - 3 r^2 \mu - 14 r \mu^2 + 
    10 \mu^3)\cosh\beta)\sinh\beta}{12\sqrt{2}\mu r^2\qty(r+2\mu \sinh^2\frac{\beta}{2})^3},\label{eq:GMGHS}
\end{align}
with $g'_{ty}$ and $B'$ vanishing. Such solutions were constructed numerically in Refs.~\cite{Kanti:1995vq,Ohta:2012ih,Herdeiro:2021gbw}. In principle, the corrections are also known from the ten-dimensional perspective as the equal charge limit of two-charge heterotic black holes~\cite{Massai:2023cis}. Still, as we have seen, there are many non-trivial field redefinitions involved in identifying the four-dimensional fields from the ten-dimensional ones, so, that is to say, the solution of~\cite{Massai:2023cis} differs from ours in choice of field redefinition frame. In the extremal limit, we expect that \eqref{eq:GMGHS} should match the two-equal charge limit of~\cite{Cai:2025yyv}, although it is difficult to make a direct comparison as it seems that the two solutions differ by an (as yet unknown) coordinate transformation. This is due to the fact that the extremal solution of~\cite{Cai:2025yyv} in the field redefinition frame~\eqref{eq:BdR4dGaugeFields} does not take an isotropic form.

\section{Thermodynamics and Multipole moments}\label{sec:multipoles}
We now turn our attention to the experimentally observable properties of the Kerr-Sen black hole. In particular, we will now compute the thermodynamic quantities and multipole moments. All expressions in this section should be understood to be in the Einstein frame, unless specified otherwise.

\subsection{Two-derivative Kerr-Sen}
The two-derivative mass, angular momentum, and electric charge of the Kerr-Sen black hole can be obtained from the asymptotic behavior of the metric and gauge field and are given by
\begin{align}
    M^{(0)}&=\mu\cosh^2\frac{\beta}{2},\qquad\quad\! J^{(0)}=\mu a\cosh^2\frac{\beta}{2},\qquad Q^{(0)}=\frac{\mu}{2\sqrt{2}}\sinh\beta.\label{eq:twoDerivCharges}
\end{align}
These relations can be inverted as
\begin{equation}
    \mu=M^{(0)}-\frac{2(Q^{(0)})^2}{M^{(0)}},\qquad a=\frac{J^{(0)}}{M^{(0)}},\qquad \sinh\beta=\frac{2\sqrt{2}Q^{(0)}M^{(0)}}{(M^{(0)})^2-2(Q^{(0)})^2}.\label{eq:twoDerivChargesInverse}
\end{equation}
The horizon is a Killing horizon for the vector $\xi=\partial_t+\Omega_H\partial_y$. We may compute the horizon velocity via
\begin{equation}
    \Omega_H=-\qty(\frac{g_{ty}}{g_{yy}}\bigg\vert_{r=r_+}-\frac{g_{ty}}{g_{yy}}\bigg\vert_{r=\infty}),\label{eq:horVel}
\end{equation}
which, for the two-derivative solution, reads
\begin{equation}
    \Omega_{H}^{(0)}=\frac{a}{r_+^2+a^2}\sech^2\frac{\beta}{2}.
\end{equation}
We compute the temperature via
\begin{equation}
    T=\frac{\kappa}{2\pi},\qquad \kappa^2=-\frac{g^{\mu\nu}\partial_\mu\xi^2\partial_\nu\xi^2}{4\xi^2}\bigg\vert_{r=r_+},\label{eq:temp}
\end{equation}
which gives us, to leading order, that the temperature is
\begin{equation}
    T^{(0)}=\frac{r_+^2-a^2}{4\pi r_+(r_+^2+a^2)}\sech^2\frac{\beta}{2}.
\end{equation}
The electric potential
\begin{equation}
    \Phi_e=-\xi^\mu\mathcal A_\mu\big\vert^{r=\infty}_{r=r_+}\,,\label{eq:potential}
\end{equation}
reads, for the two-derivative solution,
\begin{equation}
   \Phi_{e}^{(0)}=\sqrt{2}\tanh\frac{\beta}{2}. 
\end{equation}
Finally, the two-derivative entropy is just the area of the horizon,
\begin{equation}
    S^{(0)}=\pi(r_+^2+a^2)\cosh^2\frac{\beta}{2}.
\end{equation}

Next, we turn our attention to the multipole moments. There are three approaches to computing multipole moments due to Geroch-Hansen~\mbox{\cite{Geroch:1970cc,Geroch:1970cd,Hansen:1974zz}}, Thorne~\cite{Thorne:1980ru}, and using covariant phase space~\cite{Compere:2017wrj}, which have been shown to be equivalent~\cite{Cano:2022wwo,Compere:2017wrj}. The Geroch-Hansen approach is elegant and manifestly coordinate invariant, but is also rather involved. We will instead make use of Thorne's formalism~\cite{Thorne:1980ru}, which requires that we perform a coordinate transformation from Boyer-Lindquist coordinates \((r, x)\) to asymptotically Cartesian mass-centered (ACMC-\(\infty\)) coordinates \((r_S, x_S)\) defined by~\cite{Cano:2022wwo,Bena:2020uup}
\be
r_S\sqrt{1-x_S^2}=\sqrt{r^2+a^2}\sqrt{1-x^2},\qquad r_Sx_S=rx\ ,
\label{ACMC leading}
\ee
in terms of which, the metric in the far zone takes the form~\cite{Thorne:1980ru}
\bea
g_{tt}&=&-1+\frac{2M}{r}+\sum_{\ell\geq1}^{\infty}\frac{2}{r^{\ell+1}}\left({M}_\ell P_\ell+\sum_{\ell'<\ell}c_{\ell\ell'}^{(tt)}P_{\ell'}\right),\nn\\
g_{ty}&=&-2r(1-x^2)\left[\sum_{\ell\geq1}^{\infty}\frac{1}{r^{\ell+1}}\left(\frac{{\mathcal{S}}_\ell}{\ell}P'_\ell
+\sum_{\ell'<\ell}c_{\ell\ell'}^{(ty)}P_{\ell'}'\right)
\right],\cr
g_{rr}&=&1+\sum_{\ell\geq0}^{\infty}\frac{1}{r^{\ell+1}}\sum_{\ell'\leq\ell}c_{\ell\ell'}^{(rr)}P_{\ell'},\quad
g_{xx}=\frac{r^2}{1-x^2}\left[1+\sum_{\ell\geq0}^{\infty}\frac{1}{r^{\ell+1}}\sum_{\ell'\leq\ell}c_{\ell\ell'}^{(xx)}P_{\ell'}
\right],\nn\\
g_{yy}&=&r^2(1-x^2)\left[1+\sum_{\ell\geq0}^{\infty}\frac{1}{r^{\ell+1}}\sum_{\ell'\leq\ell}c_{\ell\ell'}^{(yy)}P_{\ell'}
\right],\quad g_{rx}=r\left[\sum_{\ell\geq0}^{\infty}\frac{1}{r^{\ell+1}}\sum_{\ell'\leq\ell}c_{\ell\ell'}^{(rx)}P_{\ell'}'
\right]\ ,
\label{AC-N}
\eea
where $P_\ell$ are the Legendre polynomials as a function of $x$ and $P_\ell'$ denotes the derivative with respect to $x$. Note that, after performing the change of coordinates \eqref{ACMC leading}, we have dropped the $S$ subscript for notational convenience. Here, $M_\ell$ and $\mathcal S_\ell$ are the mass and current multipoles, respectively. In contrast, the $c_{\ell\ell'}^{(ij)}$ coefficients are gauge-dependent and hence non-physical~\cite{Thorne:1980ru}. The two-derivative multipole moments for the Kerr-Sen solution can be read off as
\begin{align}
    M_{2\ell}^{(0)}&=(-a^2)^\ell M^{(0)},\qquad M_{2\ell+1}^{(0)}=0,\nn\\
    \mathcal{S}_{2\ell+1}^{(0)}&=(-a^2)^\ell J^{(0)},\qquad\quad\ \mathcal{S}_{2\ell}^{(0)}=0.
\end{align}
In particular, $M_{0}^{(0)}=M^{(0)}$ is the mass and $\mathcal S_{1}^{(0)}=J^{(0)}$ is the angular momentum. Notably, these relations are independent of $\beta$ for fixed $M^{(0)}$ and $J^{(0)}$, which indicates that, at the two-derivative level, we cannot distinguish a Kerr black hole from a Kerr-Sen black hole through the observation of the multipole moments. Note that these are also the same gravitational multipole moments as for the Kerr-Newmann black hole in Einstein-Maxwell theory~\mbox{\cite{Sotiriou:2004ud,Bena:2020uup,Fodor:2020fnq}}. Therefore, we need to check the contributions from the low-energy effective theory to determine whether corrections to the multipole moments of rotating black holes come from the \(\alpha'\) corrections in string theory.

We may likewise also expand the gauge field into multipole moments as
\begin{align}
    \mathcal A_t&=-\sum_{\ell\geq0}^{\infty}\frac{4}{r^{\ell+1}}\Big({\mathcal{Q}}_\ell P_\ell+\sum_{\ell'<\ell}c_{\ell\ell'}^{(t)}P_{\ell'}\Big)\, ,
\nn\\
\mathcal A_y&=\sum_{\ell\geq0}^{\infty}\left\{\frac{4x}{r^{2\ell}}\Big({\mathcal{P}}_{2\ell} P_{2\ell}+\sum_{\ell'<\ell}c_{2\ell,2\ell'}^{(y,1)}P_{2\ell'}\Big)+\frac{4(1-x^2)}{r^{2\ell}}\sum_{\ell'<\ell}c_{2\ell,2\ell'}^{(y,2)}P_{2\ell'}\right.\nn\\
&\qquad\qquad\left.-\frac{1-x^2}{r^{2\ell+1}}\frac{4}{2\ell+1}\Big({\mathcal{P}}_{2\ell+1} P_{2\ell+1}'+\sum_{\ell'<\ell}(c_{2\ell+1,2\ell'+1}^{(y,1)}P_{2\ell'+1}'
+c_{2\ell+1,2\ell'+1}^{(y,2)}P_{2\ell'+1})\Big)
\right\}.\label{ACMC ele}
\end{align}
The $\mathcal Q_\ell$ are interpreted as electric multipole moments and, for the two-derivative solution, take the form
\begin{equation}
    \mathcal Q_{2\ell}^{(0)}=(-a^2)^\ell Q^{(0)},\qquad \mathcal Q_{2\ell+1}^{(0)}=0,
\end{equation}
whereas the $\mathcal P_\ell$ are interpreted as magnetic multipole moments\footnote{This can be seen by dualizing the gauge field. One then sees that the multipole moments $\tilde{\mathcal Q}_\ell$ and $\tilde{\mathcal P}_\ell$ of the dualized gauge field satisfy $\tilde{\mathcal Q}_\ell=\mathcal P_\ell$ and $\tilde{\mathcal P}_\ell=\mathcal Q_\ell$~\cite{Ma:2024ulp}.} and take the form
\begin{equation}
    \mathcal{P}_{2\ell+1}^{(0)}=-a(-a^2)^\ell Q^{(0)},\qquad \mathcal{P}_{2\ell}^{(0)}=0.
\end{equation}
These electromagnetic multipole moments also match those of the Kerr-Newman black hole.

\subsection{Four-derivative Kerr-Sen}
 For convenience, we will use the notation $s_\beta=\sinh\frac{\beta}{2}$ and $c_\beta=\cosh\frac{\beta}{2}$. Upon taking the four-derivative corrections~\eqref{eq:KerrSen4der} to the metric into account, we find that the mass becomes
 \begin{equation}
     M=M^{(0)}+\alpha'\frac{s_{\beta}^{2}}{8\mu}\qty(1-\frac{\chi^{2}}{4}-\frac{\chi^{4}}{8}-\frac{5\chi^{6}}{64}-\frac{7\chi^{8}}{128}-\frac{21\chi^{10}}{512})+\mathcal O(\chi^{12}),
 \end{equation}
 while the angular momentum becomes
 \begin{equation}
     J=J^{(0)}-\frac{5\alpha'}{32}s_{\beta}^{2}\,\chi\,\qty(1-\frac{\chi^{2}}{10}-\frac{3\chi^{4}}{80}-\frac{3\chi^{6}}{160}-\frac{7\chi^{8}}{640})+\mathcal O(\chi^{11}),\label{eq:angMomCorr}
 \end{equation}
 and the charge becomes
  \begin{equation}
     Q=Q^{(0)}+\alpha'\frac{c_{\beta}s_{\beta}}{8\sqrt{2}\mu}\qty(1-\frac{\chi^{2}}{4}-\frac{\chi^{4}}{8}-\frac{5\chi^{6}}{64}-\frac{7\chi^{8}}{128}-\frac{21\chi^{10}}{512})+\mathcal O(\chi^{12}).
 \end{equation}
Clearly, the charges receive $\alpha'$ corrections. If we were to solve for the solution from scratch, we would have imposed that the charges are unmodified as a boundary condition. Equivalently, we may shift the constants of integration
\begin{align}
    \mu&\to\mu+\frac{\alpha's_{\beta}^{2}}{8\mu c_{\beta}^{2}}\qty(1-\frac{\chi^{2}}{4}-\frac{\chi^{4}}{8}-\frac{5\chi^{6}}{64}-\frac{7\chi^{8}}{128}-\frac{21\chi^{10}}{512})+\mathcal{O}(\chi^{12}),\nn\\
    \chi&\to\chi+\frac{5\alpha'\chi s_{\beta}^{2}}{32\mu^{2}c_{\beta}^{2}}\qty(1-\frac{\chi^{2}}{10}-\frac{3\chi^{4}}{80}-\frac{3\chi^{6}}{160}-\frac{7\chi^{8}}{640})+\mathcal{O}(\chi^{11}),\nn\\
    \beta&\to\beta-\frac{\alpha's_{\beta}}{4\mu^{2}c_{\beta}}\qty(1-\frac{\chi^{2}}{4}-\frac{\chi^{4}}{8}-\frac{5\chi^{6}}{64}-\frac{7\chi^{8}}{128}-\frac{21\chi^{10}}{512})+\mathcal{O}(\chi^{12}),\label{eq:paramshift}
\end{align}
to fix the charges to be
\bea
M=M^{(0)}+\mathcal{O}(\alpha'^{2}),\qquad J=J^{(0)}+\mathcal{O}(\alpha'^{2}),\qquad Q=Q^{(0)}+\mathcal{O}(\alpha'^{2}).
\eea

From~\eqref{eq:horVel} and taking the shift~\eqref{eq:paramshift} into account, the horizon velocity becomes
\begin{equation}
    \Omega_{H}=\Omega_{H}^{(0)}+\alpha'\frac{5as_{\beta}^{2}}{64\mu^{4}c_{\beta}^{4}}\qty(1+\frac{\chi^{2}}{5}+\frac{11\chi^{4}}{80}+\frac{\chi^{6}}{8}+\frac{77\chi^{8}}{640}+\mathcal{O}(\chi^{10}))\,.\label{eq:horVelCorr}
\end{equation}
Notice that both \eqref{eq:angMomCorr} and \eqref{eq:horVelCorr} vanish when $\chi=0$, as we would expect.
From~\eqref{eq:temp}, we find that the four-derivative correction to temperature is given by
\begin{equation}
    T=T^{(0)}+\frac{5s_{\beta}^{2}\alpha'}{256\pi\mu^{3}c_{\beta}^{4}}\qty(1-\frac{5\chi^{2}}{4}-\frac{11\chi^{4}}{16}-\frac{\chi^{6}}{2}-\frac{263\chi^{8}}{640}-\frac{1839\chi^{10}}{5120}+\mathcal{O}(\chi^{12}))\,.
\end{equation}
From~\eqref{eq:potential}, the electric potential gets corrected to
\begin{equation}
   \Phi_{e}=\Phi_{e}^{(0)}-\frac{5s_{\beta}\alpha'}{16\sqrt{2}\mu^{2}c_{\beta}^{3}}\qty(1-\frac{\chi^{2}}{2}-\frac{19\chi^{4}}{80}-\frac{23\chi^{6}}{160}-\frac{63\chi^{8}}{640}-\frac{93\chi^{10}}{1280}+\mathcal{O}(\chi^{12}))\,.
\end{equation}
Note that all of the above corrections disappear as $\beta\to0$, which is consistent with the fact that the Kerr metric receives no corrections in the Einstein frame and remains uncharged under the gauge field. Note that these quantities can be written as closed-form expressions in terms of $\Lambda$, but they are generally unenlightening, so we have omitted them.

Finally, we wish to obtain the four-derivative entropy. Na\"ively, we would like to apply the Iyer-Wald prescription~\mbox{\cite{Wald:1993nt,Iyer:1994ys,Jacobson:1993vj}}
\begin{equation}
    S=-\frac{1}{8}\int_{\mathfrak B}\dd[2]x\sqrt{h}\,\pdv{\mathcal L}{R_{\hat\mu\hat\nu\hat\rho\hat\sigma}}\epsilon_{\hat\mu\hat\nu}\epsilon_{\hat\rho\hat\sigma},
\end{equation}
where $h$ is the induced metric on the the bifurcation horizon $\mathfrak B$ and $\epsilon\propto\dd t\land\dd r$ is a unit binormal such that $\epsilon^2=-2$. However, this only applies to theories where every field is a tensor, \emph{i.e.}, to those without gauge symmetries. In particular, the Lorentz-Chern-Simons terms appearing in $\tilde H$ are especially problematic~\cite{Elgood:2020svt,Elgood:2020mdx,Elgood:2020nls}. For our action~\eqref{eq:BdR4dGaugeFields}, the entropy should instead be given by~\cite{Elgood:2020nls}\footnote{Note that we are working in the string frame here.}
\begin{equation}
    S=\frac{1}{4}\int_{\mathfrak B}\dd[2]x\sqrt{h}\,e^{-2\phi}\qty[1-\frac{\alpha'}{16}\qty(2R^{\hat\mu\hat\nu\hat\rho\hat\sigma}(\Omega_-)-2H^{\hat\nu\hat\rho\hat\lambda}H^{\hat\mu\hat\sigma}{}_{\hat\lambda}-\mathcal F^{\hat\mu\hat\nu}\mathcal F^{\hat\rho\hat\sigma})\epsilon_{\hat\mu\hat\nu}\epsilon_{\hat\rho\hat\sigma}+\frac{\alpha'}{8}H^{\hat\mu\hat\nu\hat\rho}\Pi_{\hat\mu}\epsilon_{\hat\nu\hat\rho}],\label{eq:OrtinEntropy}
\end{equation}
where $\Pi$ is the vertical Lorentz momentum map
associated to $\epsilon$ and defined by
\begin{equation}
    \dd\Pi\big\vert_{\mathfrak B}=R(\Omega_-)^{\hat a\hat b}\epsilon_{\hat a\hat b}\big\vert_{\mathfrak B}.\label{eq:Pi}
\end{equation}
This expression for the entropy almost matches that of the Iyer-Wald prescription, except for the coefficient of the last term in~\eqref{eq:OrtinEntropy}. Unfortunately, it is difficult to integrate \eqref{eq:Pi} to obtain $\Pi$. One approach to get around this would be to dualize $H$ to an axion $\varphi$, which is difficult given the complexity of the four-derivative corrections. Instead, one can integrate the first law
\begin{equation}
    \dd M=T\dd S+\Phi_e\dd Q+\Omega_H\dd J,\label{eq:firstLaw}
\end{equation}
in order to obtain the entropy
\begin{equation}
    S=S^{(0)}+\alpha'\qty[\mathfrak{s}+\frac{5}{8}s_{\beta}^{2}\pi\qty(1-\frac{\chi^{2}}{4}-\frac{7\chi^{4}}{40}-\frac{9\chi^{6}}{64}-\frac{77\chi^{8}}{640}-\frac{273\chi^{10}}{2560})+\mathcal{O}(\chi^{12})],
\end{equation}
where $\mathfrak s$ is a constant of integration. When $\beta\to 0$, we may use the Iyer-Wald prescription to compute~\cite{Cano:2021rey}
\begin{equation}
    S\big\vert_{\beta=0}=S^{(0)}\big\vert_{\beta=0}+\frac{\alpha'\pi}{2}.
\end{equation}
This is a consequence of the fact that for Ricci flat solutions,\footnote{This is true for the horizon, where extrinsic curvature vanishes.}
\begin{equation}
    R^{\hat\mu\hat\nu\hat\rho\hat\sigma}\epsilon_{\hat\mu\hat\nu}\epsilon_{\hat\rho\hat\sigma}=\mathcal R,
\end{equation}
where $\mathcal R$ is the induced Ricci scalar on the horizon. Thus, the four-derivative part of the entropy~\eqref{eq:OrtinEntropy} becomes proportional to the Euler characteristic, $\chi(\mathfrak B)=2$. This allows us to fix
\begin{equation}
    \mathfrak s=\frac{\pi}{2}.
\end{equation}

As a check of our results, we may compute the on-shell action $I$, which is proportional to the Gibbs free energy $G$, 
\begin{equation}
    I=T^{-1} G.
\end{equation}
Before the parameter shift~\eqref{eq:paramshift}, this gives
\begin{equation}
    G=\frac{\mu}{2}-\frac{\alpha'}{16\mu}\qty(1-\frac{\chi^{2}}{4}-\frac{\chi^{4}}{8}-\frac{5\chi^{6}}{64}-\frac{7\chi^{8}}{128}-\frac{21\chi^{10}}{512}),\label{eq:Gibbs}
\end{equation}
which is independent of $\beta$, as we would expect for an $O(2,1)$ invariant action. Note that this must be calculated using the full four-derivative solution, as the Reall-Santos procedure~\cite{Reall:2019sah} cannot be applied, which was first noticed in~\cite{Ma:2022gtm}. This is because the Lorentz-Chern-Simons form $\omega_{3L}$ transforms anomalously under local Lorentz transformations; normally, this transformation would be absorbed by an anomalous transformation of $B$, but using only the two-derivative value of $B$ results in a frame ambiguity for $\omega_{3L}$.

After the parameter redefinition~\eqref{eq:paramshift}, we get
\begin{equation}
    G=\frac{\mu}{2}-\frac{\alpha'}{16\mu c_{\beta}^{2}}\qty(1-\frac{\chi^{2}}{4}-\frac{\chi^{4}}{8}-\frac{5\chi^{6}}{64}-\frac{7\chi^{8}}{128}-\frac{21\chi^{10}}{512}).
\end{equation}
Note that after shifting the integration constants, we do indeed pick up $\beta$ dependence as the shift complicates the interpretation of $\beta$ as simply being the boost parameter. One can check that this free energy satisfies
\begin{equation}
    G=M-TS-\Phi_e Q-\Omega_H J,
\end{equation}
which provides a check of the computed thermodynamic quantities.
 
We now compute the multipole moments for the full four-derivative solution \eqref{eq:KerrSen4der} and apply the shift \eqref{eq:paramshift}. We write the multipole moments as
\begin{align}
    M_\ell=M_{\ell}^{(0)}+\alpha'\delta M_\ell,\qquad \mathcal S_\ell=\mathcal S_{\ell}^{(0)}+\alpha'\delta\mathcal S_\ell,\qquad  \mathcal Q_\ell=\mathcal Q_{\ell}^{(0)}+\alpha'\delta\mathcal Q_\ell,\qquad \mathcal P_\ell=\mathcal P_{\ell}^{(0)}+\alpha'\delta\mathcal P_\ell.
\end{align}
We find that the first several gravitational multipole moments are given by
\begin{align}
    \delta M_{2} & =-\frac{143}{240}\mu\chi^{2}s_{\beta}^{2}\qty(1-\frac{265\chi^{2}}{2002}-\frac{395\chi^{4}}{6864}-\frac{1655\chi^{6}}{50336}-\frac{5145\chi^{8}}{237952})+\mathcal O(\chi^{12}),\nn\\
    \delta M_{4} & =\frac{2293\mu^{3}\chi^{4}s_{\beta}^{2}}{1960}\qty(1-\frac{1981\chi^{2}}{13758}-\frac{77707\chi^{4}}{1210704}-\frac{391265\chi^{6}}{10492768})+\mathcal O(\chi^{12}),\nn \\
    \delta M_{6} & =-\frac{135001\mu^{5}\chi^{6}s_{\beta}^{2}}{77616}\qty(1-\frac{443975\chi^{2}}{2970022}-\frac{20776161\chi^{4}}{308882288})+\mathcal O(\chi^{12}),\nn\\
    \delta M_{8} & =\frac{13715861\mu^{7}\chi^{8}s_{\beta}^{2}}{5945940}\qty(1-\frac{54470591\chi^{2}}{356612386})+\mathcal O(\chi^{12}),\nn\\
    \delta\mathcal{S}_{3}&=-\frac{11}{20}\mu^{2}\chi^{3}s_{\beta}^{2}\qty(1-\frac{219\chi^{2}}{1232}-\frac{43\chi^{4}}{528}-\frac{743\chi^{6}}{15488})+\mathcal O(\chi^{11}),\nn \\
    \delta\mathcal{S}_{5} & =\frac{557}{504}\mu^{4}\chi^{5}s_{\beta}^{2}\qty(1-\frac{1171\chi^{2}}{6684}-\frac{983\chi^{4}}{12254})+\mathcal O(\chi^{11}),\nn\\
    \delta\mathcal{S}_{7}&=-\frac{28529\mu^{6}\chi^{7}s_{\beta}^{2}}{17160}\qty(1-\frac{1307935\chi^{2}}{7531656})+\mathcal O(\chi^{11}),\nn\\
    \delta\mathcal{S}_{9}&=\frac{8307\mu^{8}\chi^{9}s_{\beta}^{2}}{3740}+\mathcal O(\chi^{11}).
\end{align}
We find that the first several electromagnetic multipole moments are given by
\begin{align}
    \delta\mathcal{Q}_{2}&=-\frac{\mu\chi^{2}s_{\beta}}{960\sqrt{2}c_{\beta}}\left(286\qty(c_{\beta}^{2}+s_{\beta}^{2})-254-\frac{5}{14}\chi^{2}\qty(106\qty(c_{\beta}^{2}+s_{\beta}^{2})-146)\right.\nn\\
    &\kern7em -\frac{5}{48}\chi^{4}\qty(158\qty(c_{\beta}^{2}+s_{\beta}^{2})-238)-\frac{5}{352}\chi^{6}\qty(662\qty(c_{\beta}^{2}+s_{\beta}^{2})-1054)\nn\\
    &\kern7em\left.-\frac{105\chi^{8}}{1664}\qty(98\qty(c_{\beta}^{2}+s_{\beta}^{2})-162)\right)+\mathcal O(\chi^{12}),\nn \\
    \delta\mathcal{Q}_{4} & =\frac{\mu^{3}\chi^{4}s_{\beta}}{3920\sqrt{2}c_{\beta}}\left(2293\qty(c_{\beta}^{2}+s_{\beta}^{2})-2117-\frac{7}{6}\chi^{2}\qty(283\qty(c_{\beta}^{2}+s_{\beta}^{2})-347)\right.\nonumber \\
    & \left.\kern4em-\frac{7}{528}\chi^{4}\qty(-14309+11101\qty(c_{\beta}^{2}+s_{\beta}^{2}))-\frac{245\chi^{6}}{4576}\qty(-2121+1597\qty(c_{\beta}^{2}+s_{\beta}^{2}))\right)+\mathcal O(\chi^{12}),\nn\\
    \delta\mathcal{Q}_{6} & =-\frac{\mu^{5}\chi^{6}s_{\beta}}{155232\sqrt{2}c_{\beta}}\left(135001\qty(c_{\beta}^{2}+s_{\beta}^{2})-126953-\frac{7}{22}\chi^{2}\qty(-73789+63425(c_{\beta}^{2}+s_{\beta}^{2}))\right.\nn\\
    &\left. \kern8em\ -\frac{21}{2288}\chi^{4}\qty(-1190837+989341\qty(c_{\beta}^{2}+s_{\beta}^{2}))\right)+\mathcal O(\chi^{12}),\nn\\
    \delta\mathcal{Q}_{8} & =-\frac{13040869\mu^{7}\chi^{8}s_{\beta}}{11891880\sqrt{2}c_{\beta}}\qty(1-\frac{13715861}{13040869}\qty(c_{\beta}^{2}+s_{\beta}^{2})-\frac{61475239\chi^{2}}{339062594}\qty(1-\frac{7781513}{8782177}\qty(c_{\beta}^{2}+s_{\beta}^{2})))\nn\\
    &\quad+\mathcal O(\chi^{12}),\nn\\
    \delta\mathcal{P}_{1} & =\frac{9\chi s_{\beta}}{32\sqrt{2}c_{\beta}}\qty(1-\frac{\chi^{2}}{6}-\frac{11\chi^{4}}{144}-\frac{13\chi^{6}}{288}-\frac{35\chi^{8}}{1152})+\mathcal O(\chi^{11}),\nn\\
    \delta\mathcal{P}_{3} & =\frac{\mu^{2}\chi^{3}s_{\beta}}{320\sqrt{2}c_{\beta}}\left(-182+88\qty(c_{\beta}^{2}+s_{\beta}^{2})-\frac{3}{14}\chi^{2}\qty(-137+73\qty(c_{\beta}^{2}+s_{\beta}^{2}))\right.\nn\\
    &\left. \kern6em-\frac{1}{24}\chi^{4}\qty(-323+172\qty(c_{\beta}^{2}+s_{\beta}^{2}))-\frac{1}{176}\chi^{6}\qty(-1402+743\qty(c_{\beta}^{2}+s_{\beta}^{2}))\right)+\mathcal O(\chi^{11}),\nn\\
    \delta\mathcal{P}_{5} & =-\frac{\mu^{4}\chi^{5}s_{\beta}}{2016\sqrt{2}c_{\beta}}\left(-1721+1114\qty(c_{\beta}^{2}+s_{\beta}^{2})-\frac{1}{6}\chi^{2}\qty(-1664+1171\qty(c_{\beta}^{2}+s_{\beta}^{2}))\right.\nn\\
    &\left.\kern7em\  -\frac{1}{176}\chi^{4}\qty(-22387+15728\qty(c_{\beta}^{2}+s_{\beta}^{2}))\right)+\mathcal O(\chi^{11}),\nn\\
    \delta\mathcal{P}_{7} & =\frac{\mu^{6}\chi^{7}s_{\beta}}{68640\sqrt{2}c_{\beta}}\qty(57058\qty(c_{\beta}^{2}+s_{\beta}^{2})-78077-\frac{5}{132}\chi^{2}\qty(-333007+261587\qty(c_{\beta}^{2}+s_{\beta}^{2})))+\mathcal O(\chi^{11}),\nn\\
    \delta\mathcal{P}_{9}&=\frac{42507\mu^{8}\chi^{9}s_{\beta}}{29920\sqrt{2}c_{\beta}}\qty(1-\frac{3692}{4723}\qty(c_{\beta}^{2}+s_{\beta}^{2}))+\mathcal O(\chi^{11}).
\end{align}
As before, $M_0$, $\mathcal{S}_1$, and $\mathcal{Q}_0$ are the mass, angular momentum, and electric charge, respectively, and so we have not duplicated them in the multipole formulae. 

If we view the gauge field $\mathcal A$ as corresponding to a dark photon living in the hidden sector, then we would not be able to measure the electromagnetic multipole moments, so the Kerr-Sen black hole would appear to simply be a rotating black hole. In this scenario, it would be natural to compare with the Kerr black hole (${\beta=0}$), for which we notice that the corrections to the multipole moments all vanish. This would unequivocally distinguish the two solutions in experimental observations. 

\subsection{Comparison with Einstein-Maxwell theory}
Alternatively, we might view the gauge field as being the $U(1)$ Maxwell field of the Standard Model. Of course, we are unlikely to find a charged black hole in the night sky, but it is nevertheless interesting to compare with Einstein-Maxwell theory. In this scenario, we must compare with a charged black hole. The unique stationary axisymmetric solution to heterotic supergravity is given by the Kerr-Sen solution, whereas in Einstein-Maxwell theory it is given by the Kerr-Newman solution, which, at the two-derivative level, is given by
\begin{align}
    \dd s_4^2&=-\frac{\mathring\Delta}{\Sigma}\left(\dd t-a(1-x^2)\dd y\right)^2+\Sigma\qty(\frac{\dd r^2}{\mathring\Delta}+\frac{\dd x^2}{1-x^2})+\frac{1-x^2}{\Sigma}\left(a\,\dd t-(r^2+a^2)\dd y\right)^2,\nn\\
    \mathcal A&=-\frac{2Qr}{\Sigma}\qty(\dd t-a(1-x^2)\dd y),
\end{align}
where
\begin{equation}
    \mathring\Delta=r^2-2M r+a^2+Q^2,\qquad \Sigma=r^2+a^2x^2.
\end{equation}
This describes a charged, rotating, stationary, axisymmetric black hole with mass $M$, electric charge $Q$, and angular momentum $J=aM$. In particular, this reduces to the Kerr solution~\eqref{Kerr} when $Q=0$. However, note that this is not a solution of heterotic supergravity (except when $Q=0$) due to the non-trivial coupling of the dilaton with the gauge field.

Up to field redefinitions, the most general four-derivative action for Einstein-Maxwell theory is given by\footnote{Following the parameterization in~\cite{Ma:2024ulp}.}
\begin{align}
    e^{-1}\mathcal L_{\mathrm{EM}}&=R-\frac{1}{4}\mathcal F^2+\alpha_1R_{\hat\mu\hat\nu\hat\rho\hat\sigma}^2+\alpha_2R_{\hat\mu\hat\nu\hat\rho\hat\sigma}\mathcal F^{\hat\mu\hat\nu}\mathcal F^{\hat\rho\hat\sigma}+\frac{\alpha_3}{4} \mathcal F^4+\frac{2\alpha_0 - 16\alpha_1 - 8\alpha_2 - 9\alpha_3}{64}(\mathcal F^2)^2\nn\\
    &\quad+\beta_1R_{\hat\mu\hat\nu\hat\rho\hat\sigma}\mathcal F^{\hat\mu\hat\nu}\widetilde{\mathcal F}^{\hat\rho\hat\sigma}+\frac{\beta_0}{4}\widetilde{\mathcal F}^{\hat\mu\hat\nu}\mathcal F_{\hat\mu\hat\nu}\mathcal F^2,
\end{align}
where $\widetilde{\mathcal F}^{\hat\mu\hat\nu}=\frac{1}{2}\epsilon^{\hat\mu\hat\nu\hat\rho\hat\sigma}\mathcal F_{\hat\rho\hat\sigma}$. Note that the second line consists entirely of parity-odd terms. The full four-derivative corrections to the Kerr-Newman metric are not known in closed form but were computed perturbatively in 
\begin{equation}
    \chi_a=\frac{a}{M},\qquad \chi_Q=\frac{Q}{M},
\end{equation} 
by Ref.~\cite{Ma:2024ulp}. Since $\alpha'$ is just a length scale, for the sake of comparison, we define
\begin{equation}
    \alpha_i=\alpha'\bar\alpha_i,\qquad \beta_j=\alpha'\bar\beta_j.
\end{equation}
The first few multipole moments were found to be
\begin{align}
    \delta M_2^{(\mathrm{KN})}&=\bar\alpha_2 M \chi_a^2 \chi_Q^2-\frac{M \chi_a^2 \chi_Q^2}{300}\left[30 \bar\alpha_2 \chi_a^2+\left(8 \bar\alpha_0-76 \bar\alpha_1-203 \bar\alpha_2-19 \bar\alpha_3\right) \chi_Q^2\right]+\mathcal{O}(\chi^8),\nn\\
    \delta M_3^{(\mathrm{KN})}&=-\frac{23}{25} \bar\beta_1 M^2 \chi_a^3 \chi_Q^2+\frac{M^2 \chi_a^3 \chi_Q^2}{700}\left[91 \bar\beta_1 \chi_a^2+\qty(142 \bar\beta_0 -391 \bar\beta_1) \chi_Q^2\right]+\mathcal{O}(\chi^8),\nn\\
    \delta \mathcal{S}_2^{(\mathrm{KN})}&= \bar\beta_1 M \chi_a^2 \chi_Q^2-\frac{M \chi_a^2 \chi_Q^2}{300}\left[30 \bar\beta_1 \chi_a^2+\qty(70 \bar\beta_0-159 \bar\beta_1)\chi_Q^2\right] \nn\\
    &\quad -\frac{M \chi_a^2 \chi_Q^2}{560}\left[21 \bar\beta_1 \chi_a^4+\qty(-28\bar\beta_0+15 \bar\beta_1 )\chi_a^2 \chi_Q^2+\qty(140 \bar\beta_0 -168 \bar\beta_1)\chi_Q^4\right]+\mathcal{O}(\chi^8),\nn\\
    \delta \mathcal{S}_3^{(\mathrm{KN})}&=\frac{23}{25} \bar\alpha_2 M^2 \chi_a^3 \chi_Q^2-\frac{M^2 \chi_a^3 \chi_Q^2}{4900}\left[637\bar\alpha_2 \chi_a^2+\left(102 \bar\alpha_0-1172 \bar\alpha_1-3501 \bar\alpha_2-293 \bar\alpha_3\right) \chi_Q^2\right]\nn\\
    &\quad+\mathcal{O}(\chi^8),\nn\\
    \delta \mathcal{Q}_1^{(\mathrm{KN})}&= -\frac{1}{2} \bar\beta_1 \chi_a \chi_Q+\frac{\chi_a \chi_Q}{120}\left[6 \bar\beta_1 \chi_a^2+(3 \bar\beta_1+14 \bar\beta_0) \chi_Q^2\right]+\frac{\chi_a \chi_Q}{480}\left[4 \bar\beta_0 \chi_Q^2\left(7 \chi_Q^2-3 \chi_a^2\right)\right. \nn\\
    & \left.\quad+3 \bar\beta_1\left(3 \chi_a^4-2 \chi_a^2 \chi_Q^2+4 \chi_Q^4\right)\right]+\frac{\chi_a \chi_Q}{640}\left[\bar\beta_1\left(9 \chi_a^4 \chi_Q^2-12 \chi_a^2 \chi_Q^4+6 \chi_a^6+10 \chi_Q^6\right)\right.\nn\\
    & \left.\quad+2 \bar\beta_0\left(10 \chi_Q^6-3 \chi_a^4 \chi_Q^2\right)\right]+\mathcal{O}(\chi^8),\nn\\
    \delta \mathcal{Q}_2^{(\mathrm{KN})}&=\frac{3}{50} \bar\alpha_2 M \chi_a^2 \chi_Q+\frac{M \chi_a^2 \chi_Q}{4200}\left[\bar\alpha_2\left(90 \chi_a^2+1519 \chi_Q^2\right)-49\left(\bar\alpha_0+4 \bar\alpha_1+\bar\alpha_3\right) \chi_Q^2\right] \nn\\
    &\quad+\frac{M \chi_a^2 \chi_Q}{94080}\left[980 \bar\alpha_2 \chi_a^4-3\left(197 \bar\alpha_0+1016 \bar\alpha_1+3288 \bar\alpha_2+254 \bar\alpha_3\right) \chi_a^2 \chi_Q^2\right. \nn\\
    & \left.\qquad\qquad\quad\ \ -392\left(4 \bar\alpha_0-20 \bar\alpha_1-64 \bar\alpha_2-4 \bar\alpha_3\right) \chi_Q^4\right]+\mathcal{O}(\chi^8),\nn\\
    \delta \mathcal{P}_1^{(\mathrm{KN})}&= -\frac{1}{2} \bar\alpha_2 \chi_a \chi_Q+\frac{\chi_a \chi_Q}{120}\left[6\bar\alpha_2 \chi_a^2+\left(\bar\alpha_0-20 \bar\alpha_1-13\bar\alpha_2-5 \bar\alpha_3\right) \chi_Q^2\right]+\frac{\chi_a \chi_Q}{3360}\left[63 \bar\alpha_2 \chi_a^4\right. \nn\\
    &\quad \left.-6\left(4 \bar\alpha_0+4 \bar\alpha_1+27 \bar\alpha_2+\bar\alpha_3\right) \chi_a^2 \chi_Q^2+14\left(\bar\alpha_0-20 \bar\alpha_1-10 \bar\alpha_2-5 \bar\alpha_3\right) \chi_Q^4\right]+\mathcal{O}(\chi^8),\nn\\
    \delta\mathcal{P}_{2}^{(\mathrm{KN})}& =-\frac{3}{50}\bar\beta_{1}M\chi_{a}^{2}\chi_{Q}-\frac{M\chi_{a}^{2}\chi_{Q}}{4200}\Big[90\bar\beta_{1}\chi_{a}^{2}-49(2\bar\beta_{0}-39\bar\beta_{1})\chi_{Q}^{2}\Big] \nn\\
    &\quad-\frac{M\chi_{a}^{2}\chi_{Q}}{3360}\Big[35\bar\beta_{1}\chi_{a}^{4}-3(10\bar\beta_{0}+53\bar\beta_{1})\chi_{a}^{2}\chi_{Q}^{2}-28(13\bar\beta_{0}-30\bar\beta_{1})\chi_{Q}^{4}\Big]+\mathcal{O}(\chi^8).\label{eq:KNmultipoles}
\end{align}
These multipole moments were shown to be invariant under field redefinitions. Note that $\bar\beta_2$ does not appear explicitly. Clearly, the $\bar\beta_i$ must vanish to be equivalent to the Kerr-Sen case, so we will focus on the parity-even sector from here on.

In order to compare our results with these, we must rewrite them in terms of physical invariants (\emph{i.e.}, the mass $M$ and electric charge $Q$) instead of the integration constants $\mu$ and $\beta$, and then expand in $\chi_a$ and $\chi_Q$. Note that $a=J/M$ for the Kerr-Sen and Kerr-Newman solutions. From~\eqref{eq:twoDerivChargesInverse}, we see that
\begin{equation}
    \sinh\beta=\frac{2\sqrt{2}\chi_Q}{1-2\chi_Q^2}.
\end{equation}
Given that $\chi=a/\mu$, we also have the relation
\begin{equation}
    \chi=\frac{\chi_a}{1-2\chi_Q^2}.
\end{equation}
Finally, using the relation \eqref{eq:twoDerivCharges} for the mass, we deduce that
\begin{equation}
    \mu=\frac{2M}{1+\sqrt{1+8\chi_Q}}.
\end{equation}
Upon doing this, we find that
\begin{align}
    \delta M_2&=-\frac{143}{120}M\chi_a^2\chi_Q^2\qty(1-\frac{265\chi_a^2}{2002})\qty(1-4\chi_Q^2)+\mathcal O(\chi_a^5,\chi_Q^5),\nn\\
    \delta\mathcal S_3&=-\frac{11}{10}M^2 \chi_a^3\chi_Q^2\qty(1-\frac{219\chi_a^2}{1232})\qty(1-6\chi_Q^2)+\mathcal O(\chi_a^7,\chi_Q^5),\nn\\
    \delta\mathcal Q_2&=-\frac{1}{30}M \chi_a^2\chi_Q\qty[1+\frac{25\chi_a^2}{56}+\frac{127\chi_Q^2}{4}\qty(1-\frac{365\chi_a^2}{1778})]+\mathcal O(\chi_a^5,\chi_Q^5),\nn\\
    \delta\mathcal P_1&=\frac{9}{32}\chi_a\chi_Q\qty(1-\frac{\chi_a^2}{6})\qty(1-2\chi_Q^2)+\mathcal O(\chi_a^5,\chi_Q^5),
\end{align}
These can never agree with \eqref{eq:KNmultipoles}. This can be easily seen by comparing the mass multipoles
\begin{align}
    \delta M_2^{(\mathrm{KN})}&=\bar\alpha_2M\chi_a^2\chi_Q^2\qty(1-\frac{1}{10}\chi_a^2)+\mathcal O(\chi_a^5,\chi_Q^3),\nn\\
    \delta M_2&=-\frac{143}{120}M\chi_a^2\chi_Q^2\qty(1-\frac{265}{2002}\chi_a^2)+\mathcal O(\chi_a^5,\chi_Q^3),
\end{align}
or the current multipoles
\begin{align}
    \delta\mathcal S_3^{(\mathrm{KN})}&=\frac{23}{25}\bar\alpha_2M^2\chi_a^3\chi_Q^2\qty(1-\frac{13}{92}\chi_a^2)+\mathcal O(\chi_a^6,\chi_Q^3),\nn\\
    \delta\mathcal S_3&=-\frac{11}{10}M^2\chi_a^3\chi_Q^2\qty(1-\frac{219}{1232}\chi_a^2)+\mathcal O(\chi_a^6,\chi_Q^3),
\end{align}
or the electric multipoles
\begin{align}
    \delta\mathcal Q_2^{(\mathrm{KN})}&=\frac{3}{50}\bar\alpha_2M\chi_a^2\chi_Q\qty(1+\frac{5}{14}\chi_a^2)+\mathcal O(\chi_a^5,\chi_Q^3),\nn\\
    \delta\mathcal Q_2&=-\frac{1}{30}M\chi_a^2\chi_Q\qty(1+\frac{25}{56}\chi_a^2)+\mathcal O(\chi_a^5,\chi_Q^3),
\end{align}
or the magnetic multipoles
\begin{align}
    \delta\mathcal P_1^{(\mathrm{KN})}&=-\frac{1}{2}\bar\alpha_2\chi_a\chi_Q\qty(1-\frac{1}{10}\chi_a^2)+\mathcal O(\chi_a^3,\chi_Q^3),\nn\\
    \delta\mathcal P_1&=\frac{9}{32}\chi_a\chi_Q\qty(1-\frac{1}{6}\chi_a^2)+\mathcal O(\chi_a^3,\chi_Q^3).
\end{align}
Clearly, even if we choose $\bar\alpha_2$ to fit the leading order behavior, we can never fit the subleading behavior of any of the multipoles. Looking at the leading order terms, we note that even just fitting the leading order $\delta M_2$, $\delta\mathcal S_3$, $\delta\mathcal Q_2$, and $\delta\mathcal P_1$ requires that $\bar\alpha_2=-143/120$, $-55/46$, $-5/9$, and $-9/16$, respectively, which cannot be simultaneously satisfied. That is, at best, we can match one of the multipole moments to leading order. Hence, we can (in principle) experimentally distinguish Kerr-Sen and Kerr-Newman black holes.

\section{Discussion}\label{sec:disc}
In this paper, we have obtained the four-derivative corrected Kerr-Sen solution and computed the associated multipole moments. In particular, the four-derivative corrections to the multipole moments are distinct from those of the Kerr solution in heterotic supergravity. They are also distinct from the multipole moments of the Kerr-Newman solution of Einstein-Maxwell theory, so we hope that sufficiently sensitive experiments could test these differences. 

At the level of solution generation, it would be interesting to extend this procedure to more general $O(d+p,d)$ transformations of the Kerr solution (or other seed solutions), as in~\cite{Cvetic:1995sz,Cvetic:1995kv,Cvetic:1996xz}. In principle, this is straightforward, but in practice, given the complexity of the solution we have obtained for the particular choice of $O(2,1)$ transformation, we should expect that more general $O(d+p,d)$ boosts will lead to much more complicated correction terms. One exception is the static limit ($a\to 0$), which took a considerably simpler form \eqref{eq:GMGHS}. It would also be interesting to extend our results to order $\alpha'^2$. Various authors have studied the $O(d,d)$ covariant formulation of the action to this order~\cite{Baron:2020xel,Hronek:2021nqk,Hronek:2022dyr}, and, in principle, the necessary field redefinitions have been computed in~\cite{Hronek:2022dyr}.

It would be interesting to investigate the extremal Kerr-Sen black hole. In principle, this could be done by starting with an expansion of the Kerr solution around $\chi = 1$. However, in practice, since this corresponds to a very highly spinning black hole, the corrections to the Kerr solution are no longer a finite series in $x$, which seems to make the solution intractable. However, one approach is to consider the near-horizon extremal geometry, as in~\cite{Chen:2018jed,Cano:2023dyg,Cassani:2023vsa,Cano:2024tcr}. The added constraints from the enhanced $SL(2,\mathbb R)\times U(1)$ isometry at the horizon turn the problem of solving for the scalars into an ordinary differential equation in $x$. We will treat this case in a future work.

It is also interesting to note that it was conjectured in~\cite{Jayaprakash:2024xlr} that the $O(d,d)$ symmetry of the action \eqref{eq:OddActionEff3d} could be extended to $O(d+p,d)$ symmetry by allowing the $O(d)_+$ indices to run from 1 to $d+p$ instead. That is, given gauge fields $\mathcal F^{\mathfrak a}$, the authors of Ref.~\cite{Jayaprakash:2024xlr} suggested directly promoting
    \begin{equation}
        \hat F^{(-)\,a}=F^{(-)\,a},\qquad \hat F^{(+)\,\hat a}=\begin{pmatrix}
            F^{(+)\,a}\\ \mathcal F^{\mathfrak a}\label{eq:JimGuess}
        \end{pmatrix},
    \end{equation}
    where $\hat a=\{a,\mathfrak a\}$. Given our discussion in Section~\ref{sec:Odd} and our more detailed results in Appendix~\ref{app:gaugeFields}, it seems that the natural way to extend the symmetry is actually to promote
    \begin{equation}
        \hat F^{(\pm)\,\hat a}=\begin{pmatrix}
            F^{(\pm)\,a}\\ \frac{1}{\sqrt{2}}\mathcal F^{\mathfrak a}\label{eq:nattyGuess}
        \end{pmatrix},
    \end{equation}
    symmetrically. This corresponds to the statement that $O(d+p,d)$ is the diagonal subgroup of $O(d+p,d+p)$. One may then perform a change of basis to bring this into the form \eqref{eq:JimGuess}, 
    \begin{equation}
    \mathcal H\to V\mathcal H V^T,\qquad \mathbb A\to V\mathbb A,\qquad \eta\to V\eta V^T,\qquad V=\begin{pmatrix}
        \openone&\quad0&\quad0&\quad0\\
        0&\quad\frac{1}{\sqrt{2}}\openone&\quad 0&\frac{1}{\sqrt{2}}\openone\\
        0&\quad0&\quad\openone&\quad 0\\
        0&\quad-\frac{1}{\sqrt{2}}\openone&\quad0&\quad\frac{1}{\sqrt{2}}\openone
    \end{pmatrix}.
\end{equation}
    which does not affect the form of the $O(d+p,d)$ covariant action since $V^TV=\openone$. Note that this is just the inverse of the transformation~\eqref{eq:VtransH}. Naturally, this does change the form of $\mathcal H$, $\mathbb A$, and $\eta$ from that found in Eq.~\eqref{eq:HafterRot} to that found in Eq.~\eqref{eq:HbeforeRot}. This is to say that both choices \eqref{eq:JimGuess} and \eqref{eq:nattyGuess} are valid as long as $\mathcal H$ and $\eta$ are chosen appropriately.

\let\oldaddcontentsline\addcontentsline
\renewcommand{\addcontentsline}[3]{}
\begin{acknowledgments}
    We would like to thank James Liu and Leoplodo Pando Zayas for discussions. This work is supported by the National Key Research and Development Program No. 2022YFE0134300 and the National Natural Science Foundation of China (NSFC) under Grants No. 12175164 and No. 12247103. L.M.~is also supported in part by National Natural Science Foundation of China (NSFC) grant No.~12447138, Postdoctoral Fellowship Program of CPSF Grant No.~GZC20241211 and the China Postdoctoral Science Foundation under Grant No.~2024M762338.
\end{acknowledgments}
\let\addcontentsline\oldaddcontentsline

\appendix
\section{Adding $U(1)$ gauge fields}\label{app:gaugeFields}
In this appendix, we show that heterotic supergravity with $p$ abelian gauge fields can be obtained from the consistent truncation of heterotic supergravity without gauge fields on~$T^p$.

It was found in~\cite{Liu:2023fqq} that the reduction of $(D+p)$-dimensional heterotic supergravity (without gauge fields) 
\begin{align}
     e^{-1}\mathcal L&=e^{-2\Phi}\left[R+4(\partial\Phi)^2-\frac{1}{12}\tilde H^2+\frac{\alpha'}{8}R(\Omega_-)_{MNPQ}^2\right],
\end{align}
on a torus $T^p$
\begin{align}
    \dd s_{D+p}^2&=g_{\mu\nu}\dd x^\mu\dd x^\nu+g_{ij}\eta^i\eta^j,\qquad \eta^i=\dd y^i+A^i_\mu\dd x^\mu,\nn\\
    B&=\frac{1}{2}b_{\mu\nu}\dd x^\mu\land\dd x^\nu+B_{\mu i}\dd x^\mu\land\eta^i+\frac{1}{2}b_{ij}\eta^i\land\eta^j,\nn\\
    \phi&=\Phi-\frac{1}{4}\log\det g_{ij},
\end{align}
with field strengths
\begin{equation}
    F^i=\dd A^i,\qquad G_i=\dd B_i,\qquad \tilde G_i=G_i-b_{ij}F^j,
\end{equation}
can be consistently truncated such that 
\begin{equation}
    \tilde G_i=F^i,\qquad g_{ij}=\delta_{ij},\qquad b_{ij}=0,
\end{equation}
and leads to the action
\begin{align}
     e^{-1}\mathcal L&=e^{-2\phi}\left\{R+4(\partial\phi)^2-\frac{1}{12}\tilde H^2-\frac{1}{4}(F^i)^2\right.\nn\\
    &\qquad\qquad\left.+\frac{\alpha'}{8}\qty[R(\omega_-)_{\mu\nu ab}^2-R(\omega_-)_{\mu\nu\rho\sigma}F^{i\mu\nu}F^{i\rho\sigma}-\frac{1}{2}F_{\mu\nu}^iF_{\nu\rho}^iF_{\rho\sigma}^jF_{\sigma\mu}^j+\frac{1}{2}F_{\mu\nu}^iF_{\nu\rho}^jF_{\rho\sigma}^iF_{\sigma\mu}^j]\right\}.\label{eq:reducedTorus}
\end{align}

We can now ask how the $O(d+p,d)$ transformation rules appear. Doing a truncated reduction on $T^p$ followed by an untruncated reduction on $T^d$ gives a metric and a $B$ field
\begin{align}
    \dd s^2_{D+p}&=g_{\mu\nu}\dd x^\mu\dd x^\nu+g_{ij}\eta^i\eta^j+\delta_{\mathfrak a\mathfrak b}\xi^{\mathfrak a}\xi^{\mathfrak b},\qquad \eta^i=\dd y^i+A^i,\qquad \xi^{\mathfrak a}=\dd z^{\mathfrak a}+\frac{1}{\sqrt{2}}\mathcal A^{\mathfrak a},\\
    B&=\frac{1}{2}b_{\mu\nu}\dd x^\mu\land\dd x^\nu+B_{\mu i}\,\dd x^i\land\eta^i+\frac{1}{2}b_{ij}\eta^i\land\eta^j+ \frac{1}{\sqrt{2}}\mathcal A^{\mathfrak a}\land\xi^{\mathfrak a}.
\end{align}
However, it is important to note that we must also reduce the gauge field that arises from the first $T^p$ reduction
\begin{equation}
    \mathcal A^{\mathfrak a}=\mathcal A_\mu^{\mathfrak a}\,\dd x^\mu+\mathcal A_i^{\mathfrak a}\,\eta^i,
\end{equation}
which means that when re-expressed as a proper $T^{d+p}$ reduction, we have
\begin{align}
    g_{\hat i\hat j}&=\begin{pmatrix}
        g_{ij}+\frac{1}{2}\mathcal A_i^{\mathfrak a}\mathcal A_j^{\mathfrak a}&\quad \frac{1}{\sqrt{2}}\mathcal A_i^{\mathfrak b}\\ \frac{1}{\sqrt{2}}\mathcal A_j^{\mathfrak a}&\quad\delta_{\mathfrak a\mathfrak b}
    \end{pmatrix},\qquad A_\mu^{\hat i}=\begin{pmatrix}
        A_\mu^i\\ \frac{1}{\sqrt{2}}\mathcal A_\mu^a
    \end{pmatrix},\nn\\
    b_{\hat i\hat j}&=\begin{pmatrix}
        b_{ij}&\quad\frac{1}{\sqrt{2}}\mathcal A_i^{\mathfrak b}\\ -\frac{1}{\sqrt{2}}\mathcal A_j^{\mathfrak a}&\quad 0
    \end{pmatrix},\qquad B_{\mu\hat i}=\begin{pmatrix}
        B_{\mu i}+\frac{1}{2}\mathcal A_\mu^{\mathfrak a} \mathcal A_i^{\mathfrak a}\\
        \frac{1}{\sqrt{2}}\mathcal A_\mu^{\mathfrak a}
    \end{pmatrix},
\end{align}
where $\hat i=\{i,\mathfrak a\}$. The first equation is, of course, just the Kaluza-Klein metric for reducing $g_{\hat i\hat j}$ on $T^p$. Computing the generalized metric and field strengths gives
\begin{align}
    \mathcal H&=\begin{pmatrix}
        \quad g_{ij}+c_{ki}g^{kl}c_{lj}+\mathcal A_i^{\mathfrak a}\mathcal A_j^{\mathfrak a}&\quad \frac{1}{\sqrt{2}}(c_{ki}g^{kl}\mathcal A_l^{\mathfrak b}+\mathcal A_i^{\mathfrak b})&\quad -g^{jk}c_{ki}&\quad \frac{1}{\sqrt{2}}(c_{ki}g^{kl}\mathcal A_l^{\mathfrak b}+\mathcal A_i^{\mathfrak b})\\
        \frac{1}{\sqrt{2}}(c_{kj}g^{kl}\mathcal A_l^{\mathfrak a}+\mathcal A_j^{\mathfrak a})&\quad \delta^{\mathfrak a\mathfrak b}+\frac{1}{2}\mathcal A_k^{\mathfrak a}g^{kl}\mathcal A^{\mathfrak b}_l&\quad-\frac{1}{\sqrt{2}}g^{jk}\mathcal A_k^{\mathfrak a}&\quad \frac{1}{2}\mathcal A_k^{\mathfrak a}g^{kl}\mathcal A^{\mathfrak b}_l\\
        -g^{ik}c_{kj}&\quad-\frac{1}{\sqrt{2}}g^{ik}\mathcal A_k^{\mathfrak b}&g^{ij}&\quad -\frac{1}{\sqrt{2}}g^{ik}\mathcal A_k^{\mathfrak b}\\
        \frac{1}{\sqrt{2}}(c_{kj}g^{kl}\mathcal A_l^{\mathfrak a}+\mathcal A_j^{\mathfrak a})&\quad\frac{1}{2}\mathcal A_k^{\mathfrak a}g^{kl}\mathcal A^{\mathfrak b}_l&\quad -\frac{1}{\sqrt{2}}g^{jk}\mathcal A_k^{\mathfrak a}&\quad \delta_{\mathfrak a\mathfrak b}+\frac{1}{2}\mathcal A_k^{\mathfrak a}g^{kl}\mathcal A^{\mathfrak b}_l
    \end{pmatrix},\nn\\
    \mathbb A_\mu&=\begin{pmatrix}
        A_\mu^i\\ \frac{1}{\sqrt{2}}\mathcal A_\mu^{\mathfrak a}\\B_{\mu i}+\frac{1}{2}\mathcal A_\mu^{\mathfrak a}\mathcal A_i^{\mathfrak a}\\
        \frac{1}{\sqrt{2}}\mathcal A_\mu^{\mathfrak a}
    \end{pmatrix},\label{eq:HbeforeRot}
\end{align}
where
\begin{equation}
    c_{ij}=b_{ij}+\frac{1}{2}\mathcal A_i^{\mathfrak a}\mathcal A_j^{\mathfrak a}.
\end{equation}
In order to make the $O(d+p,d)$ subgroup manifest, we do a change of basis
\begin{equation}
    \mathcal H\to V\mathcal H V^T,\qquad \mathbb A\to V\mathbb A,\qquad \eta\to V\eta V^T,\qquad V=\begin{pmatrix}
        \openone&\quad0&\quad0&\quad0\\
        0&\quad\frac{1}{\sqrt{2}}\openone&\quad 0&-\frac{1}{\sqrt{2}}\openone\\
        0&\quad0&\quad\openone&\quad 0\\
        0&\quad\frac{1}{\sqrt{2}}\openone&\quad0&\quad\frac{1}{\sqrt{2}}\openone
    \end{pmatrix}.\label{eq:VtransH}
\end{equation}
to get
\begin{align}
    \mathcal H&=\begin{pmatrix}
        \quad g_{ij}+c_{ki}g^{kl}c_{lj}+\mathcal A_i^{\mathfrak a}\mathcal A_j^{\mathfrak a}&\quad 0&\quad -g^{jk}c_{ki}&\quad c_{ki}g^{kl}\mathcal A_l^{\mathfrak b}+\mathcal A_i^{\mathfrak b}\\
        0&\quad \delta^{\mathfrak a\mathfrak b}&\quad0&\quad 0\\
        -g^{ik}c_{kj}&\quad0&g^{ij}&\quad -g^{ik}\mathcal A_k^{\mathfrak b}\\
        c_{kj}g^{kl}\mathcal A_l^{\mathfrak a}+\mathcal A_j^{\mathfrak a}&\quad0&\quad -g^{jk}\mathcal A_k^{\mathfrak a}&\quad \delta_{\mathfrak a\mathfrak b}+\mathcal A_k^{\mathfrak a}g^{kl}\mathcal A^{\mathfrak b}_l
    \end{pmatrix},\nn\\
    \mathbb A_\mu&=\begin{pmatrix}
        A_\mu^i\\ 0\\B_{\mu i}+\frac{1}{2}\mathcal A_\mu^{\mathfrak a}\mathcal A_i^{\mathfrak a}\\
        \mathcal A_\mu^{\mathfrak a}
    \end{pmatrix},\qquad\eta=\begin{pmatrix}
        0&\quad0&\quad\delta_i{}^{j}&\quad0\\
        0&\quad-\delta^{\mathfrak a\mathfrak b}&\quad0&\quad0\\
        \delta^{i}{}_j&\quad0&\quad0&\quad0\\
        0&\quad0&\quad0&\quad\delta_{\mathfrak a\mathfrak b}
    \end{pmatrix}.\label{eq:HafterRot}
\end{align}
It is then clear that we get an $O(d+p,d)$ coset by erasing the second row/column:
\begin{align}
    \mathcal H&=\begin{pmatrix}
        \quad g_{ij}+c_{ki}g^{kl}c_{lj}+\mathcal A_i^{\mathfrak a}\mathcal A_j^{\mathfrak a}&\quad -g^{jk}c_{ki}&\quad c_{ki}g^{kl}\mathcal A_l^{\mathfrak b}+\mathcal A_i^{\mathfrak b}\\
        -g^{ik}c_{kj}&g^{ij}&\quad -g^{ik}\mathcal A_k^{\mathfrak b}\\
        c_{kj}g^{kl}\mathcal A_l^{\mathfrak a}+\mathcal A_j^{\mathfrak a}&\quad -g^{jk}\mathcal A_k^{\mathfrak a}&\quad \delta_{\mathfrak a\mathfrak b}+\mathcal A_k^{\mathfrak a}g^{kl}\mathcal A^{\mathfrak a}_l
    \end{pmatrix},\nn\\
    \mathbb A_\mu&=\begin{pmatrix}
        A_\mu^i\\B_{\mu i}+\frac{1}{2}\mathcal A_\mu^{\mathfrak a}\mathcal A_i^{\mathfrak a}\\
        \mathcal A_\mu^{\mathfrak a}
    \end{pmatrix},\qquad
    \eta=\begin{pmatrix}
        0&\quad\delta_i{}^{j}&\quad0\\
        \delta^i{}_{j}&\quad0&\quad0\\
        0&\quad0&\quad\delta_{\mathfrak a\mathfrak b}
    \end{pmatrix}.
\end{align}
This precisely matches the results of~\cite{Maharana:1992my,Sen:1994fa,Hohm:2011ex} for the case of abelian gauge fields, where ${\kappa_{ab}=\delta_{ab}}$.

\section{$O(d,d)$ invariance of $O(d)_-\times O(d)_+$ actions}\label{app:JimTranslate}
In this Appendix, we discuss how to rewrite the $O(d)_-\times O(d)_+$ invariant Lagrangian \eqref{eq:OddActionEff3d} in an explicitly $O(d,d)$ invariant way. Although we consider $d=2$ in the main text, the results presented here are indeed valid for any $d$. We follow the conventions of~\cite{Jayaprakash:2024xlr}, as well as much of the content.

We begin by recalling the definition of the coset representative and introducing the generalized vielbein
\begin{equation}
    \mathcal H_{MN}=\begin{pmatrix}
        g_{ij}-b_{ik}g^{kl}b_{lj} & \quad b_{ik}g^{kj}\\
        -g^{ik}b_{kj} & \quad g^{ij}
    \end{pmatrix}=\frac{1}{2}\mathcal V_M^A\mathcal V_N^A,\qquad
    \mathcal V_M^A=\begin{pmatrix}
        e^a_\mu+e^{aj}b_{ji}&\quad -e^{ai}\\
        e^a_\mu-e^{aj}b_{ji}&\quad e^{ai}
    \end{pmatrix}.\label{eq:cosetApp}
\end{equation}
Note that we are using late capital Latin letters $M,N,P,...$ for generalized (curved) $O(d,d)$ indices and early capital Latin letters $A,B,C,...$ for generalized (rigid) $O(d)_-\times O(d)_+$ indices. This differs from the main text, although the two conventions are never used simultaneously, so we hope that this avoids confusion.

The generalized vielbein applied to the coset gives
\begin{equation}
    \mathcal H_{AB}\equiv \mathcal V^M_A\mathcal H_{MN}\mathcal V^N_B=\frac{1}{2}\begin{pmatrix}
        \openone&\quad 0\\0&\quad\openone
    \end{pmatrix},
\end{equation}
while applied to $\eta$ it gives
\begin{equation}
    \eta_{MN}=\begin{pmatrix}
        0&\quad\openone\\ \openone&\quad0
    \end{pmatrix},\qquad \eta_{AB}=\frac{1}{2}\begin{pmatrix}
        \openone&\quad 0\\0&\quad-\openone
    \end{pmatrix}.
\end{equation}
All indices are raised and lowered with $\eta$. One can then define the projectors
\begin{equation}
    \mathfrak P^{(\pm)}\equiv\mathcal H\mp\eta,\qquad \mathfrak P^{(-)}_{AB}=\begin{pmatrix}
        \openone&\quad 0\\0&\quad 0
    \end{pmatrix},\qquad \mathfrak P^{(+)}_{AB}=\begin{pmatrix}
        0&\quad 0\\0&\quad \openone
    \end{pmatrix},
\end{equation}
which are clearly $O(d,d)$ covariant.

The field strengths are given by
\begin{equation}
    \mathbb F^M=\begin{pmatrix}
        F^i\\G_i
    \end{pmatrix},\qquad \mathcal V_M^A\mathbb F^M=\begin{pmatrix}
        F^{(-)\,a}\\F^{(+)\,a}
    \end{pmatrix}.
\end{equation}
Note that $G_i$ rather than $\tilde G_i$ appears in $\mathbb F$ such that it can be written $\mathbb F=\dd\mathbb A$. From here, we immediately see that
\begin{equation}
    F^{(\pm)\,a}_{\mu\nu}F^{(\pm)\,a}_{\rho\sigma}=\mathfrak P^{(\pm)}_{AB}\mathbb F^A_{\mu\nu}\mathbb F^B_{\rho\sigma}=\mathfrak P^{(\pm)}_{MN}\mathbb F^M_{\mu\nu}\mathbb F^N_{\rho\sigma}.
\end{equation}
The first equality embeds the $O(d)_\pm$ invariant expression into an $O(d)_-\times O(d)_+$ one, and the second actually expresses it in an $O(d,d)$ invariant form. We note that we can neither build $F^{(+)\,a}_{\mu\nu}F^{(-)\,a}_{\rho\sigma}$ out of $O(d,d)$ invariant objects nor ${O(d)_-\times O(d)_+}$ ones.

The scalar terms can be built from the Maurer-Cartan form
\begin{equation}
    L^A{}_B=\dd\mathcal V_M^A\mathcal V^M_B=\begin{pmatrix}
        Q^{(--)}&\quad P^{(-+)}\\P^{(+-)}&\quad Q^{(++)}
    \end{pmatrix},\label{eq:MaurerCartan}
\end{equation}
satisfying
\begin{equation}
    \dd L^A{}_B+L^A{}_C\land L^C{}_B=0.
\end{equation}
Thus, we can find that, for example,
\begin{align}
    F^{(\pm)\,a}_{\mu\nu}P^{(\pm\mp)}_{\rho ab}F^{(\mp)\,b}_{\sigma\lambda}&=\mathfrak P^{(\pm)}_{AB}\mathfrak P^{(\mp)}_{CD}\mathbb F^A_{\mu\nu}L_{\rho}{}^{BC}\mathbb F^D_{\sigma\lambda},\nn\\
    P^{(+-)\,ab}_\mu P^{(-+)\,bc}_\nu P^{(+-)\,cd}_\rho P^{(-+)\,da}_\sigma &=\mathfrak P^{(-)}_{BC}\mathfrak P^{(+)}_{CD}\mathfrak P^{(-)}_{DE}\mathfrak P^{(+)}_{FG}\mathfrak P^{(-)}_{HA}L_\mu{}^{AB}L_\nu{}^{CD}L_\rho{}^{EF} L_\sigma{}^{GH}.
\end{align}
Indeed, any non-differentiated $(\pm)$ term can be obtained by substituting in $\mathfrak P^{(\pm)}_{MN}$ contracted with appropriate factors of $\mathbb F$ and $L$.

The only terms that remain to be dealt with are ones with derivatives. These are quite easy in tangent space since $\mathfrak P^{(\pm)}_{AB}$ is constant; however, we have to convert the rigid indices to curved indices, which can be done via
\begin{align}
    \partial_\mu\mathcal V_M^A&=\partial_\mu\mathcal V_N^A\mathcal H^{NP}\mathcal H_{PM}=\frac{1}{4}\partial_\mu\mathcal V_N^A\mathcal V^N_B\mathcal V^P_B\mathcal V_P^C\mathcal V_M^C=\frac{1}{4}L_\mu{}^A{}_B\mathcal V_M^B=\frac{1}{4}L_\mu{}^A{}_M,
\end{align}
and similarly
\begin{equation}
    \partial_\mu\mathcal V_A^M=\mathcal H_{AB}\partial_\mu\qty(\mathcal H^{MN}\mathcal V_N^B)=\mathcal V_{NA}\partial_\mu\mathcal H^{MN}+\frac{1}{4}\mathcal H^{MN}L_{\mu AN},
\end{equation}
where we have used \eqref{eq:cosetApp} and \eqref{eq:MaurerCartan}. For example, as a slight abuse of notation, we find that
\begin{align}
    \mathcal D_\mu^{(-)}\mathbb F^A_{\nu\rho}&=\mathcal V^A_M\qty(\mathcal D_\mu^{(-)}\mathbb F^M_{\nu\rho}+\frac{1}{4}L_\mu{}^M{}_N\mathbb F^N_{\nu\rho}),\nn\\
    \mathcal D_\mu^{\prime(-)}\mathbb F^A_{\nu\rho}&=\mathcal V^A_M\qty(\mathcal D_\mu^{\prime(-)}\mathbb F^M_{\nu\rho}+\frac{1}{4}L_\mu{}^M{}_N\mathbb F^N_{\nu\rho}),\nn\\
    \mathcal D^{(-)}_\mu L_\nu{}^{AB}&=\mathcal V^A_M\mathcal V^B_N\qty(\mathcal D^{(-)}_\mu L_\nu{}^{MN}+\frac{1}{4}L_\mu{}^M{}_PL_\nu{}^{PN}+\frac{1}{4}L_\mu{}^N{}_PL_\nu{}^{MP}).
\end{align}
This suggests that we define new derivatives $\mathfrak D^{(-)}$ and $\mathfrak D^{\prime(-)}$ in analogy to $\mathcal D^{(-)}$ and $\mathcal D^{\prime (-)}$, covariant with respect to the additional $\tfrac{1}{4}L$. Thus, dealing with derivatives amounts to a replacement $\mathcal D\to\mathfrak D$. Hence, with these identities, we can always write any ${O(d)_-\times O(d)_+}$ invariant expression in an $O(d,d)$ invariant form.

\let\oldaddcontentsline\addcontentsline
\renewcommand{\addcontentsline}[3]{}
\bibliography{cite}

\begin{thebibliography}{130}%
\makeatletter
\providecommand \@ifxundefined [1]{%
 \@ifx{#1\undefined}
}%
\providecommand \@ifnum [1]{%
 \ifnum #1\expandafter \@firstoftwo
 \else \expandafter \@secondoftwo
 \fi
}%
\providecommand \@ifx [1]{%
 \ifx #1\expandafter \@firstoftwo
 \else \expandafter \@secondoftwo
 \fi
}%
\providecommand \natexlab [1]{#1}%
\providecommand \enquote  [1]{``#1''}%
\providecommand \bibnamefont  [1]{#1}%
\providecommand \bibfnamefont [1]{#1}%
\providecommand \citenamefont [1]{#1}%
\providecommand \href@noop [0]{\@secondoftwo}%
\providecommand \href [0]{\begingroup \@sanitize@url \@href}%
\providecommand \@href[1]{\@@startlink{#1}\@@href}%
\providecommand \@@href[1]{\endgroup#1\@@endlink}%
\providecommand \@sanitize@url [0]{\catcode `\\12\catcode `\$12\catcode `\&12\catcode `\#12\catcode `\^12\catcode `\_12\catcode `\%12\relax}%
\providecommand \@@startlink[1]{}%
\providecommand \@@endlink[0]{}%
\providecommand \url  [0]{\begingroup\@sanitize@url \@url }%
\providecommand \@url [1]{\endgroup\@href {#1}{\urlprefix }}%
\providecommand \urlprefix  [0]{URL }%
\providecommand \Eprint [0]{\href }%
\providecommand \doibase [0]{https://doi.org/}%
\providecommand \selectlanguage [0]{\@gobble}%
\providecommand \bibinfo  [0]{\@secondoftwo}%
\providecommand \bibfield  [0]{\@secondoftwo}%
\providecommand \translation [1]{[#1]}%
\providecommand \BibitemOpen [0]{}%
\providecommand \bibitemStop [0]{}%
\providecommand \bibitemNoStop [0]{.\EOS\space}%
\providecommand \EOS [0]{\spacefactor3000\relax}%
\providecommand \BibitemShut  [1]{\csname bibitem#1\endcsname}%
\let\auto@bib@innerbib\@empty
\bibitem [{\citenamefont {Veneziano}(1991)}]{Veneziano:1991ek}%
  \BibitemOpen
  \bibfield  {author} {\bibinfo {author} {\bibfnamefont {G.}~\bibnamefont {Veneziano}},\ }\bibfield  {title} {\bibinfo {title} {{Scale factor duality for classical and quantum strings}},\ }\href {https://doi.org/10.1016/0370-2693(91)90055-U} {\bibfield  {journal} {\bibinfo  {journal} {Phys. Lett. B}\ }\textbf {\bibinfo {volume} {265}},\ \bibinfo {pages} {287} (\bibinfo {year} {1991})}\BibitemShut {NoStop}%
\bibitem [{\citenamefont {Meissner}\ and\ \citenamefont {Veneziano}(1991)}]{Meissner:1991zj}%
  \BibitemOpen
  \bibfield  {author} {\bibinfo {author} {\bibfnamefont {K.~A.}\ \bibnamefont {Meissner}}\ and\ \bibinfo {author} {\bibfnamefont {G.}~\bibnamefont {Veneziano}},\ }\bibfield  {title} {\bibinfo {title} {{Symmetries of cosmological superstring vacua}},\ }\href {https://doi.org/10.1016/0370-2693(91)90520-Z} {\bibfield  {journal} {\bibinfo  {journal} {Phys. Lett. B}\ }\textbf {\bibinfo {volume} {267}},\ \bibinfo {pages} {33} (\bibinfo {year} {1991})}\BibitemShut {NoStop}%
\bibitem [{\citenamefont {Sen}(1991)}]{Sen:1991zi}%
  \BibitemOpen
  \bibfield  {author} {\bibinfo {author} {\bibfnamefont {A.}~\bibnamefont {Sen}},\ }\bibfield  {title} {\bibinfo {title} {{O(d) x O(d) symmetry of the space of cosmological solutions in string theory, scale factor duality and two-dimensional black holes}},\ }\href {https://doi.org/10.1016/0370-2693(91)90090-D} {\bibfield  {journal} {\bibinfo  {journal} {Phys. Lett. B}\ }\textbf {\bibinfo {volume} {271}},\ \bibinfo {pages} {295} (\bibinfo {year} {1991})}\BibitemShut {NoStop}%
\bibitem [{\citenamefont {Sen}(1992{\natexlab{a}})}]{Sen:1991cn}%
  \BibitemOpen
  \bibfield  {author} {\bibinfo {author} {\bibfnamefont {A.}~\bibnamefont {Sen}},\ }\bibfield  {title} {\bibinfo {title} {{Twisted black p-brane solutions in string theory}},\ }\href {https://doi.org/10.1016/0370-2693(92)90300-S} {\bibfield  {journal} {\bibinfo  {journal} {Phys. Lett. B}\ }\textbf {\bibinfo {volume} {274}},\ \bibinfo {pages} {34} (\bibinfo {year} {1992}{\natexlab{a}})},\ \Eprint {https://arxiv.org/abs/hep-th/9108011} {arXiv:hep-th/9108011} \BibitemShut {NoStop}%
\bibitem [{\citenamefont {Gasperini}\ \emph {et~al.}(1991)\citenamefont {Gasperini}, \citenamefont {Maharana},\ and\ \citenamefont {Veneziano}}]{Gasperini:1991qy}%
  \BibitemOpen
  \bibfield  {author} {\bibinfo {author} {\bibfnamefont {M.}~\bibnamefont {Gasperini}}, \bibinfo {author} {\bibfnamefont {J.}~\bibnamefont {Maharana}},\ and\ \bibinfo {author} {\bibfnamefont {G.}~\bibnamefont {Veneziano}},\ }\bibfield  {title} {\bibinfo {title} {{From trivial to nontrivial conformal string backgrounds via O(d,d) transformations}},\ }\href {https://doi.org/10.1016/0370-2693(91)91831-F} {\bibfield  {journal} {\bibinfo  {journal} {Phys. Lett. B}\ }\textbf {\bibinfo {volume} {272}},\ \bibinfo {pages} {277} (\bibinfo {year} {1991})}\BibitemShut {NoStop}%
\bibitem [{\citenamefont {Hassan}\ and\ \citenamefont {Sen}(1992)}]{Hassan:1991mq}%
  \BibitemOpen
  \bibfield  {author} {\bibinfo {author} {\bibfnamefont {S.~F.}\ \bibnamefont {Hassan}}\ and\ \bibinfo {author} {\bibfnamefont {A.}~\bibnamefont {Sen}},\ }\bibfield  {title} {\bibinfo {title} {{Twisting classical solutions in heterotic string theory}},\ }\href {https://doi.org/10.1016/0550-3213(92)90336-A} {\bibfield  {journal} {\bibinfo  {journal} {Nucl. Phys. B}\ }\textbf {\bibinfo {volume} {375}},\ \bibinfo {pages} {103} (\bibinfo {year} {1992})},\ \Eprint {https://arxiv.org/abs/hep-th/9109038} {arXiv:hep-th/9109038} \BibitemShut {NoStop}%
\bibitem [{\citenamefont {Sen}(1992{\natexlab{b}})}]{Sen:1992ua}%
  \BibitemOpen
  \bibfield  {author} {\bibinfo {author} {\bibfnamefont {A.}~\bibnamefont {Sen}},\ }\bibfield  {title} {\bibinfo {title} {{Rotating charged black hole solution in heterotic string theory}},\ }\href {https://doi.org/10.1103/PhysRevLett.69.1006} {\bibfield  {journal} {\bibinfo  {journal} {Phys. Rev. Lett.}\ }\textbf {\bibinfo {volume} {69}},\ \bibinfo {pages} {1006} (\bibinfo {year} {1992}{\natexlab{b}})},\ \Eprint {https://arxiv.org/abs/hep-th/9204046} {arXiv:hep-th/9204046} \BibitemShut {NoStop}%
\bibitem [{\citenamefont {Cvetic}\ and\ \citenamefont {Youm}(1995)}]{Cvetic:1995sz}%
  \BibitemOpen
  \bibfield  {author} {\bibinfo {author} {\bibfnamefont {M.}~\bibnamefont {Cvetic}}\ and\ \bibinfo {author} {\bibfnamefont {D.}~\bibnamefont {Youm}},\ }\bibfield  {title} {\bibinfo {title} {{All the four-dimensional static, spherically symmetric solutions of Abelian Kaluza-Klein theory}},\ }\href {https://doi.org/10.1103/PhysRevLett.75.4165} {\bibfield  {journal} {\bibinfo  {journal} {Phys. Rev. Lett.}\ }\textbf {\bibinfo {volume} {75}},\ \bibinfo {pages} {4165} (\bibinfo {year} {1995})},\ \Eprint {https://arxiv.org/abs/hep-th/9503082} {arXiv:hep-th/9503082} \BibitemShut {NoStop}%
\bibitem [{\citenamefont {Cvetic}\ and\ \citenamefont {Youm}(1996{\natexlab{a}})}]{Cvetic:1995kv}%
  \BibitemOpen
  \bibfield  {author} {\bibinfo {author} {\bibfnamefont {M.}~\bibnamefont {Cvetic}}\ and\ \bibinfo {author} {\bibfnamefont {D.}~\bibnamefont {Youm}},\ }\bibfield  {title} {\bibinfo {title} {{All the static spherically symmetric black holes of heterotic string on a six torus}},\ }\href {https://doi.org/10.1016/0550-3213(96)00219-2} {\bibfield  {journal} {\bibinfo  {journal} {Nucl. Phys. B}\ }\textbf {\bibinfo {volume} {472}},\ \bibinfo {pages} {249} (\bibinfo {year} {1996}{\natexlab{a}})},\ \Eprint {https://arxiv.org/abs/hep-th/9512127} {arXiv:hep-th/9512127} \BibitemShut {NoStop}%
\bibitem [{\citenamefont {Cvetic}\ and\ \citenamefont {Youm}(1996{\natexlab{b}})}]{Cvetic:1996xz}%
  \BibitemOpen
  \bibfield  {author} {\bibinfo {author} {\bibfnamefont {M.}~\bibnamefont {Cvetic}}\ and\ \bibinfo {author} {\bibfnamefont {D.}~\bibnamefont {Youm}},\ }\bibfield  {title} {\bibinfo {title} {{General rotating five-dimensional black holes of toroidally compactified heterotic string}},\ }\href {https://doi.org/10.1016/0550-3213(96)00355-0} {\bibfield  {journal} {\bibinfo  {journal} {Nucl. Phys. B}\ }\textbf {\bibinfo {volume} {476}},\ \bibinfo {pages} {118} (\bibinfo {year} {1996}{\natexlab{b}})},\ \Eprint {https://arxiv.org/abs/hep-th/9603100} {arXiv:hep-th/9603100} \BibitemShut {NoStop}%
\bibitem [{\citenamefont {Rogatko}(2010)}]{Rogatko:2010hf}%
  \BibitemOpen
  \bibfield  {author} {\bibinfo {author} {\bibfnamefont {M.}~\bibnamefont {Rogatko}},\ }\bibfield  {title} {\bibinfo {title} {{Uniqueness Theorem for Stationary Axisymmetric Black Holes in Einstein-Maxwell-axion-dilaton Gravity}},\ }\href {https://doi.org/10.1103/PhysRevD.82.044017} {\bibfield  {journal} {\bibinfo  {journal} {Phys. Rev. D}\ }\textbf {\bibinfo {volume} {82}},\ \bibinfo {pages} {044017} (\bibinfo {year} {2010})},\ \Eprint {https://arxiv.org/abs/1007.4374} {arXiv:1007.4374 [hep-th]} \BibitemShut {NoStop}%
\bibitem [{\citenamefont {Fierz}(1956)}]{Fierz:1956zz}%
  \BibitemOpen
  \bibfield  {author} {\bibinfo {author} {\bibfnamefont {M.}~\bibnamefont {Fierz}},\ }\bibfield  {title} {\bibinfo {title} {{On the physical interpretation of P.Jordan's extended theory of gravitation}},\ }\href@noop {} {\bibfield  {journal} {\bibinfo  {journal} {Helv. Phys. Acta}\ }\textbf {\bibinfo {volume} {29}},\ \bibinfo {pages} {128} (\bibinfo {year} {1956})}\BibitemShut {NoStop}%
\bibitem [{\citenamefont {Brans}\ and\ \citenamefont {Dicke}(1961)}]{Brans:1961sx}%
  \BibitemOpen
  \bibfield  {author} {\bibinfo {author} {\bibfnamefont {C.}~\bibnamefont {Brans}}\ and\ \bibinfo {author} {\bibfnamefont {R.~H.}\ \bibnamefont {Dicke}},\ }\bibfield  {title} {\bibinfo {title} {{Mach's principle and a relativistic theory of gravitation}},\ }\href {https://doi.org/10.1103/PhysRev.124.925} {\bibfield  {journal} {\bibinfo  {journal} {Phys. Rev.}\ }\textbf {\bibinfo {volume} {124}},\ \bibinfo {pages} {925} (\bibinfo {year} {1961})}\BibitemShut {NoStop}%
\bibitem [{\citenamefont {Damour}\ and\ \citenamefont {Nordtvedt}(1993)}]{Damour:1992kf}%
  \BibitemOpen
  \bibfield  {author} {\bibinfo {author} {\bibfnamefont {T.}~\bibnamefont {Damour}}\ and\ \bibinfo {author} {\bibfnamefont {K.}~\bibnamefont {Nordtvedt}},\ }\bibfield  {title} {\bibinfo {title} {{General relativity as a cosmological attractor of tensor scalar theories}},\ }\href {https://doi.org/10.1103/PhysRevLett.70.2217} {\bibfield  {journal} {\bibinfo  {journal} {Phys. Rev. Lett.}\ }\textbf {\bibinfo {volume} {70}},\ \bibinfo {pages} {2217} (\bibinfo {year} {1993})}\BibitemShut {NoStop}%
\bibitem [{\citenamefont {Damour}\ and\ \citenamefont {Polyakov}(1994)}]{Damour:1994zq}%
  \BibitemOpen
  \bibfield  {author} {\bibinfo {author} {\bibfnamefont {T.}~\bibnamefont {Damour}}\ and\ \bibinfo {author} {\bibfnamefont {A.~M.}\ \bibnamefont {Polyakov}},\ }\bibfield  {title} {\bibinfo {title} {{The String dilaton and a least coupling principle}},\ }\href {https://doi.org/10.1016/0550-3213(94)90143-0} {\bibfield  {journal} {\bibinfo  {journal} {Nucl. Phys. B}\ }\textbf {\bibinfo {volume} {423}},\ \bibinfo {pages} {532} (\bibinfo {year} {1994})},\ \Eprint {https://arxiv.org/abs/hep-th/9401069} {arXiv:hep-th/9401069} \BibitemShut {NoStop}%
\bibitem [{\citenamefont {Catena}\ \emph {et~al.}(2007)\citenamefont {Catena}, \citenamefont {Pietroni},\ and\ \citenamefont {Scarabello}}]{Catena:2006bd}%
  \BibitemOpen
  \bibfield  {author} {\bibinfo {author} {\bibfnamefont {R.}~\bibnamefont {Catena}}, \bibinfo {author} {\bibfnamefont {M.}~\bibnamefont {Pietroni}},\ and\ \bibinfo {author} {\bibfnamefont {L.}~\bibnamefont {Scarabello}},\ }\bibfield  {title} {\bibinfo {title} {{Einstein and Jordan reconciled: a frame-invariant approach to scalar-tensor cosmology}},\ }\href {https://doi.org/10.1103/PhysRevD.76.084039} {\bibfield  {journal} {\bibinfo  {journal} {Phys. Rev. D}\ }\textbf {\bibinfo {volume} {76}},\ \bibinfo {pages} {084039} (\bibinfo {year} {2007})},\ \Eprint {https://arxiv.org/abs/astro-ph/0604492} {arXiv:astro-ph/0604492} \BibitemShut {NoStop}%
\bibitem [{\citenamefont {Damour}(1996)}]{Damour:1996xx}%
  \BibitemOpen
  \bibfield  {author} {\bibinfo {author} {\bibfnamefont {T.}~\bibnamefont {Damour}},\ }\bibfield  {title} {\bibinfo {title} {{Gravitation, experiment and cosmology}},\ }in\ \href@noop {} {\emph {\bibinfo {booktitle} {{Les Houches Summer School on Gravitation and Quantizations, Session 57}}}}\ (\bibinfo {year} {1996})\ \Eprint {https://arxiv.org/abs/gr-qc/9606079} {arXiv:gr-qc/9606079} \BibitemShut {NoStop}%
\bibitem [{\citenamefont {Bartolo}\ and\ \citenamefont {Pietroni}(2000)}]{Bartolo:1999sq}%
  \BibitemOpen
  \bibfield  {author} {\bibinfo {author} {\bibfnamefont {N.}~\bibnamefont {Bartolo}}\ and\ \bibinfo {author} {\bibfnamefont {M.}~\bibnamefont {Pietroni}},\ }\bibfield  {title} {\bibinfo {title} {{Scalar tensor gravity and quintessence}},\ }\href {https://doi.org/10.1103/PhysRevD.61.023518} {\bibfield  {journal} {\bibinfo  {journal} {Phys. Rev. D}\ }\textbf {\bibinfo {volume} {61}},\ \bibinfo {pages} {023518} (\bibinfo {year} {2000})},\ \Eprint {https://arxiv.org/abs/hep-ph/9908521} {arXiv:hep-ph/9908521} \BibitemShut {NoStop}%
\bibitem [{\citenamefont {Esposito-Farese}\ and\ \citenamefont {Polarski}(2001)}]{Esposito-Farese:2000pbo}%
  \BibitemOpen
  \bibfield  {author} {\bibinfo {author} {\bibfnamefont {G.}~\bibnamefont {Esposito-Farese}}\ and\ \bibinfo {author} {\bibfnamefont {D.}~\bibnamefont {Polarski}},\ }\bibfield  {title} {\bibinfo {title} {{Scalar tensor gravity in an accelerating universe}},\ }\href {https://doi.org/10.1103/PhysRevD.63.063504} {\bibfield  {journal} {\bibinfo  {journal} {Phys. Rev. D}\ }\textbf {\bibinfo {volume} {63}},\ \bibinfo {pages} {063504} (\bibinfo {year} {2001})},\ \Eprint {https://arxiv.org/abs/gr-qc/0009034} {arXiv:gr-qc/0009034} \BibitemShut {NoStop}%
\bibitem [{\citenamefont {Gannouji}\ \emph {et~al.}(2006)\citenamefont {Gannouji}, \citenamefont {Polarski}, \citenamefont {Ranquet},\ and\ \citenamefont {Starobinsky}}]{Gannouji:2006jm}%
  \BibitemOpen
  \bibfield  {author} {\bibinfo {author} {\bibfnamefont {R.}~\bibnamefont {Gannouji}}, \bibinfo {author} {\bibfnamefont {D.}~\bibnamefont {Polarski}}, \bibinfo {author} {\bibfnamefont {A.}~\bibnamefont {Ranquet}},\ and\ \bibinfo {author} {\bibfnamefont {A.~A.}\ \bibnamefont {Starobinsky}},\ }\bibfield  {title} {\bibinfo {title} {{Scalar-Tensor Models of Normal and Phantom Dark Energy}},\ }\href {https://doi.org/10.1088/1475-7516/2006/09/016} {\bibfield  {journal} {\bibinfo  {journal} {JCAP}\ }\textbf {\bibinfo {volume} {09}},\ \bibinfo {pages} {016}},\ \Eprint {https://arxiv.org/abs/astro-ph/0606287} {arXiv:astro-ph/0606287} \BibitemShut {NoStop}%
\bibitem [{\citenamefont {Garcia-Bellido}\ and\ \citenamefont {Wands}(1995)}]{Garcia-Bellido:1995him}%
  \BibitemOpen
  \bibfield  {author} {\bibinfo {author} {\bibfnamefont {J.}~\bibnamefont {Garcia-Bellido}}\ and\ \bibinfo {author} {\bibfnamefont {D.}~\bibnamefont {Wands}},\ }\bibfield  {title} {\bibinfo {title} {{Constraints from inflation on scalar - tensor gravity theories}},\ }\href {https://doi.org/10.1103/PhysRevD.52.6739} {\bibfield  {journal} {\bibinfo  {journal} {Phys. Rev. D}\ }\textbf {\bibinfo {volume} {52}},\ \bibinfo {pages} {6739} (\bibinfo {year} {1995})},\ \Eprint {https://arxiv.org/abs/gr-qc/9506050} {arXiv:gr-qc/9506050} \BibitemShut {NoStop}%
\bibitem [{\citenamefont {Sonner}\ and\ \citenamefont {Townsend}(2006)}]{Sonner:2006yn}%
  \BibitemOpen
  \bibfield  {author} {\bibinfo {author} {\bibfnamefont {J.}~\bibnamefont {Sonner}}\ and\ \bibinfo {author} {\bibfnamefont {P.~K.}\ \bibnamefont {Townsend}},\ }\bibfield  {title} {\bibinfo {title} {{Recurrent acceleration in dilaton-axion cosmology}},\ }\href {https://doi.org/10.1103/PhysRevD.74.103508} {\bibfield  {journal} {\bibinfo  {journal} {Phys. Rev. D}\ }\textbf {\bibinfo {volume} {74}},\ \bibinfo {pages} {103508} (\bibinfo {year} {2006})},\ \Eprint {https://arxiv.org/abs/hep-th/0608068} {arXiv:hep-th/0608068} \BibitemShut {NoStop}%
\bibitem [{\citenamefont {Catena}\ and\ \citenamefont {Moller}(2008)}]{Catena:2007jf}%
  \BibitemOpen
  \bibfield  {author} {\bibinfo {author} {\bibfnamefont {R.}~\bibnamefont {Catena}}\ and\ \bibinfo {author} {\bibfnamefont {J.}~\bibnamefont {Moller}},\ }\bibfield  {title} {\bibinfo {title} {{Axion-dilaton cosmology and dark energy}},\ }\href {https://doi.org/10.1088/1475-7516/2008/03/012} {\bibfield  {journal} {\bibinfo  {journal} {JCAP}\ }\textbf {\bibinfo {volume} {03}},\ \bibinfo {pages} {012}},\ \Eprint {https://arxiv.org/abs/0709.1931} {arXiv:0709.1931 [hep-ph]} \BibitemShut {NoStop}%
\bibitem [{\citenamefont {Linde}(1991)}]{Linde:1991km}%
  \BibitemOpen
  \bibfield  {author} {\bibinfo {author} {\bibfnamefont {A.~D.}\ \bibnamefont {Linde}},\ }\bibfield  {title} {\bibinfo {title} {{Axions in inflationary cosmology}},\ }\href {https://doi.org/10.1016/0370-2693(91)90130-I} {\bibfield  {journal} {\bibinfo  {journal} {Phys. Lett. B}\ }\textbf {\bibinfo {volume} {259}},\ \bibinfo {pages} {38} (\bibinfo {year} {1991})}\BibitemShut {NoStop}%
\bibitem [{\citenamefont {Catena}\ \emph {et~al.}(2004{\natexlab{a}})\citenamefont {Catena}, \citenamefont {Fornengo}, \citenamefont {Masiero}, \citenamefont {Pietroni},\ and\ \citenamefont {Rosati}}]{Catena:2004ba}%
  \BibitemOpen
  \bibfield  {author} {\bibinfo {author} {\bibfnamefont {R.}~\bibnamefont {Catena}}, \bibinfo {author} {\bibfnamefont {N.}~\bibnamefont {Fornengo}}, \bibinfo {author} {\bibfnamefont {A.}~\bibnamefont {Masiero}}, \bibinfo {author} {\bibfnamefont {M.}~\bibnamefont {Pietroni}},\ and\ \bibinfo {author} {\bibfnamefont {F.}~\bibnamefont {Rosati}},\ }\bibfield  {title} {\bibinfo {title} {{Dark matter relic abundance and scalar - tensor dark energy}},\ }\href {https://doi.org/10.1103/PhysRevD.70.063519} {\bibfield  {journal} {\bibinfo  {journal} {Phys. Rev. D}\ }\textbf {\bibinfo {volume} {70}},\ \bibinfo {pages} {063519} (\bibinfo {year} {2004}{\natexlab{a}})},\ \Eprint {https://arxiv.org/abs/astro-ph/0403614} {arXiv:astro-ph/0403614} \BibitemShut {NoStop}%
\bibitem [{\citenamefont {Catena}\ \emph {et~al.}(2004{\natexlab{b}})\citenamefont {Catena}, \citenamefont {Pietroni},\ and\ \citenamefont {Scarabello}}]{Catena:2004pz}%
  \BibitemOpen
  \bibfield  {author} {\bibinfo {author} {\bibfnamefont {R.}~\bibnamefont {Catena}}, \bibinfo {author} {\bibfnamefont {M.}~\bibnamefont {Pietroni}},\ and\ \bibinfo {author} {\bibfnamefont {L.}~\bibnamefont {Scarabello}},\ }\bibfield  {title} {\bibinfo {title} {{Dynamical relaxation of the dark matter to baryon ratio}},\ }\href {https://doi.org/10.1103/PhysRevD.70.103526} {\bibfield  {journal} {\bibinfo  {journal} {Phys. Rev. D}\ }\textbf {\bibinfo {volume} {70}},\ \bibinfo {pages} {103526} (\bibinfo {year} {2004}{\natexlab{b}})},\ \Eprint {https://arxiv.org/abs/astro-ph/0407646} {arXiv:astro-ph/0407646} \BibitemShut {NoStop}%
\bibitem [{\citenamefont {Schelke}\ \emph {et~al.}(2006)\citenamefont {Schelke}, \citenamefont {Catena}, \citenamefont {Fornengo}, \citenamefont {Masiero},\ and\ \citenamefont {Pietroni}}]{Schelke:2006eg}%
  \BibitemOpen
  \bibfield  {author} {\bibinfo {author} {\bibfnamefont {M.}~\bibnamefont {Schelke}}, \bibinfo {author} {\bibfnamefont {R.}~\bibnamefont {Catena}}, \bibinfo {author} {\bibfnamefont {N.}~\bibnamefont {Fornengo}}, \bibinfo {author} {\bibfnamefont {A.}~\bibnamefont {Masiero}},\ and\ \bibinfo {author} {\bibfnamefont {M.}~\bibnamefont {Pietroni}},\ }\bibfield  {title} {\bibinfo {title} {{Constraining pre Big-Bang-Nucleosynthesis Expansion using Cosmic Antiprotons}},\ }\href {https://doi.org/10.1103/PhysRevD.74.083505} {\bibfield  {journal} {\bibinfo  {journal} {Phys. Rev. D}\ }\textbf {\bibinfo {volume} {74}},\ \bibinfo {pages} {083505} (\bibinfo {year} {2006})},\ \Eprint {https://arxiv.org/abs/hep-ph/0605287} {arXiv:hep-ph/0605287} \BibitemShut {NoStop}%
\bibitem [{\citenamefont {Campbell}\ \emph {et~al.}(1993)\citenamefont {Campbell}, \citenamefont {Kaloper}, \citenamefont {Madden},\ and\ \citenamefont {Olive}}]{Campbell:1992hc}%
  \BibitemOpen
  \bibfield  {author} {\bibinfo {author} {\bibfnamefont {B.~A.}\ \bibnamefont {Campbell}}, \bibinfo {author} {\bibfnamefont {N.}~\bibnamefont {Kaloper}}, \bibinfo {author} {\bibfnamefont {R.}~\bibnamefont {Madden}},\ and\ \bibinfo {author} {\bibfnamefont {K.~A.}\ \bibnamefont {Olive}},\ }\bibfield  {title} {\bibinfo {title} {{Physical properties of four-dimensional superstring gravity black hole solutions}},\ }\href {https://doi.org/10.1016/0550-3213(93)90620-5} {\bibfield  {journal} {\bibinfo  {journal} {Nucl. Phys. B}\ }\textbf {\bibinfo {volume} {399}},\ \bibinfo {pages} {137} (\bibinfo {year} {1993})},\ \Eprint {https://arxiv.org/abs/hep-th/9301129} {arXiv:hep-th/9301129} \BibitemShut {NoStop}%
\bibitem [{\citenamefont {Dastan}\ \emph {et~al.}(2016)\citenamefont {Dastan}, \citenamefont {Saffari},\ and\ \citenamefont {Soroushfar}}]{Dastan:2016bfy}%
  \BibitemOpen
  \bibfield  {author} {\bibinfo {author} {\bibfnamefont {S.}~\bibnamefont {Dastan}}, \bibinfo {author} {\bibfnamefont {R.}~\bibnamefont {Saffari}},\ and\ \bibinfo {author} {\bibfnamefont {S.}~\bibnamefont {Soroushfar}},\ }\bibfield  {title} {\bibinfo {title} {{Shadow of a Kerr-Sen dilaton-axion Black Hole}},\ }\Eprint {https://arxiv.org/abs/1610.09477} {arXiv:1610.09477 [gr-qc]}  (\bibinfo {year} {2016})\BibitemShut {NoStop}%
\bibitem [{\citenamefont {Cvetic}\ \emph {et~al.}(2017)\citenamefont {Cvetic}, \citenamefont {Gibbons},\ and\ \citenamefont {Pope}}]{Cvetic:2017zde}%
  \BibitemOpen
  \bibfield  {author} {\bibinfo {author} {\bibfnamefont {M.}~\bibnamefont {Cvetic}}, \bibinfo {author} {\bibfnamefont {G.~W.}\ \bibnamefont {Gibbons}},\ and\ \bibinfo {author} {\bibfnamefont {C.~N.}\ \bibnamefont {Pope}},\ }\bibfield  {title} {\bibinfo {title} {{STU Black Holes and SgrA}},\ }\href {https://doi.org/10.1088/1475-7516/2017/08/016} {\bibfield  {journal} {\bibinfo  {journal} {JCAP}\ }\textbf {\bibinfo {volume} {08}},\ \bibinfo {pages} {016}},\ \Eprint {https://arxiv.org/abs/1705.05740} {arXiv:1705.05740 [gr-qc]} \BibitemShut {NoStop}%
\bibitem [{\citenamefont {Guo}\ \emph {et~al.}(2020)\citenamefont {Guo}, \citenamefont {Song},\ and\ \citenamefont {Yan}}]{Guo:2019lur}%
  \BibitemOpen
  \bibfield  {author} {\bibinfo {author} {\bibfnamefont {M.}~\bibnamefont {Guo}}, \bibinfo {author} {\bibfnamefont {S.}~\bibnamefont {Song}},\ and\ \bibinfo {author} {\bibfnamefont {H.}~\bibnamefont {Yan}},\ }\bibfield  {title} {\bibinfo {title} {{Observational signature of a near-extremal Kerr-Sen black hole in the heterotic string theory}},\ }\href {https://doi.org/10.1103/PhysRevD.101.024055} {\bibfield  {journal} {\bibinfo  {journal} {Phys. Rev. D}\ }\textbf {\bibinfo {volume} {101}},\ \bibinfo {pages} {024055} (\bibinfo {year} {2020})},\ \Eprint {https://arxiv.org/abs/1911.04796} {arXiv:1911.04796 [gr-qc]} \BibitemShut {NoStop}%
\bibitem [{\citenamefont {Narang}\ \emph {et~al.}(2020)\citenamefont {Narang}, \citenamefont {Mohanty},\ and\ \citenamefont {Kumar}}]{Narang:2020bgo}%
  \BibitemOpen
  \bibfield  {author} {\bibinfo {author} {\bibfnamefont {A.}~\bibnamefont {Narang}}, \bibinfo {author} {\bibfnamefont {S.}~\bibnamefont {Mohanty}},\ and\ \bibinfo {author} {\bibfnamefont {A.}~\bibnamefont {Kumar}},\ }\bibfield  {title} {\bibinfo {title} {{Test of Kerr-Sen metric with black hole observations}},\ }\Eprint {https://arxiv.org/abs/2002.12786} {arXiv:2002.12786 [gr-qc]}  (\bibinfo {year} {2020})\BibitemShut {NoStop}%
\bibitem [{\citenamefont {Xavier}\ \emph {et~al.}(2020)\citenamefont {Xavier}, \citenamefont {Cunha}, \citenamefont {Crispino},\ and\ \citenamefont {Herdeiro}}]{Xavier:2020egv}%
  \BibitemOpen
  \bibfield  {author} {\bibinfo {author} {\bibfnamefont {S.~V. M. C.~B.}\ \bibnamefont {Xavier}}, \bibinfo {author} {\bibfnamefont {P.~V.~P.}\ \bibnamefont {Cunha}}, \bibinfo {author} {\bibfnamefont {L.~C.~B.}\ \bibnamefont {Crispino}},\ and\ \bibinfo {author} {\bibfnamefont {C.~A.~R.}\ \bibnamefont {Herdeiro}},\ }\bibfield  {title} {\bibinfo {title} {{Shadows of charged rotating black holes: Kerr\textendash{}Newman versus Kerr\textendash{}Sen}},\ }\href {https://doi.org/10.1142/S0218271820410059} {\bibfield  {journal} {\bibinfo  {journal} {Int. J. Mod. Phys. D}\ }\textbf {\bibinfo {volume} {29}},\ \bibinfo {pages} {2041005} (\bibinfo {year} {2020})},\ \Eprint {https://arxiv.org/abs/2003.14349} {arXiv:2003.14349 [gr-qc]} \BibitemShut {NoStop}%
\bibitem [{\citenamefont {Shahzadi}\ \emph {et~al.}(2022)\citenamefont {Shahzadi}, \citenamefont {Kolo\v{s}}, \citenamefont {Stuchl\'\i{}k},\ and\ \citenamefont {Habib}}]{Shahzadi:2022rzq}%
  \BibitemOpen
  \bibfield  {author} {\bibinfo {author} {\bibfnamefont {M.}~\bibnamefont {Shahzadi}}, \bibinfo {author} {\bibfnamefont {M.}~\bibnamefont {Kolo\v{s}}}, \bibinfo {author} {\bibfnamefont {Z.}~\bibnamefont {Stuchl\'\i{}k}},\ and\ \bibinfo {author} {\bibfnamefont {Y.}~\bibnamefont {Habib}},\ }\bibfield  {title} {\bibinfo {title} {{Testing alternative theories of gravity by fitting the hot-spot data of Sgr~A*}},\ }\href {https://doi.org/10.1140/epjc/s10052-022-10347-4} {\bibfield  {journal} {\bibinfo  {journal} {Eur. Phys. J. C}\ }\textbf {\bibinfo {volume} {82}},\ \bibinfo {pages} {407} (\bibinfo {year} {2022})},\ \Eprint {https://arxiv.org/abs/2201.04442} {arXiv:2201.04442 [gr-qc]} \BibitemShut {NoStop}%
\bibitem [{\citenamefont {Della~Monica}\ \emph {et~al.}(2023)\citenamefont {Della~Monica}, \citenamefont {de~Martino},\ and\ \citenamefont {de~Laurentis}}]{DellaMonica:2023ydm}%
  \BibitemOpen
  \bibfield  {author} {\bibinfo {author} {\bibfnamefont {R.}~\bibnamefont {Della~Monica}}, \bibinfo {author} {\bibfnamefont {I.}~\bibnamefont {de~Martino}},\ and\ \bibinfo {author} {\bibfnamefont {M.}~\bibnamefont {de~Laurentis}},\ }\bibfield  {title} {\bibinfo {title} {{Testing space\textendash{}time geometries and theories of gravity at the Galactic centre with pulsar\textquoteright{}s time delay}},\ }\href {https://doi.org/10.1093/mnras/stad2125} {\bibfield  {journal} {\bibinfo  {journal} {Mon. Not. Roy. Astron. Soc.}\ }\textbf {\bibinfo {volume} {524}},\ \bibinfo {pages} {3782} (\bibinfo {year} {2023})},\ \Eprint {https://arxiv.org/abs/2305.18178} {arXiv:2305.18178 [gr-qc]} \BibitemShut {NoStop}%
\bibitem [{\citenamefont {Sahoo}\ \emph {et~al.}(2024)\citenamefont {Sahoo}, \citenamefont {Yadav},\ and\ \citenamefont {Banerjee}}]{Sahoo:2023czj}%
  \BibitemOpen
  \bibfield  {author} {\bibinfo {author} {\bibfnamefont {S.~K.}\ \bibnamefont {Sahoo}}, \bibinfo {author} {\bibfnamefont {N.}~\bibnamefont {Yadav}},\ and\ \bibinfo {author} {\bibfnamefont {I.}~\bibnamefont {Banerjee}},\ }\bibfield  {title} {\bibinfo {title} {{Imprints of Einstein-Maxwell-dilaton-axion gravity in the observed shadows of Sgr A* and M87*}},\ }\href {https://doi.org/10.1103/PhysRevD.109.044008} {\bibfield  {journal} {\bibinfo  {journal} {Phys. Rev. D}\ }\textbf {\bibinfo {volume} {109}},\ \bibinfo {pages} {044008} (\bibinfo {year} {2024})},\ \Eprint {https://arxiv.org/abs/2305.14870} {arXiv:2305.14870 [gr-qc]} \BibitemShut {NoStop}%
\bibitem [{\citenamefont {Feng}\ \emph {et~al.}(2025)\citenamefont {Feng}, \citenamefont {Yang},\ and\ \citenamefont {Chen}}]{Feng:2024iqj}%
  \BibitemOpen
  \bibfield  {author} {\bibinfo {author} {\bibfnamefont {H.}~\bibnamefont {Feng}}, \bibinfo {author} {\bibfnamefont {R.-J.}\ \bibnamefont {Yang}},\ and\ \bibinfo {author} {\bibfnamefont {W.-Q.}\ \bibnamefont {Chen}},\ }\bibfield  {title} {\bibinfo {title} {{Thin accretion disk and shadow of Kerr\textendash{}Sen black hole in Einstein\textendash{}Maxwell-dilaton\textendash{}axion gravity}},\ }\href {https://doi.org/10.1016/j.astropartphys.2024.103075} {\bibfield  {journal} {\bibinfo  {journal} {Astropart. Phys.}\ }\textbf {\bibinfo {volume} {166}},\ \bibinfo {pages} {103075} (\bibinfo {year} {2025})},\ \Eprint {https://arxiv.org/abs/2403.18541} {arXiv:2403.18541 [gr-qc]} \BibitemShut {NoStop}%
\bibitem [{\citenamefont {Bena}\ and\ \citenamefont {Mayerson}(2020)}]{Bena:2020see}%
  \BibitemOpen
  \bibfield  {author} {\bibinfo {author} {\bibfnamefont {I.}~\bibnamefont {Bena}}\ and\ \bibinfo {author} {\bibfnamefont {D.~R.}\ \bibnamefont {Mayerson}},\ }\bibfield  {title} {\bibinfo {title} {{Multipole Ratios: A New Window into Black Holes}},\ }\href {https://doi.org/10.1103/PhysRevLett.125.221602} {\bibfield  {journal} {\bibinfo  {journal} {Phys. Rev. Lett.}\ }\textbf {\bibinfo {volume} {125}},\ \bibinfo {pages} {221602} (\bibinfo {year} {2020})},\ \Eprint {https://arxiv.org/abs/2006.10750} {arXiv:2006.10750 [hep-th]} \BibitemShut {NoStop}%
\bibitem [{\citenamefont {Bianchi}\ \emph {et~al.}(2020)\citenamefont {Bianchi}, \citenamefont {Consoli}, \citenamefont {Grillo}, \citenamefont {Morales}, \citenamefont {Pani},\ and\ \citenamefont {Raposo}}]{Bianchi:2020bxa}%
  \BibitemOpen
  \bibfield  {author} {\bibinfo {author} {\bibfnamefont {M.}~\bibnamefont {Bianchi}}, \bibinfo {author} {\bibfnamefont {D.}~\bibnamefont {Consoli}}, \bibinfo {author} {\bibfnamefont {A.}~\bibnamefont {Grillo}}, \bibinfo {author} {\bibfnamefont {J.~F.}\ \bibnamefont {Morales}}, \bibinfo {author} {\bibfnamefont {P.}~\bibnamefont {Pani}},\ and\ \bibinfo {author} {\bibfnamefont {G.}~\bibnamefont {Raposo}},\ }\bibfield  {title} {\bibinfo {title} {{Distinguishing fuzzballs from black holes through their multipolar structure}},\ }\href {https://doi.org/10.1103/PhysRevLett.125.221601} {\bibfield  {journal} {\bibinfo  {journal} {Phys. Rev. Lett.}\ }\textbf {\bibinfo {volume} {125}},\ \bibinfo {pages} {221601} (\bibinfo {year} {2020})},\ \Eprint {https://arxiv.org/abs/2007.01743} {arXiv:2007.01743 [hep-th]} \BibitemShut {NoStop}%
\bibitem [{\citenamefont {Bianchi}\ \emph {et~al.}(2021)\citenamefont {Bianchi}, \citenamefont {Consoli}, \citenamefont {Grillo}, \citenamefont {Morales}, \citenamefont {Pani},\ and\ \citenamefont {Raposo}}]{Bianchi:2020miz}%
  \BibitemOpen
  \bibfield  {author} {\bibinfo {author} {\bibfnamefont {M.}~\bibnamefont {Bianchi}}, \bibinfo {author} {\bibfnamefont {D.}~\bibnamefont {Consoli}}, \bibinfo {author} {\bibfnamefont {A.}~\bibnamefont {Grillo}}, \bibinfo {author} {\bibfnamefont {J.~F.}\ \bibnamefont {Morales}}, \bibinfo {author} {\bibfnamefont {P.}~\bibnamefont {Pani}},\ and\ \bibinfo {author} {\bibfnamefont {G.}~\bibnamefont {Raposo}},\ }\bibfield  {title} {\bibinfo {title} {{The multipolar structure of fuzzballs}},\ }\href {https://doi.org/10.1007/JHEP01(2021)003} {\bibfield  {journal} {\bibinfo  {journal} {JHEP}\ }\textbf {\bibinfo {volume} {01}},\ \bibinfo {pages} {003}},\ \Eprint {https://arxiv.org/abs/2008.01445} {arXiv:2008.01445 [hep-th]} \BibitemShut {NoStop}%
\bibitem [{\citenamefont {Bah}\ \emph {et~al.}(2021)\citenamefont {Bah}, \citenamefont {Bena}, \citenamefont {Heidmann}, \citenamefont {Li},\ and\ \citenamefont {Mayerson}}]{Bah:2021jno}%
  \BibitemOpen
  \bibfield  {author} {\bibinfo {author} {\bibfnamefont {I.}~\bibnamefont {Bah}}, \bibinfo {author} {\bibfnamefont {I.}~\bibnamefont {Bena}}, \bibinfo {author} {\bibfnamefont {P.}~\bibnamefont {Heidmann}}, \bibinfo {author} {\bibfnamefont {Y.}~\bibnamefont {Li}},\ and\ \bibinfo {author} {\bibfnamefont {D.~R.}\ \bibnamefont {Mayerson}},\ }\bibfield  {title} {\bibinfo {title} {{Gravitational footprints of black holes and their microstate geometries}},\ }\href {https://doi.org/10.1007/JHEP10(2021)138} {\bibfield  {journal} {\bibinfo  {journal} {JHEP}\ }\textbf {\bibinfo {volume} {10}},\ \bibinfo {pages} {138}},\ \Eprint {https://arxiv.org/abs/2104.10686} {arXiv:2104.10686 [hep-th]} \BibitemShut {NoStop}%
\bibitem [{\citenamefont {Amaro-Seoane}\ \emph {et~al.}(2017)\citenamefont {Amaro-Seoane} \emph {et~al.}}]{LISA:2017pwj}%
  \BibitemOpen
  \bibfield  {author} {\bibinfo {author} {\bibfnamefont {P.}~\bibnamefont {Amaro-Seoane}} \emph {et~al.} (\bibinfo {collaboration} {LISA}),\ }\bibfield  {title} {\bibinfo {title} {{Laser Interferometer Space Antenna}},\ }\Eprint {https://arxiv.org/abs/1702.00786} {arXiv:1702.00786 [astro-ph.IM]}  (\bibinfo {year} {2017})\BibitemShut {NoStop}%
\bibitem [{\citenamefont {Barack}\ and\ \citenamefont {Cutler}(2007)}]{Barack:2006pq}%
  \BibitemOpen
  \bibfield  {author} {\bibinfo {author} {\bibfnamefont {L.}~\bibnamefont {Barack}}\ and\ \bibinfo {author} {\bibfnamefont {C.}~\bibnamefont {Cutler}},\ }\bibfield  {title} {\bibinfo {title} {{Using LISA EMRI sources to test off-Kerr deviations in the geometry of massive black holes}},\ }\href {https://doi.org/10.1103/PhysRevD.75.042003} {\bibfield  {journal} {\bibinfo  {journal} {Phys. Rev. D}\ }\textbf {\bibinfo {volume} {75}},\ \bibinfo {pages} {042003} (\bibinfo {year} {2007})},\ \Eprint {https://arxiv.org/abs/gr-qc/0612029} {arXiv:gr-qc/0612029} \BibitemShut {NoStop}%
\bibitem [{\citenamefont {Cardoso}\ and\ \citenamefont {Gualtieri}(2016)}]{Cardoso:2016ryw}%
  \BibitemOpen
  \bibfield  {author} {\bibinfo {author} {\bibfnamefont {V.}~\bibnamefont {Cardoso}}\ and\ \bibinfo {author} {\bibfnamefont {L.}~\bibnamefont {Gualtieri}},\ }\bibfield  {title} {\bibinfo {title} {{Testing the black hole \textquoteleft{}no-hair\textquoteright{} hypothesis}},\ }\href {https://doi.org/10.1088/0264-9381/33/17/174001} {\bibfield  {journal} {\bibinfo  {journal} {Class. Quant. Grav.}\ }\textbf {\bibinfo {volume} {33}},\ \bibinfo {pages} {174001} (\bibinfo {year} {2016})},\ \Eprint {https://arxiv.org/abs/1607.03133} {arXiv:1607.03133 [gr-qc]} \BibitemShut {NoStop}%
\bibitem [{\citenamefont {Babak}\ \emph {et~al.}(2017)\citenamefont {Babak}, \citenamefont {Gair}, \citenamefont {Sesana}, \citenamefont {Barausse}, \citenamefont {Sopuerta}, \citenamefont {Berry}, \citenamefont {Berti}, \citenamefont {Amaro-Seoane}, \citenamefont {Petiteau},\ and\ \citenamefont {Klein}}]{Babak:2017tow}%
  \BibitemOpen
  \bibfield  {author} {\bibinfo {author} {\bibfnamefont {S.}~\bibnamefont {Babak}}, \bibinfo {author} {\bibfnamefont {J.}~\bibnamefont {Gair}}, \bibinfo {author} {\bibfnamefont {A.}~\bibnamefont {Sesana}}, \bibinfo {author} {\bibfnamefont {E.}~\bibnamefont {Barausse}}, \bibinfo {author} {\bibfnamefont {C.~F.}\ \bibnamefont {Sopuerta}}, \bibinfo {author} {\bibfnamefont {C.~P.~L.}\ \bibnamefont {Berry}}, \bibinfo {author} {\bibfnamefont {E.}~\bibnamefont {Berti}}, \bibinfo {author} {\bibfnamefont {P.}~\bibnamefont {Amaro-Seoane}}, \bibinfo {author} {\bibfnamefont {A.}~\bibnamefont {Petiteau}},\ and\ \bibinfo {author} {\bibfnamefont {A.}~\bibnamefont {Klein}},\ }\bibfield  {title} {\bibinfo {title} {{Science with the space-based interferometer LISA. V: Extreme mass-ratio inspirals}},\ }\href {https://doi.org/10.1103/PhysRevD.95.103012} {\bibfield  {journal} {\bibinfo  {journal} {Phys. Rev. D}\ }\textbf {\bibinfo {volume} {95}},\ \bibinfo {pages} {103012} (\bibinfo {year} {2017})},\ \Eprint
  {https://arxiv.org/abs/1703.09722} {arXiv:1703.09722 [gr-qc]} \BibitemShut {NoStop}%
\bibitem [{\citenamefont {Ryan}(1995)}]{Ryan:1995wh}%
  \BibitemOpen
  \bibfield  {author} {\bibinfo {author} {\bibfnamefont {F.~D.}\ \bibnamefont {Ryan}},\ }\bibfield  {title} {\bibinfo {title} {{Gravitational waves from the inspiral of a compact object into a massive, axisymmetric body with arbitrary multipole moments}},\ }\href {https://doi.org/10.1103/PhysRevD.52.5707} {\bibfield  {journal} {\bibinfo  {journal} {Phys. Rev. D}\ }\textbf {\bibinfo {volume} {52}},\ \bibinfo {pages} {5707} (\bibinfo {year} {1995})}\BibitemShut {NoStop}%
\bibitem [{\citenamefont {Krishnendu}\ \emph {et~al.}(2019)\citenamefont {Krishnendu}, \citenamefont {Mishra},\ and\ \citenamefont {Arun}}]{Krishnendu:2018nqa}%
  \BibitemOpen
  \bibfield  {author} {\bibinfo {author} {\bibfnamefont {N.~V.}\ \bibnamefont {Krishnendu}}, \bibinfo {author} {\bibfnamefont {C.~K.}\ \bibnamefont {Mishra}},\ and\ \bibinfo {author} {\bibfnamefont {K.~G.}\ \bibnamefont {Arun}},\ }\bibfield  {title} {\bibinfo {title} {{Spin-induced deformations and tests of binary black hole nature using third-generation detectors}},\ }\href {https://doi.org/10.1103/PhysRevD.99.064008} {\bibfield  {journal} {\bibinfo  {journal} {Phys. Rev. D}\ }\textbf {\bibinfo {volume} {99}},\ \bibinfo {pages} {064008} (\bibinfo {year} {2019})},\ \Eprint {https://arxiv.org/abs/1811.00317} {arXiv:1811.00317 [gr-qc]} \BibitemShut {NoStop}%
\bibitem [{\citenamefont {Rasheed}(1995)}]{Rasheed:1995zv}%
  \BibitemOpen
  \bibfield  {author} {\bibinfo {author} {\bibfnamefont {D.}~\bibnamefont {Rasheed}},\ }\bibfield  {title} {\bibinfo {title} {{The Rotating dyonic black holes of Kaluza-Klein theory}},\ }\href {https://doi.org/10.1016/0550-3213(95)00396-A} {\bibfield  {journal} {\bibinfo  {journal} {Nucl. Phys. B}\ }\textbf {\bibinfo {volume} {454}},\ \bibinfo {pages} {379} (\bibinfo {year} {1995})},\ \Eprint {https://arxiv.org/abs/hep-th/9505038} {arXiv:hep-th/9505038} \BibitemShut {NoStop}%
\bibitem [{\citenamefont {Larsen}(2000)}]{Larsen:1999pp}%
  \BibitemOpen
  \bibfield  {author} {\bibinfo {author} {\bibfnamefont {F.}~\bibnamefont {Larsen}},\ }\bibfield  {title} {\bibinfo {title} {{Rotating Kaluza-Klein black holes}},\ }\href {https://doi.org/10.1016/S0550-3213(00)00064-X} {\bibfield  {journal} {\bibinfo  {journal} {Nucl. Phys. B}\ }\textbf {\bibinfo {volume} {575}},\ \bibinfo {pages} {211} (\bibinfo {year} {2000})},\ \Eprint {https://arxiv.org/abs/hep-th/9909102} {arXiv:hep-th/9909102} \BibitemShut {NoStop}%
\bibitem [{\citenamefont {Bena}\ and\ \citenamefont {Mayerson}(2021)}]{Bena:2020uup}%
  \BibitemOpen
  \bibfield  {author} {\bibinfo {author} {\bibfnamefont {I.}~\bibnamefont {Bena}}\ and\ \bibinfo {author} {\bibfnamefont {D.~R.}\ \bibnamefont {Mayerson}},\ }\bibfield  {title} {\bibinfo {title} {{Black Holes Lessons from Multipole Ratios}},\ }\href {https://doi.org/10.1007/JHEP03(2021)114} {\bibfield  {journal} {\bibinfo  {journal} {JHEP}\ }\textbf {\bibinfo {volume} {03}},\ \bibinfo {pages} {114}},\ \Eprint {https://arxiv.org/abs/2007.09152} {arXiv:2007.09152 [hep-th]} \BibitemShut {NoStop}%
\bibitem [{\citenamefont {Hohm}\ \emph {et~al.}(2015)\citenamefont {Hohm}, \citenamefont {Sen},\ and\ \citenamefont {Zwiebach}}]{Hohm:2014sxa}%
  \BibitemOpen
  \bibfield  {author} {\bibinfo {author} {\bibfnamefont {O.}~\bibnamefont {Hohm}}, \bibinfo {author} {\bibfnamefont {A.}~\bibnamefont {Sen}},\ and\ \bibinfo {author} {\bibfnamefont {B.}~\bibnamefont {Zwiebach}},\ }\bibfield  {title} {\bibinfo {title} {{Heterotic Effective Action and Duality Symmetries Revisited}},\ }\href {https://doi.org/10.1007/JHEP02(2015)079} {\bibfield  {journal} {\bibinfo  {journal} {JHEP}\ }\textbf {\bibinfo {volume} {02}},\ \bibinfo {pages} {079}},\ \Eprint {https://arxiv.org/abs/1411.5696} {arXiv:1411.5696 [hep-th]} \BibitemShut {NoStop}%
\bibitem [{\citenamefont {Godazgar}\ and\ \citenamefont {Godazgar}(2013)}]{Godazgar:2013bja}%
  \BibitemOpen
  \bibfield  {author} {\bibinfo {author} {\bibfnamefont {H.}~\bibnamefont {Godazgar}}\ and\ \bibinfo {author} {\bibfnamefont {M.}~\bibnamefont {Godazgar}},\ }\bibfield  {title} {\bibinfo {title} {{Duality completion of higher derivative corrections}},\ }\href {https://doi.org/10.1007/JHEP09(2013)140} {\bibfield  {journal} {\bibinfo  {journal} {JHEP}\ }\textbf {\bibinfo {volume} {09}},\ \bibinfo {pages} {140}},\ \Eprint {https://arxiv.org/abs/1306.4918} {arXiv:1306.4918 [hep-th]} \BibitemShut {NoStop}%
\bibitem [{\citenamefont {Marques}\ and\ \citenamefont {Nunez}(2015)}]{Marques:2015vua}%
  \BibitemOpen
  \bibfield  {author} {\bibinfo {author} {\bibfnamefont {D.}~\bibnamefont {Marques}}\ and\ \bibinfo {author} {\bibfnamefont {C.~A.}\ \bibnamefont {Nunez}},\ }\bibfield  {title} {\bibinfo {title} {{T-duality and \ensuremath{\alpha}'-corrections}},\ }\href {https://doi.org/10.1007/JHEP10(2015)084} {\bibfield  {journal} {\bibinfo  {journal} {JHEP}\ }\textbf {\bibinfo {volume} {10}},\ \bibinfo {pages} {084}},\ \Eprint {https://arxiv.org/abs/1507.00652} {arXiv:1507.00652 [hep-th]} \BibitemShut {NoStop}%
\bibitem [{\citenamefont {Hohm}\ and\ \citenamefont {Zwiebach}(2016)}]{Hohm:2015doa}%
  \BibitemOpen
  \bibfield  {author} {\bibinfo {author} {\bibfnamefont {O.}~\bibnamefont {Hohm}}\ and\ \bibinfo {author} {\bibfnamefont {B.}~\bibnamefont {Zwiebach}},\ }\bibfield  {title} {\bibinfo {title} {{T-duality Constraints on Higher Derivatives Revisited}},\ }\href {https://doi.org/10.1007/JHEP04(2016)101} {\bibfield  {journal} {\bibinfo  {journal} {JHEP}\ }\textbf {\bibinfo {volume} {04}},\ \bibinfo {pages} {101}},\ \Eprint {https://arxiv.org/abs/1510.00005} {arXiv:1510.00005 [hep-th]} \BibitemShut {NoStop}%
\bibitem [{\citenamefont {Garousi}(2019{\natexlab{a}})}]{Garousi:2019wgz}%
  \BibitemOpen
  \bibfield  {author} {\bibinfo {author} {\bibfnamefont {M.~R.}\ \bibnamefont {Garousi}},\ }\bibfield  {title} {\bibinfo {title} {{Four-derivative couplings via the $T$-duality invariance constraint}},\ }\href {https://doi.org/10.1103/PhysRevD.99.126005} {\bibfield  {journal} {\bibinfo  {journal} {Phys. Rev. D}\ }\textbf {\bibinfo {volume} {99}},\ \bibinfo {pages} {126005} (\bibinfo {year} {2019}{\natexlab{a}})},\ \Eprint {https://arxiv.org/abs/1904.11282} {arXiv:1904.11282 [hep-th]} \BibitemShut {NoStop}%
\bibitem [{\citenamefont {Garousi}(2019{\natexlab{b}})}]{Garousi:2019mca}%
  \BibitemOpen
  \bibfield  {author} {\bibinfo {author} {\bibfnamefont {M.~R.}\ \bibnamefont {Garousi}},\ }\bibfield  {title} {\bibinfo {title} {{Effective action of bosonic string theory at order $\alpha'^2 $}},\ }\href {https://doi.org/10.1140/epjc/s10052-019-7357-4} {\bibfield  {journal} {\bibinfo  {journal} {Eur. Phys. J. C}\ }\textbf {\bibinfo {volume} {79}},\ \bibinfo {pages} {827} (\bibinfo {year} {2019}{\natexlab{b}})},\ \Eprint {https://arxiv.org/abs/1907.06500} {arXiv:1907.06500 [hep-th]} \BibitemShut {NoStop}%
\bibitem [{\citenamefont {Garousi}(2021)}]{Garousi:2020gio}%
  \BibitemOpen
  \bibfield  {author} {\bibinfo {author} {\bibfnamefont {M.~R.}\ \bibnamefont {Garousi}},\ }\bibfield  {title} {\bibinfo {title} {{Effective action of type II superstring theories at order $\alpha'^{3}$: NS-NS couplings}},\ }\href {https://doi.org/10.1007/JHEP02(2021)157} {\bibfield  {journal} {\bibinfo  {journal} {JHEP}\ }\textbf {\bibinfo {volume} {02}},\ \bibinfo {pages} {157}},\ \Eprint {https://arxiv.org/abs/2011.02753} {arXiv:2011.02753 [hep-th]} \BibitemShut {NoStop}%
\bibitem [{\citenamefont {Codina}\ \emph {et~al.}(2021)\citenamefont {Codina}, \citenamefont {Hohm},\ and\ \citenamefont {Marques}}]{Codina:2020kvj}%
  \BibitemOpen
  \bibfield  {author} {\bibinfo {author} {\bibfnamefont {T.}~\bibnamefont {Codina}}, \bibinfo {author} {\bibfnamefont {O.}~\bibnamefont {Hohm}},\ and\ \bibinfo {author} {\bibfnamefont {D.}~\bibnamefont {Marques}},\ }\bibfield  {title} {\bibinfo {title} {{String Dualities at Order $\alpha'^{\,3}$}},\ }\href {https://doi.org/10.1103/PhysRevLett.126.171602} {\bibfield  {journal} {\bibinfo  {journal} {Phys. Rev. Lett.}\ }\textbf {\bibinfo {volume} {126}},\ \bibinfo {pages} {171602} (\bibinfo {year} {2021})},\ \Eprint {https://arxiv.org/abs/2012.15677} {arXiv:2012.15677 [hep-th]} \BibitemShut {NoStop}%
\bibitem [{\citenamefont {David}\ and\ \citenamefont {Liu}(2022)}]{David:2021jqn}%
  \BibitemOpen
  \bibfield  {author} {\bibinfo {author} {\bibfnamefont {M.}~\bibnamefont {David}}\ and\ \bibinfo {author} {\bibfnamefont {J.~T.}\ \bibnamefont {Liu}},\ }\bibfield  {title} {\bibinfo {title} {{T duality and hints of generalized geometry in string \ensuremath{\alpha}' corrections}},\ }\href {https://doi.org/10.1103/PhysRevD.106.106008} {\bibfield  {journal} {\bibinfo  {journal} {Phys. Rev. D}\ }\textbf {\bibinfo {volume} {106}},\ \bibinfo {pages} {106008} (\bibinfo {year} {2022})},\ \Eprint {https://arxiv.org/abs/2108.04370} {arXiv:2108.04370 [hep-th]} \BibitemShut {NoStop}%
\bibitem [{\citenamefont {David}\ and\ \citenamefont {Liu}(2023)}]{David:2022jcl}%
  \BibitemOpen
  \bibfield  {author} {\bibinfo {author} {\bibfnamefont {M.}~\bibnamefont {David}}\ and\ \bibinfo {author} {\bibfnamefont {J.~T.}\ \bibnamefont {Liu}},\ }\bibfield  {title} {\bibinfo {title} {{T-duality building blocks for \ensuremath{\alpha}' string corrections}},\ }\href {https://doi.org/10.1103/PhysRevD.107.046008} {\bibfield  {journal} {\bibinfo  {journal} {Phys. Rev. D}\ }\textbf {\bibinfo {volume} {107}},\ \bibinfo {pages} {046008} (\bibinfo {year} {2023})},\ \Eprint {https://arxiv.org/abs/2210.16593} {arXiv:2210.16593 [hep-th]} \BibitemShut {NoStop}%
\bibitem [{\citenamefont {Ozkan}\ \emph {et~al.}(2024)\citenamefont {Ozkan}, \citenamefont {Pang},\ and\ \citenamefont {Sezgin}}]{Ozkan:2024euj}%
  \BibitemOpen
  \bibfield  {author} {\bibinfo {author} {\bibfnamefont {M.}~\bibnamefont {Ozkan}}, \bibinfo {author} {\bibfnamefont {Y.}~\bibnamefont {Pang}},\ and\ \bibinfo {author} {\bibfnamefont {E.}~\bibnamefont {Sezgin}},\ }\bibfield  {title} {\bibinfo {title} {{Higher derivative supergravities in diverse dimensions}},\ }\href {https://doi.org/10.1016/j.physrep.2024.07.002} {\bibfield  {journal} {\bibinfo  {journal} {Phys. Rept.}\ }\textbf {\bibinfo {volume} {1086}},\ \bibinfo {pages} {1} (\bibinfo {year} {2024})},\ \Eprint {https://arxiv.org/abs/2401.08945} {arXiv:2401.08945 [hep-th]} \BibitemShut {NoStop}%
\bibitem [{\citenamefont {Siegel}(1993)}]{Siegel:1993th}%
  \BibitemOpen
  \bibfield  {author} {\bibinfo {author} {\bibfnamefont {W.}~\bibnamefont {Siegel}},\ }\bibfield  {title} {\bibinfo {title} {{Superspace duality in low-energy superstrings}},\ }\href {https://doi.org/10.1103/PhysRevD.48.2826} {\bibfield  {journal} {\bibinfo  {journal} {Phys. Rev. D}\ }\textbf {\bibinfo {volume} {48}},\ \bibinfo {pages} {2826} (\bibinfo {year} {1993})},\ \Eprint {https://arxiv.org/abs/hep-th/9305073} {arXiv:hep-th/9305073} \BibitemShut {NoStop}%
\bibitem [{\citenamefont {Hull}\ and\ \citenamefont {Zwiebach}(2009)}]{Hull:2009mi}%
  \BibitemOpen
  \bibfield  {author} {\bibinfo {author} {\bibfnamefont {C.}~\bibnamefont {Hull}}\ and\ \bibinfo {author} {\bibfnamefont {B.}~\bibnamefont {Zwiebach}},\ }\bibfield  {title} {\bibinfo {title} {{Double Field Theory}},\ }\href {https://doi.org/10.1088/1126-6708/2009/09/099} {\bibfield  {journal} {\bibinfo  {journal} {JHEP}\ }\textbf {\bibinfo {volume} {09}},\ \bibinfo {pages} {099}},\ \Eprint {https://arxiv.org/abs/0904.4664} {arXiv:0904.4664 [hep-th]} \BibitemShut {NoStop}%
\bibitem [{\citenamefont {Hohm}\ \emph {et~al.}(2010{\natexlab{a}})\citenamefont {Hohm}, \citenamefont {Hull},\ and\ \citenamefont {Zwiebach}}]{Hohm:2010jy}%
  \BibitemOpen
  \bibfield  {author} {\bibinfo {author} {\bibfnamefont {O.}~\bibnamefont {Hohm}}, \bibinfo {author} {\bibfnamefont {C.}~\bibnamefont {Hull}},\ and\ \bibinfo {author} {\bibfnamefont {B.}~\bibnamefont {Zwiebach}},\ }\bibfield  {title} {\bibinfo {title} {{Background independent action for double field theory}},\ }\href {https://doi.org/10.1007/JHEP07(2010)016} {\bibfield  {journal} {\bibinfo  {journal} {JHEP}\ }\textbf {\bibinfo {volume} {07}},\ \bibinfo {pages} {016}},\ \Eprint {https://arxiv.org/abs/1003.5027} {arXiv:1003.5027 [hep-th]} \BibitemShut {NoStop}%
\bibitem [{\citenamefont {Hohm}\ \emph {et~al.}(2010{\natexlab{b}})\citenamefont {Hohm}, \citenamefont {Hull},\ and\ \citenamefont {Zwiebach}}]{Hohm:2010pp}%
  \BibitemOpen
  \bibfield  {author} {\bibinfo {author} {\bibfnamefont {O.}~\bibnamefont {Hohm}}, \bibinfo {author} {\bibfnamefont {C.}~\bibnamefont {Hull}},\ and\ \bibinfo {author} {\bibfnamefont {B.}~\bibnamefont {Zwiebach}},\ }\bibfield  {title} {\bibinfo {title} {{Generalized metric formulation of double field theory}},\ }\href {https://doi.org/10.1007/JHEP08(2010)008} {\bibfield  {journal} {\bibinfo  {journal} {JHEP}\ }\textbf {\bibinfo {volume} {08}},\ \bibinfo {pages} {008}},\ \Eprint {https://arxiv.org/abs/1006.4823} {arXiv:1006.4823 [hep-th]} \BibitemShut {NoStop}%
\bibitem [{\citenamefont {Hohm}\ and\ \citenamefont {Kwak}(2011)}]{Hohm:2011ex}%
  \BibitemOpen
  \bibfield  {author} {\bibinfo {author} {\bibfnamefont {O.}~\bibnamefont {Hohm}}\ and\ \bibinfo {author} {\bibfnamefont {S.~K.}\ \bibnamefont {Kwak}},\ }\bibfield  {title} {\bibinfo {title} {{Double Field Theory Formulation of Heterotic Strings}},\ }\href {https://doi.org/10.1007/JHEP06(2011)096} {\bibfield  {journal} {\bibinfo  {journal} {JHEP}\ }\textbf {\bibinfo {volume} {06}},\ \bibinfo {pages} {096}},\ \Eprint {https://arxiv.org/abs/1103.2136} {arXiv:1103.2136 [hep-th]} \BibitemShut {NoStop}%
\bibitem [{\citenamefont {Hohm}\ \emph {et~al.}(2014)\citenamefont {Hohm}, \citenamefont {Siegel},\ and\ \citenamefont {Zwiebach}}]{Hohm:2013jaa}%
  \BibitemOpen
  \bibfield  {author} {\bibinfo {author} {\bibfnamefont {O.}~\bibnamefont {Hohm}}, \bibinfo {author} {\bibfnamefont {W.}~\bibnamefont {Siegel}},\ and\ \bibinfo {author} {\bibfnamefont {B.}~\bibnamefont {Zwiebach}},\ }\bibfield  {title} {\bibinfo {title} {{Doubled $\alpha'$-geometry}},\ }\href {https://doi.org/10.1007/JHEP02(2014)065} {\bibfield  {journal} {\bibinfo  {journal} {JHEP}\ }\textbf {\bibinfo {volume} {02}},\ \bibinfo {pages} {065}},\ \Eprint {https://arxiv.org/abs/1306.2970} {arXiv:1306.2970 [hep-th]} \BibitemShut {NoStop}%
\bibitem [{\citenamefont {Bedoya}\ \emph {et~al.}(2014)\citenamefont {Bedoya}, \citenamefont {Marques},\ and\ \citenamefont {Nunez}}]{Bedoya:2014pma}%
  \BibitemOpen
  \bibfield  {author} {\bibinfo {author} {\bibfnamefont {O.~A.}\ \bibnamefont {Bedoya}}, \bibinfo {author} {\bibfnamefont {D.}~\bibnamefont {Marques}},\ and\ \bibinfo {author} {\bibfnamefont {C.}~\bibnamefont {Nunez}},\ }\bibfield  {title} {\bibinfo {title} {{Heterotic $\alpha$'-corrections in Double Field Theory}},\ }\href {https://doi.org/10.1007/JHEP12(2014)074} {\bibfield  {journal} {\bibinfo  {journal} {JHEP}\ }\textbf {\bibinfo {volume} {12}},\ \bibinfo {pages} {074}},\ \Eprint {https://arxiv.org/abs/1407.0365} {arXiv:1407.0365 [hep-th]} \BibitemShut {NoStop}%
\bibitem [{\citenamefont {Hohm}\ and\ \citenamefont {Zwiebach}(2014)}]{Hohm:2014xsa}%
  \BibitemOpen
  \bibfield  {author} {\bibinfo {author} {\bibfnamefont {O.}~\bibnamefont {Hohm}}\ and\ \bibinfo {author} {\bibfnamefont {B.}~\bibnamefont {Zwiebach}},\ }\bibfield  {title} {\bibinfo {title} {{Double field theory at order $\alpha'$}},\ }\href {https://doi.org/10.1007/JHEP11(2014)075} {\bibfield  {journal} {\bibinfo  {journal} {JHEP}\ }\textbf {\bibinfo {volume} {11}},\ \bibinfo {pages} {075}},\ \Eprint {https://arxiv.org/abs/1407.3803} {arXiv:1407.3803 [hep-th]} \BibitemShut {NoStop}%
\bibitem [{\citenamefont {Coimbra}\ \emph {et~al.}(2014)\citenamefont {Coimbra}, \citenamefont {Minasian}, \citenamefont {Triendl},\ and\ \citenamefont {Waldram}}]{Coimbra:2014qaa}%
  \BibitemOpen
  \bibfield  {author} {\bibinfo {author} {\bibfnamefont {A.}~\bibnamefont {Coimbra}}, \bibinfo {author} {\bibfnamefont {R.}~\bibnamefont {Minasian}}, \bibinfo {author} {\bibfnamefont {H.}~\bibnamefont {Triendl}},\ and\ \bibinfo {author} {\bibfnamefont {D.}~\bibnamefont {Waldram}},\ }\bibfield  {title} {\bibinfo {title} {{Generalised geometry for string corrections}},\ }\href {https://doi.org/10.1007/JHEP11(2014)160} {\bibfield  {journal} {\bibinfo  {journal} {JHEP}\ }\textbf {\bibinfo {volume} {11}},\ \bibinfo {pages} {160}},\ \Eprint {https://arxiv.org/abs/1407.7542} {arXiv:1407.7542 [hep-th]} \BibitemShut {NoStop}%
\bibitem [{\citenamefont {Lee}(2015)}]{Lee:2015kba}%
  \BibitemOpen
  \bibfield  {author} {\bibinfo {author} {\bibfnamefont {K.}~\bibnamefont {Lee}},\ }\bibfield  {title} {\bibinfo {title} {{Quadratic \ensuremath{\alpha}'-corrections to heterotic double field theory}},\ }\href {https://doi.org/10.1016/j.nuclphysb.2015.08.013} {\bibfield  {journal} {\bibinfo  {journal} {Nucl. Phys. B}\ }\textbf {\bibinfo {volume} {899}},\ \bibinfo {pages} {594} (\bibinfo {year} {2015})},\ \Eprint {https://arxiv.org/abs/1504.00149} {arXiv:1504.00149 [hep-th]} \BibitemShut {NoStop}%
\bibitem [{\citenamefont {Baron}\ \emph {et~al.}(2017)\citenamefont {Baron}, \citenamefont {Fernandez-Melgarejo}, \citenamefont {Marques},\ and\ \citenamefont {Nunez}}]{Baron:2017dvb}%
  \BibitemOpen
  \bibfield  {author} {\bibinfo {author} {\bibfnamefont {W.~H.}\ \bibnamefont {Baron}}, \bibinfo {author} {\bibfnamefont {J.~J.}\ \bibnamefont {Fernandez-Melgarejo}}, \bibinfo {author} {\bibfnamefont {D.}~\bibnamefont {Marques}},\ and\ \bibinfo {author} {\bibfnamefont {C.}~\bibnamefont {Nunez}},\ }\bibfield  {title} {\bibinfo {title} {{The Odd story of \ensuremath{\alpha}'-corrections}},\ }\href {https://doi.org/10.1007/JHEP04(2017)078} {\bibfield  {journal} {\bibinfo  {journal} {JHEP}\ }\textbf {\bibinfo {volume} {04}},\ \bibinfo {pages} {078}},\ \Eprint {https://arxiv.org/abs/1702.05489} {arXiv:1702.05489 [hep-th]} \BibitemShut {NoStop}%
\bibitem [{\citenamefont {Lescano}\ \emph {et~al.}(2021)\citenamefont {Lescano}, \citenamefont {N\'u\~nez},\ and\ \citenamefont {Rodr\'\i{}guez}}]{Lescano:2021guc}%
  \BibitemOpen
  \bibfield  {author} {\bibinfo {author} {\bibfnamefont {E.}~\bibnamefont {Lescano}}, \bibinfo {author} {\bibfnamefont {C.~A.}\ \bibnamefont {N\'u\~nez}},\ and\ \bibinfo {author} {\bibfnamefont {J.~A.}\ \bibnamefont {Rodr\'\i{}guez}},\ }\bibfield  {title} {\bibinfo {title} {{Supersymmetry, T-duality and heterotic \ensuremath{\alpha}'-corrections}},\ }\href {https://doi.org/10.1007/JHEP07(2021)092} {\bibfield  {journal} {\bibinfo  {journal} {JHEP}\ }\textbf {\bibinfo {volume} {07}},\ \bibinfo {pages} {092}},\ \Eprint {https://arxiv.org/abs/2104.09545} {arXiv:2104.09545 [hep-th]} \BibitemShut {NoStop}%
\bibitem [{\citenamefont {Hronek}\ and\ \citenamefont {Wulff}(2021{\natexlab{a}})}]{Hronek:2020xxi}%
  \BibitemOpen
  \bibfield  {author} {\bibinfo {author} {\bibfnamefont {S.}~\bibnamefont {Hronek}}\ and\ \bibinfo {author} {\bibfnamefont {L.}~\bibnamefont {Wulff}},\ }\bibfield  {title} {\bibinfo {title} {{$O(D,D)$ and the string $\alpha'$ expansion: an obstruction}},\ }\href {https://doi.org/10.1007/JHEP04(2021)013} {\bibfield  {journal} {\bibinfo  {journal} {JHEP}\ }\textbf {\bibinfo {volume} {04}},\ \bibinfo {pages} {013}},\ \Eprint {https://arxiv.org/abs/2012.13410} {arXiv:2012.13410 [hep-th]} \BibitemShut {NoStop}%
\bibitem [{\citenamefont {Hsia}\ \emph {et~al.}(2025)\citenamefont {Hsia}, \citenamefont {Kamal},\ and\ \citenamefont {Wulff}}]{Hsia:2024kpi}%
  \BibitemOpen
  \bibfield  {author} {\bibinfo {author} {\bibfnamefont {S.~W.}\ \bibnamefont {Hsia}}, \bibinfo {author} {\bibfnamefont {A.~R.}\ \bibnamefont {Kamal}},\ and\ \bibinfo {author} {\bibfnamefont {L.}~\bibnamefont {Wulff}},\ }\bibfield  {title} {\bibinfo {title} {{No manifest T duality at order {\ensuremath{\alpha}}'3}},\ }\href {https://doi.org/10.1103/PhysRevD.111.L061904} {\bibfield  {journal} {\bibinfo  {journal} {Phys. Rev. D}\ }\textbf {\bibinfo {volume} {111}},\ \bibinfo {pages} {L061904} (\bibinfo {year} {2025})},\ \Eprint {https://arxiv.org/abs/2411.15302} {arXiv:2411.15302 [hep-th]} \BibitemShut {NoStop}%
\bibitem [{\citenamefont {Lunin}\ and\ \citenamefont {Shah}(2025)}]{Lunin:2024vsx}%
  \BibitemOpen
  \bibfield  {author} {\bibinfo {author} {\bibfnamefont {O.}~\bibnamefont {Lunin}}\ and\ \bibinfo {author} {\bibfnamefont {P.}~\bibnamefont {Shah}},\ }\bibfield  {title} {\bibinfo {title} {{Double Field Theory and {\ensuremath{\alpha}}' corrections: Explicit examples}},\ }\href {https://doi.org/10.1016/j.nuclphysb.2025.116932} {\bibfield  {journal} {\bibinfo  {journal} {Nucl. Phys. B}\ }\textbf {\bibinfo {volume} {1017}},\ \bibinfo {pages} {116932} (\bibinfo {year} {2025})},\ \Eprint {https://arxiv.org/abs/2408.04833} {arXiv:2408.04833 [hep-th]} \BibitemShut {NoStop}%
\bibitem [{\citenamefont {Eloy}\ \emph {et~al.}(2020)\citenamefont {Eloy}, \citenamefont {Hohm},\ and\ \citenamefont {Samtleben}}]{Eloy:2020dko}%
  \BibitemOpen
  \bibfield  {author} {\bibinfo {author} {\bibfnamefont {C.}~\bibnamefont {Eloy}}, \bibinfo {author} {\bibfnamefont {O.}~\bibnamefont {Hohm}},\ and\ \bibinfo {author} {\bibfnamefont {H.}~\bibnamefont {Samtleben}},\ }\bibfield  {title} {\bibinfo {title} {{Duality Invariance and Higher Derivatives}},\ }\href {https://doi.org/10.1103/PhysRevD.101.126018} {\bibfield  {journal} {\bibinfo  {journal} {Phys. Rev. D}\ }\textbf {\bibinfo {volume} {101}},\ \bibinfo {pages} {126018} (\bibinfo {year} {2020})},\ \Eprint {https://arxiv.org/abs/2004.13140} {arXiv:2004.13140 [hep-th]} \BibitemShut {NoStop}%
\bibitem [{\citenamefont {Elgood}\ and\ \citenamefont {Ortin}(2020)}]{Elgood:2020xwu}%
  \BibitemOpen
  \bibfield  {author} {\bibinfo {author} {\bibfnamefont {Z.}~\bibnamefont {Elgood}}\ and\ \bibinfo {author} {\bibfnamefont {T.}~\bibnamefont {Ortin}},\ }\bibfield  {title} {\bibinfo {title} {{T duality and Wald entropy formula in the Heterotic Superstring effective action at first-order in \ensuremath{\alpha}'}},\ }\href {https://doi.org/10.1007/JHEP10(2020)097} {\bibfield  {journal} {\bibinfo  {journal} {JHEP}\ }\textbf {\bibinfo {volume} {10}},\ \bibinfo {pages} {097}},\ \bibinfo {note} {[Erratum: JHEP 06, 105 (2021)]},\ \Eprint {https://arxiv.org/abs/2005.11272} {arXiv:2005.11272 [hep-th]} \BibitemShut {NoStop}%
\bibitem [{\citenamefont {Ortin}(2021)}]{Ortin:2020xdm}%
  \BibitemOpen
  \bibfield  {author} {\bibinfo {author} {\bibfnamefont {T.}~\bibnamefont {Ortin}},\ }\bibfield  {title} {\bibinfo {title} {{O(n, n) invariance and Wald entropy formula in the Heterotic Superstring effective action at first order in $\alpha'$}},\ }\href {https://doi.org/10.1007/JHEP01(2021)187} {\bibfield  {journal} {\bibinfo  {journal} {JHEP}\ }\textbf {\bibinfo {volume} {01}},\ \bibinfo {pages} {187}},\ \Eprint {https://arxiv.org/abs/2005.14618} {arXiv:2005.14618 [hep-th]} \BibitemShut {NoStop}%
\bibitem [{\citenamefont {Jayaprakash}\ and\ \citenamefont {Liu}(2024)}]{Jayaprakash:2024xlr}%
  \BibitemOpen
  \bibfield  {author} {\bibinfo {author} {\bibfnamefont {S.}~\bibnamefont {Jayaprakash}}\ and\ \bibinfo {author} {\bibfnamefont {J.~T.}\ \bibnamefont {Liu}},\ }\bibfield  {title} {\bibinfo {title} {{Higher derivative heterotic supergravity on a torus and supersymmetry}},\ }\href {https://doi.org/10.1007/JHEP12(2024)076} {\bibfield  {journal} {\bibinfo  {journal} {JHEP}\ }\textbf {\bibinfo {volume} {12}},\ \bibinfo {pages} {076}},\ \Eprint {https://arxiv.org/abs/2406.14600} {arXiv:2406.14600 [hep-th]} \BibitemShut {NoStop}%
\bibitem [{\citenamefont {Cano}\ and\ \citenamefont {Ruip\'erez}(2019)}]{Cano:2019ore}%
  \BibitemOpen
  \bibfield  {author} {\bibinfo {author} {\bibfnamefont {P.~A.}\ \bibnamefont {Cano}}\ and\ \bibinfo {author} {\bibfnamefont {A.}~\bibnamefont {Ruip\'erez}},\ }\bibfield  {title} {\bibinfo {title} {{Leading higher-derivative corrections to Kerr geometry}},\ }\href {https://doi.org/10.1007/JHEP05(2019)189} {\bibfield  {journal} {\bibinfo  {journal} {JHEP}\ }\textbf {\bibinfo {volume} {05}},\ \bibinfo {pages} {189}},\ \bibinfo {note} {[Erratum: JHEP 03, 187 (2020)]},\ \Eprint {https://arxiv.org/abs/1901.01315} {arXiv:1901.01315 [gr-qc]} \BibitemShut {NoStop}%
\bibitem [{\citenamefont {Cano}\ and\ \citenamefont {Ruip\'erez}(2022)}]{Cano:2021rey}%
  \BibitemOpen
  \bibfield  {author} {\bibinfo {author} {\bibfnamefont {P.~A.}\ \bibnamefont {Cano}}\ and\ \bibinfo {author} {\bibfnamefont {A.}~\bibnamefont {Ruip\'erez}},\ }\bibfield  {title} {\bibinfo {title} {{String gravity in D=4}},\ }\href {https://doi.org/10.1103/PhysRevD.105.044022} {\bibfield  {journal} {\bibinfo  {journal} {Phys. Rev. D}\ }\textbf {\bibinfo {volume} {105}},\ \bibinfo {pages} {044022} (\bibinfo {year} {2022})},\ \Eprint {https://arxiv.org/abs/2111.04750} {arXiv:2111.04750 [hep-th]} \BibitemShut {NoStop}%
\bibitem [{\citenamefont {Liu}\ and\ \citenamefont {Saskowski}(2023)}]{Liu:2023fqq}%
  \BibitemOpen
  \bibfield  {author} {\bibinfo {author} {\bibfnamefont {J.~T.}\ \bibnamefont {Liu}}\ and\ \bibinfo {author} {\bibfnamefont {R.~J.}\ \bibnamefont {Saskowski}},\ }\bibfield  {title} {\bibinfo {title} {{Consistent truncations in higher derivative supergravity}},\ }\href {https://doi.org/10.1007/JHEP09(2023)136} {\bibfield  {journal} {\bibinfo  {journal} {JHEP}\ }\textbf {\bibinfo {volume} {09}},\ \bibinfo {pages} {136}},\ \Eprint {https://arxiv.org/abs/2307.12420} {arXiv:2307.12420 [hep-th]} \BibitemShut {NoStop}%
\bibitem [{\citenamefont {Ma}\ \emph {et~al.}(2025)\citenamefont {Ma}, \citenamefont {Pang},\ and\ \citenamefont {Lu}}]{Ma:2024ulp}%
  \BibitemOpen
  \bibfield  {author} {\bibinfo {author} {\bibfnamefont {L.}~\bibnamefont {Ma}}, \bibinfo {author} {\bibfnamefont {Y.}~\bibnamefont {Pang}},\ and\ \bibinfo {author} {\bibfnamefont {H.}~\bibnamefont {Lu}},\ }\bibfield  {title} {\bibinfo {title} {{Leading higher derivative corrections to multipole moments of Kerr-Newman black hole}},\ }\href {https://doi.org/10.1007/JHEP02(2025)079} {\bibfield  {journal} {\bibinfo  {journal} {JHEP}\ }\textbf {\bibinfo {volume} {02}},\ \bibinfo {pages} {079}},\ \Eprint {https://arxiv.org/abs/2411.13639} {arXiv:2411.13639 [hep-th]} \BibitemShut {NoStop}%
\bibitem [{\citenamefont {Bergshoeff}\ and\ \citenamefont {de~Roo}(1989{\natexlab{a}})}]{Bergshoeff:1988nn}%
  \BibitemOpen
  \bibfield  {author} {\bibinfo {author} {\bibfnamefont {E.}~\bibnamefont {Bergshoeff}}\ and\ \bibinfo {author} {\bibfnamefont {M.}~\bibnamefont {de~Roo}},\ }\bibfield  {title} {\bibinfo {title} {{Supersymmetric Chern-simons Terms in Ten-dimensions}},\ }\href {https://doi.org/10.1016/0370-2693(89)91420-2} {\bibfield  {journal} {\bibinfo  {journal} {Phys. Lett. B}\ }\textbf {\bibinfo {volume} {218}},\ \bibinfo {pages} {210} (\bibinfo {year} {1989}{\natexlab{a}})}\BibitemShut {NoStop}%
\bibitem [{\citenamefont {Bergshoeff}\ and\ \citenamefont {de~Roo}(1989{\natexlab{b}})}]{Bergshoeff:1989de}%
  \BibitemOpen
  \bibfield  {author} {\bibinfo {author} {\bibfnamefont {E.~A.}\ \bibnamefont {Bergshoeff}}\ and\ \bibinfo {author} {\bibfnamefont {M.}~\bibnamefont {de~Roo}},\ }\bibfield  {title} {\bibinfo {title} {{The Quartic Effective Action of the Heterotic String and Supersymmetry}},\ }\href {https://doi.org/10.1016/0550-3213(89)90336-2} {\bibfield  {journal} {\bibinfo  {journal} {Nucl. Phys. B}\ }\textbf {\bibinfo {volume} {328}},\ \bibinfo {pages} {439} (\bibinfo {year} {1989}{\natexlab{b}})}\BibitemShut {NoStop}%
\bibitem [{\citenamefont {Sen}(1986)}]{Sen:1985tq}%
  \BibitemOpen
  \bibfield  {author} {\bibinfo {author} {\bibfnamefont {A.}~\bibnamefont {Sen}},\ }\bibfield  {title} {\bibinfo {title} {{Local Gauge and Lorentz Invariance of the Heterotic String Theory}},\ }\href {https://doi.org/10.1016/0370-2693(86)90804-X} {\bibfield  {journal} {\bibinfo  {journal} {Phys. Lett. B}\ }\textbf {\bibinfo {volume} {166}},\ \bibinfo {pages} {300} (\bibinfo {year} {1986})}\BibitemShut {NoStop}%
\bibitem [{\citenamefont {Cano}\ \emph {et~al.}(2018)\citenamefont {Cano}, \citenamefont {Meessen}, \citenamefont {Ort\'\i{}n},\ and\ \citenamefont {Ram\'\i{}rez}}]{Cano:2018qev}%
  \BibitemOpen
  \bibfield  {author} {\bibinfo {author} {\bibfnamefont {P.~A.}\ \bibnamefont {Cano}}, \bibinfo {author} {\bibfnamefont {P.}~\bibnamefont {Meessen}}, \bibinfo {author} {\bibfnamefont {T.}~\bibnamefont {Ort\'\i{}n}},\ and\ \bibinfo {author} {\bibfnamefont {P.~F.}\ \bibnamefont {Ram\'\i{}rez}},\ }\bibfield  {title} {\bibinfo {title} {{$\alpha'$-corrected black holes in String Theory}},\ }\href {https://doi.org/10.1007/JHEP05(2018)110} {\bibfield  {journal} {\bibinfo  {journal} {JHEP}\ }\textbf {\bibinfo {volume} {05}},\ \bibinfo {pages} {110}},\ \Eprint {https://arxiv.org/abs/1803.01919} {arXiv:1803.01919 [hep-th]} \BibitemShut {NoStop}%
\bibitem [{\citenamefont {Chimento}\ \emph {et~al.}(2018)\citenamefont {Chimento}, \citenamefont {Meessen}, \citenamefont {Ortin}, \citenamefont {Ramirez},\ and\ \citenamefont {Ruiperez}}]{Chimento:2018kop}%
  \BibitemOpen
  \bibfield  {author} {\bibinfo {author} {\bibfnamefont {S.}~\bibnamefont {Chimento}}, \bibinfo {author} {\bibfnamefont {P.}~\bibnamefont {Meessen}}, \bibinfo {author} {\bibfnamefont {T.}~\bibnamefont {Ortin}}, \bibinfo {author} {\bibfnamefont {P.~F.}\ \bibnamefont {Ramirez}},\ and\ \bibinfo {author} {\bibfnamefont {A.}~\bibnamefont {Ruiperez}},\ }\bibfield  {title} {\bibinfo {title} {{On a family of $\alpha'$-corrected solutions of the Heterotic Superstring effective action}},\ }\href {https://doi.org/10.1007/JHEP07(2018)080} {\bibfield  {journal} {\bibinfo  {journal} {JHEP}\ }\textbf {\bibinfo {volume} {07}},\ \bibinfo {pages} {080}},\ \Eprint {https://arxiv.org/abs/1803.04463} {arXiv:1803.04463 [hep-th]} \BibitemShut {NoStop}%
\bibitem [{\citenamefont {Cano}\ \emph {et~al.}(2019)\citenamefont {Cano}, \citenamefont {Chimento}, \citenamefont {Meessen}, \citenamefont {Ort\'\i{}n}, \citenamefont {Ram\'\i{}rez},\ and\ \citenamefont {Ruip\'erez}}]{Cano:2018brq}%
  \BibitemOpen
  \bibfield  {author} {\bibinfo {author} {\bibfnamefont {P.~A.}\ \bibnamefont {Cano}}, \bibinfo {author} {\bibfnamefont {S.}~\bibnamefont {Chimento}}, \bibinfo {author} {\bibfnamefont {P.}~\bibnamefont {Meessen}}, \bibinfo {author} {\bibfnamefont {T.}~\bibnamefont {Ort\'\i{}n}}, \bibinfo {author} {\bibfnamefont {P.~F.}\ \bibnamefont {Ram\'\i{}rez}},\ and\ \bibinfo {author} {\bibfnamefont {A.}~\bibnamefont {Ruip\'erez}},\ }\bibfield  {title} {\bibinfo {title} {{Beyond the near-horizon limit: Stringy corrections to Heterotic Black Holes}},\ }\href {https://doi.org/10.1007/JHEP02(2019)192} {\bibfield  {journal} {\bibinfo  {journal} {JHEP}\ }\textbf {\bibinfo {volume} {02}},\ \bibinfo {pages} {192}},\ \Eprint {https://arxiv.org/abs/1808.03651} {arXiv:1808.03651 [hep-th]} \BibitemShut {NoStop}%
\bibitem [{\citenamefont {Cano}\ \emph {et~al.}(2020)\citenamefont {Cano}, \citenamefont {Chimento}, \citenamefont {Linares}, \citenamefont {Ort\'\i{}n},\ and\ \citenamefont {Ram\'\i{}rez}}]{Cano:2019ycn}%
  \BibitemOpen
  \bibfield  {author} {\bibinfo {author} {\bibfnamefont {P.~A.}\ \bibnamefont {Cano}}, \bibinfo {author} {\bibfnamefont {S.}~\bibnamefont {Chimento}}, \bibinfo {author} {\bibfnamefont {R.}~\bibnamefont {Linares}}, \bibinfo {author} {\bibfnamefont {T.}~\bibnamefont {Ort\'\i{}n}},\ and\ \bibinfo {author} {\bibfnamefont {P.~F.}\ \bibnamefont {Ram\'\i{}rez}},\ }\bibfield  {title} {\bibinfo {title} {{$\alpha'$ corrections of Reissner-Nordstr\"om black holes}},\ }\href {https://doi.org/10.1007/JHEP02(2020)031} {\bibfield  {journal} {\bibinfo  {journal} {JHEP}\ }\textbf {\bibinfo {volume} {02}},\ \bibinfo {pages} {031}},\ \Eprint {https://arxiv.org/abs/1910.14324} {arXiv:1910.14324 [hep-th]} \BibitemShut {NoStop}%
\bibitem [{\citenamefont {Pani}\ and\ \citenamefont {Cardoso}(2009)}]{Pani:2009wy}%
  \BibitemOpen
  \bibfield  {author} {\bibinfo {author} {\bibfnamefont {P.}~\bibnamefont {Pani}}\ and\ \bibinfo {author} {\bibfnamefont {V.}~\bibnamefont {Cardoso}},\ }\bibfield  {title} {\bibinfo {title} {{Are black holes in alternative theories serious astrophysical candidates? The Case for Einstein-Dilaton-Gauss-Bonnet black holes}},\ }\href {https://doi.org/10.1103/PhysRevD.79.084031} {\bibfield  {journal} {\bibinfo  {journal} {Phys. Rev. D}\ }\textbf {\bibinfo {volume} {79}},\ \bibinfo {pages} {084031} (\bibinfo {year} {2009})},\ \Eprint {https://arxiv.org/abs/0902.1569} {arXiv:0902.1569 [gr-qc]} \BibitemShut {NoStop}%
\bibitem [{\citenamefont {Yunes}\ and\ \citenamefont {Pretorius}(2009)}]{Yunes:2009hc}%
  \BibitemOpen
  \bibfield  {author} {\bibinfo {author} {\bibfnamefont {N.}~\bibnamefont {Yunes}}\ and\ \bibinfo {author} {\bibfnamefont {F.}~\bibnamefont {Pretorius}},\ }\bibfield  {title} {\bibinfo {title} {{Dynamical Chern-Simons Modified Gravity. I. Spinning Black Holes in the Slow-Rotation Approximation}},\ }\href {https://doi.org/10.1103/PhysRevD.79.084043} {\bibfield  {journal} {\bibinfo  {journal} {Phys. Rev. D}\ }\textbf {\bibinfo {volume} {79}},\ \bibinfo {pages} {084043} (\bibinfo {year} {2009})},\ \Eprint {https://arxiv.org/abs/0902.4669} {arXiv:0902.4669 [gr-qc]} \BibitemShut {NoStop}%
\bibitem [{\citenamefont {Konno}\ \emph {et~al.}(2009)\citenamefont {Konno}, \citenamefont {Matsuyama},\ and\ \citenamefont {Tanda}}]{Konno:2009kg}%
  \BibitemOpen
  \bibfield  {author} {\bibinfo {author} {\bibfnamefont {K.}~\bibnamefont {Konno}}, \bibinfo {author} {\bibfnamefont {T.}~\bibnamefont {Matsuyama}},\ and\ \bibinfo {author} {\bibfnamefont {S.}~\bibnamefont {Tanda}},\ }\bibfield  {title} {\bibinfo {title} {{Rotating black hole in extended Chern-Simons modified gravity}},\ }\href {https://doi.org/10.1143/PTP.122.561} {\bibfield  {journal} {\bibinfo  {journal} {Prog. Theor. Phys.}\ }\textbf {\bibinfo {volume} {122}},\ \bibinfo {pages} {561} (\bibinfo {year} {2009})},\ \Eprint {https://arxiv.org/abs/0902.4767} {arXiv:0902.4767 [gr-qc]} \BibitemShut {NoStop}%
\bibitem [{\citenamefont {Yagi}\ \emph {et~al.}(2012)\citenamefont {Yagi}, \citenamefont {Yunes},\ and\ \citenamefont {Tanaka}}]{Yagi:2012ya}%
  \BibitemOpen
  \bibfield  {author} {\bibinfo {author} {\bibfnamefont {K.}~\bibnamefont {Yagi}}, \bibinfo {author} {\bibfnamefont {N.}~\bibnamefont {Yunes}},\ and\ \bibinfo {author} {\bibfnamefont {T.}~\bibnamefont {Tanaka}},\ }\bibfield  {title} {\bibinfo {title} {{Slowly Rotating Black Holes in Dynamical Chern-Simons Gravity: Deformation Quadratic in the Spin}},\ }\href {https://doi.org/10.1103/PhysRevD.86.044037} {\bibfield  {journal} {\bibinfo  {journal} {Phys. Rev. D}\ }\textbf {\bibinfo {volume} {86}},\ \bibinfo {pages} {044037} (\bibinfo {year} {2012})},\ \bibinfo {note} {[Erratum: Phys.Rev.D 89, 049902 (2014)]},\ \Eprint {https://arxiv.org/abs/1206.6130} {arXiv:1206.6130 [gr-qc]} \BibitemShut {NoStop}%
\bibitem [{\citenamefont {Mignemi}\ and\ \citenamefont {Stewart}(1993)}]{Mignemi:1992pm}%
  \BibitemOpen
  \bibfield  {author} {\bibinfo {author} {\bibfnamefont {S.}~\bibnamefont {Mignemi}}\ and\ \bibinfo {author} {\bibfnamefont {N.~R.}\ \bibnamefont {Stewart}},\ }\bibfield  {title} {\bibinfo {title} {{Dilaton axion hair for slowly rotating Kerr black holes}},\ }\href {https://doi.org/10.1016/0370-2693(93)91824-7} {\bibfield  {journal} {\bibinfo  {journal} {Phys. Lett. B}\ }\textbf {\bibinfo {volume} {298}},\ \bibinfo {pages} {299} (\bibinfo {year} {1993})},\ \Eprint {https://arxiv.org/abs/hep-th/9206018} {arXiv:hep-th/9206018} \BibitemShut {NoStop}%
\bibitem [{\citenamefont {Campbell}\ \emph {et~al.}(1992)\citenamefont {Campbell}, \citenamefont {Kaloper},\ and\ \citenamefont {Olive}}]{Campbell:1991kz}%
  \BibitemOpen
  \bibfield  {author} {\bibinfo {author} {\bibfnamefont {B.~A.}\ \bibnamefont {Campbell}}, \bibinfo {author} {\bibfnamefont {N.}~\bibnamefont {Kaloper}},\ and\ \bibinfo {author} {\bibfnamefont {K.~A.}\ \bibnamefont {Olive}},\ }\bibfield  {title} {\bibinfo {title} {{Classical hair for Kerr-Newman black holes in string gravity}},\ }\href {https://doi.org/10.1016/0370-2693(92)91452-F} {\bibfield  {journal} {\bibinfo  {journal} {Phys. Lett. B}\ }\textbf {\bibinfo {volume} {285}},\ \bibinfo {pages} {199} (\bibinfo {year} {1992})}\BibitemShut {NoStop}%
\bibitem [{\citenamefont {Cai}\ \emph {et~al.}(2025)\citenamefont {Cai}, \citenamefont {Jayaprakash}, \citenamefont {Liu},\ and\ \citenamefont {Saskowski}}]{Cai:2025yyv}%
  \BibitemOpen
  \bibfield  {author} {\bibinfo {author} {\bibfnamefont {Y.}~\bibnamefont {Cai}}, \bibinfo {author} {\bibfnamefont {S.}~\bibnamefont {Jayaprakash}}, \bibinfo {author} {\bibfnamefont {J.~T.}\ \bibnamefont {Liu}},\ and\ \bibinfo {author} {\bibfnamefont {R.~J.}\ \bibnamefont {Saskowski}},\ }\bibfield  {title} {\bibinfo {title} {{Multicenter higher-derivative BPS black holes}},\ }\Eprint {https://arxiv.org/abs/2502.05065} {arXiv:2502.05065 [hep-th]}  (\bibinfo {year} {2025})\BibitemShut {NoStop}%
\bibitem [{\citenamefont {Gibbons}(1982)}]{Gibbons:1982ih}%
  \BibitemOpen
  \bibfield  {author} {\bibinfo {author} {\bibfnamefont {G.~W.}\ \bibnamefont {Gibbons}},\ }\bibfield  {title} {\bibinfo {title} {{Antigravitating Black Hole Solitons with Scalar Hair in N=4 Supergravity}},\ }\href {https://doi.org/10.1016/0550-3213(82)90170-5} {\bibfield  {journal} {\bibinfo  {journal} {Nucl. Phys. B}\ }\textbf {\bibinfo {volume} {207}},\ \bibinfo {pages} {337} (\bibinfo {year} {1982})}\BibitemShut {NoStop}%
\bibitem [{\citenamefont {Gibbons}\ and\ \citenamefont {Maeda}(1988)}]{Gibbons:1987ps}%
  \BibitemOpen
  \bibfield  {author} {\bibinfo {author} {\bibfnamefont {G.~W.}\ \bibnamefont {Gibbons}}\ and\ \bibinfo {author} {\bibfnamefont {K.-i.}\ \bibnamefont {Maeda}},\ }\bibfield  {title} {\bibinfo {title} {{Black Holes and Membranes in Higher Dimensional Theories with Dilaton Fields}},\ }\href {https://doi.org/10.1016/0550-3213(88)90006-5} {\bibfield  {journal} {\bibinfo  {journal} {Nucl. Phys. B}\ }\textbf {\bibinfo {volume} {298}},\ \bibinfo {pages} {741} (\bibinfo {year} {1988})}\BibitemShut {NoStop}%
\bibitem [{\citenamefont {Garfinkle}\ \emph {et~al.}(1991)\citenamefont {Garfinkle}, \citenamefont {Horowitz},\ and\ \citenamefont {Strominger}}]{Garfinkle:1990qj}%
  \BibitemOpen
  \bibfield  {author} {\bibinfo {author} {\bibfnamefont {D.}~\bibnamefont {Garfinkle}}, \bibinfo {author} {\bibfnamefont {G.~T.}\ \bibnamefont {Horowitz}},\ and\ \bibinfo {author} {\bibfnamefont {A.}~\bibnamefont {Strominger}},\ }\bibfield  {title} {\bibinfo {title} {{Charged black holes in string theory}},\ }\href {https://doi.org/10.1103/PhysRevD.43.3140} {\bibfield  {journal} {\bibinfo  {journal} {Phys. Rev. D}\ }\textbf {\bibinfo {volume} {43}},\ \bibinfo {pages} {3140} (\bibinfo {year} {1991})},\ \bibinfo {note} {[Erratum: Phys.Rev.D 45, 3888 (1992)]}\BibitemShut {NoStop}%
\bibitem [{\citenamefont {Kanti}\ \emph {et~al.}(1996)\citenamefont {Kanti}, \citenamefont {Mavromatos}, \citenamefont {Rizos}, \citenamefont {Tamvakis},\ and\ \citenamefont {Winstanley}}]{Kanti:1995vq}%
  \BibitemOpen
  \bibfield  {author} {\bibinfo {author} {\bibfnamefont {P.}~\bibnamefont {Kanti}}, \bibinfo {author} {\bibfnamefont {N.~E.}\ \bibnamefont {Mavromatos}}, \bibinfo {author} {\bibfnamefont {J.}~\bibnamefont {Rizos}}, \bibinfo {author} {\bibfnamefont {K.}~\bibnamefont {Tamvakis}},\ and\ \bibinfo {author} {\bibfnamefont {E.}~\bibnamefont {Winstanley}},\ }\bibfield  {title} {\bibinfo {title} {{Dilatonic black holes in higher curvature string gravity}},\ }\href {https://doi.org/10.1103/PhysRevD.54.5049} {\bibfield  {journal} {\bibinfo  {journal} {Phys. Rev. D}\ }\textbf {\bibinfo {volume} {54}},\ \bibinfo {pages} {5049} (\bibinfo {year} {1996})},\ \Eprint {https://arxiv.org/abs/hep-th/9511071} {arXiv:hep-th/9511071} \BibitemShut {NoStop}%
\bibitem [{\citenamefont {Ohta}\ and\ \citenamefont {Torii}(2012)}]{Ohta:2012ih}%
  \BibitemOpen
  \bibfield  {author} {\bibinfo {author} {\bibfnamefont {N.}~\bibnamefont {Ohta}}\ and\ \bibinfo {author} {\bibfnamefont {T.}~\bibnamefont {Torii}},\ }\bibfield  {title} {\bibinfo {title} {{Charged Black Holes in String Theory with Gauss-Bonnet Correction in Various Dimensions}},\ }\href {https://doi.org/10.1103/PhysRevD.86.104016} {\bibfield  {journal} {\bibinfo  {journal} {Phys. Rev. D}\ }\textbf {\bibinfo {volume} {86}},\ \bibinfo {pages} {104016} (\bibinfo {year} {2012})},\ \Eprint {https://arxiv.org/abs/1208.6367} {arXiv:1208.6367 [hep-th]} \BibitemShut {NoStop}%
\bibitem [{\citenamefont {Herdeiro}\ \emph {et~al.}(2021)\citenamefont {Herdeiro}, \citenamefont {Radu},\ and\ \citenamefont {Uzawa}}]{Herdeiro:2021gbw}%
  \BibitemOpen
  \bibfield  {author} {\bibinfo {author} {\bibfnamefont {C.}~\bibnamefont {Herdeiro}}, \bibinfo {author} {\bibfnamefont {E.}~\bibnamefont {Radu}},\ and\ \bibinfo {author} {\bibfnamefont {K.}~\bibnamefont {Uzawa}},\ }\bibfield  {title} {\bibinfo {title} {{De-singularizing the extremal GMGHS black hole via higher derivatives corrections}},\ }\href {https://doi.org/10.1016/j.physletb.2021.136357} {\bibfield  {journal} {\bibinfo  {journal} {Phys. Lett. B}\ }\textbf {\bibinfo {volume} {818}},\ \bibinfo {pages} {136357} (\bibinfo {year} {2021})},\ \Eprint {https://arxiv.org/abs/2103.00884} {arXiv:2103.00884 [hep-th]} \BibitemShut {NoStop}%
\bibitem [{\citenamefont {Massai}\ \emph {et~al.}(2024)\citenamefont {Massai}, \citenamefont {Ruip\'erez},\ and\ \citenamefont {Zatti}}]{Massai:2023cis}%
  \BibitemOpen
  \bibfield  {author} {\bibinfo {author} {\bibfnamefont {S.}~\bibnamefont {Massai}}, \bibinfo {author} {\bibfnamefont {A.}~\bibnamefont {Ruip\'erez}},\ and\ \bibinfo {author} {\bibfnamefont {M.}~\bibnamefont {Zatti}},\ }\bibfield  {title} {\bibinfo {title} {{Revisiting $\alpha'$ corrections to heterotic two-charge black holes}},\ }\href {https://doi.org/10.1007/JHEP04(2024)150} {\bibfield  {journal} {\bibinfo  {journal} {JHEP}\ }\textbf {\bibinfo {volume} {04}},\ \bibinfo {pages} {150}},\ \Eprint {https://arxiv.org/abs/2311.03308} {arXiv:2311.03308 [hep-th]} \BibitemShut {NoStop}%
\bibitem [{\citenamefont {Geroch}(1970{\natexlab{a}})}]{Geroch:1970cc}%
  \BibitemOpen
  \bibfield  {author} {\bibinfo {author} {\bibfnamefont {R.~P.}\ \bibnamefont {Geroch}},\ }\bibfield  {title} {\bibinfo {title} {{Multipole moments. I. Flat space}},\ }\href {https://doi.org/10.1063/1.1665348} {\bibfield  {journal} {\bibinfo  {journal} {J. Math. Phys.}\ }\textbf {\bibinfo {volume} {11}},\ \bibinfo {pages} {1955} (\bibinfo {year} {1970}{\natexlab{a}})}\BibitemShut {NoStop}%
\bibitem [{\citenamefont {Geroch}(1970{\natexlab{b}})}]{Geroch:1970cd}%
  \BibitemOpen
  \bibfield  {author} {\bibinfo {author} {\bibfnamefont {R.~P.}\ \bibnamefont {Geroch}},\ }\bibfield  {title} {\bibinfo {title} {{Multipole moments. II. Curved space}},\ }\href {https://doi.org/10.1063/1.1665427} {\bibfield  {journal} {\bibinfo  {journal} {J. Math. Phys.}\ }\textbf {\bibinfo {volume} {11}},\ \bibinfo {pages} {2580} (\bibinfo {year} {1970}{\natexlab{b}})}\BibitemShut {NoStop}%
\bibitem [{\citenamefont {Hansen}(1974)}]{Hansen:1974zz}%
  \BibitemOpen
  \bibfield  {author} {\bibinfo {author} {\bibfnamefont {R.~O.}\ \bibnamefont {Hansen}},\ }\bibfield  {title} {\bibinfo {title} {{Multipole moments of stationary space-times}},\ }\href {https://doi.org/10.1063/1.1666501} {\bibfield  {journal} {\bibinfo  {journal} {J. Math. Phys.}\ }\textbf {\bibinfo {volume} {15}},\ \bibinfo {pages} {46} (\bibinfo {year} {1974})}\BibitemShut {NoStop}%
\bibitem [{\citenamefont {Thorne}(1980)}]{Thorne:1980ru}%
  \BibitemOpen
  \bibfield  {author} {\bibinfo {author} {\bibfnamefont {K.~S.}\ \bibnamefont {Thorne}},\ }\bibfield  {title} {\bibinfo {title} {{Multipole Expansions of Gravitational Radiation}},\ }\href {https://doi.org/10.1103/RevModPhys.52.299} {\bibfield  {journal} {\bibinfo  {journal} {Rev. Mod. Phys.}\ }\textbf {\bibinfo {volume} {52}},\ \bibinfo {pages} {299} (\bibinfo {year} {1980})}\BibitemShut {NoStop}%
\bibitem [{\citenamefont {Comp\`ere}\ \emph {et~al.}(2018)\citenamefont {Comp\`ere}, \citenamefont {Oliveri},\ and\ \citenamefont {Seraj}}]{Compere:2017wrj}%
  \BibitemOpen
  \bibfield  {author} {\bibinfo {author} {\bibfnamefont {G.}~\bibnamefont {Comp\`ere}}, \bibinfo {author} {\bibfnamefont {R.}~\bibnamefont {Oliveri}},\ and\ \bibinfo {author} {\bibfnamefont {A.}~\bibnamefont {Seraj}},\ }\bibfield  {title} {\bibinfo {title} {{Gravitational multipole moments from Noether charges}},\ }\href {https://doi.org/10.1007/JHEP05(2018)054} {\bibfield  {journal} {\bibinfo  {journal} {JHEP}\ }\textbf {\bibinfo {volume} {05}},\ \bibinfo {pages} {054}},\ \Eprint {https://arxiv.org/abs/1711.08806} {arXiv:1711.08806 [hep-th]} \BibitemShut {NoStop}%
\bibitem [{\citenamefont {Cano}\ \emph {et~al.}(2022)\citenamefont {Cano}, \citenamefont {Ganchev}, \citenamefont {Mayerson},\ and\ \citenamefont {Ruip\'erez}}]{Cano:2022wwo}%
  \BibitemOpen
  \bibfield  {author} {\bibinfo {author} {\bibfnamefont {P.~A.}\ \bibnamefont {Cano}}, \bibinfo {author} {\bibfnamefont {B.}~\bibnamefont {Ganchev}}, \bibinfo {author} {\bibfnamefont {D.~R.}\ \bibnamefont {Mayerson}},\ and\ \bibinfo {author} {\bibfnamefont {A.}~\bibnamefont {Ruip\'erez}},\ }\bibfield  {title} {\bibinfo {title} {{Black hole multipoles in higher-derivative gravity}},\ }\href {https://doi.org/10.1007/JHEP12(2022)120} {\bibfield  {journal} {\bibinfo  {journal} {JHEP}\ }\textbf {\bibinfo {volume} {12}},\ \bibinfo {pages} {120}},\ \Eprint {https://arxiv.org/abs/2208.01044} {arXiv:2208.01044 [gr-qc]} \BibitemShut {NoStop}%
\bibitem [{\citenamefont {Sotiriou}\ and\ \citenamefont {Apostolatos}(2004)}]{Sotiriou:2004ud}%
  \BibitemOpen
  \bibfield  {author} {\bibinfo {author} {\bibfnamefont {T.~P.}\ \bibnamefont {Sotiriou}}\ and\ \bibinfo {author} {\bibfnamefont {T.~A.}\ \bibnamefont {Apostolatos}},\ }\bibfield  {title} {\bibinfo {title} {{Corrected multipole moments of axisymmetric electrovacuum spacetimes}},\ }\href {https://doi.org/10.1088/0264-9381/21/24/003} {\bibfield  {journal} {\bibinfo  {journal} {Class. Quant. Grav.}\ }\textbf {\bibinfo {volume} {21}},\ \bibinfo {pages} {5727} (\bibinfo {year} {2004})},\ \Eprint {https://arxiv.org/abs/gr-qc/0407064} {arXiv:gr-qc/0407064} \BibitemShut {NoStop}%
\bibitem [{\citenamefont {Fodor}\ \emph {et~al.}(2021)\citenamefont {Fodor}, \citenamefont {Filho},\ and\ \citenamefont {Hartmann}}]{Fodor:2020fnq}%
  \BibitemOpen
  \bibfield  {author} {\bibinfo {author} {\bibfnamefont {G.}~\bibnamefont {Fodor}}, \bibinfo {author} {\bibfnamefont {E.~d. S.~C.}\ \bibnamefont {Filho}},\ and\ \bibinfo {author} {\bibfnamefont {B.}~\bibnamefont {Hartmann}},\ }\bibfield  {title} {\bibinfo {title} {{Calculation of multipole moments of axistationary electrovacuum spacetimes}},\ }\href {https://doi.org/10.1103/PhysRevD.104.064012} {\bibfield  {journal} {\bibinfo  {journal} {Phys. Rev. D}\ }\textbf {\bibinfo {volume} {104}},\ \bibinfo {pages} {064012} (\bibinfo {year} {2021})},\ \Eprint {https://arxiv.org/abs/2012.05548} {arXiv:2012.05548 [gr-qc]} \BibitemShut {NoStop}%
\bibitem [{\citenamefont {Wald}(1993)}]{Wald:1993nt}%
  \BibitemOpen
  \bibfield  {author} {\bibinfo {author} {\bibfnamefont {R.~M.}\ \bibnamefont {Wald}},\ }\bibfield  {title} {\bibinfo {title} {{Black hole entropy is the Noether charge}},\ }\href {https://doi.org/10.1103/PhysRevD.48.R3427} {\bibfield  {journal} {\bibinfo  {journal} {Phys. Rev. D}\ }\textbf {\bibinfo {volume} {48}},\ \bibinfo {pages} {R3427} (\bibinfo {year} {1993})},\ \Eprint {https://arxiv.org/abs/gr-qc/9307038} {arXiv:gr-qc/9307038} \BibitemShut {NoStop}%
\bibitem [{\citenamefont {Iyer}\ and\ \citenamefont {Wald}(1994)}]{Iyer:1994ys}%
  \BibitemOpen
  \bibfield  {author} {\bibinfo {author} {\bibfnamefont {V.}~\bibnamefont {Iyer}}\ and\ \bibinfo {author} {\bibfnamefont {R.~M.}\ \bibnamefont {Wald}},\ }\bibfield  {title} {\bibinfo {title} {{Some properties of Noether charge and a proposal for dynamical black hole entropy}},\ }\href {https://doi.org/10.1103/PhysRevD.50.846} {\bibfield  {journal} {\bibinfo  {journal} {Phys. Rev. D}\ }\textbf {\bibinfo {volume} {50}},\ \bibinfo {pages} {846} (\bibinfo {year} {1994})},\ \Eprint {https://arxiv.org/abs/gr-qc/9403028} {arXiv:gr-qc/9403028} \BibitemShut {NoStop}%
\bibitem [{\citenamefont {Jacobson}\ \emph {et~al.}(1994)\citenamefont {Jacobson}, \citenamefont {Kang},\ and\ \citenamefont {Myers}}]{Jacobson:1993vj}%
  \BibitemOpen
  \bibfield  {author} {\bibinfo {author} {\bibfnamefont {T.}~\bibnamefont {Jacobson}}, \bibinfo {author} {\bibfnamefont {G.}~\bibnamefont {Kang}},\ and\ \bibinfo {author} {\bibfnamefont {R.~C.}\ \bibnamefont {Myers}},\ }\bibfield  {title} {\bibinfo {title} {{On black hole entropy}},\ }\href {https://doi.org/10.1103/PhysRevD.49.6587} {\bibfield  {journal} {\bibinfo  {journal} {Phys. Rev. D}\ }\textbf {\bibinfo {volume} {49}},\ \bibinfo {pages} {6587} (\bibinfo {year} {1994})},\ \Eprint {https://arxiv.org/abs/gr-qc/9312023} {arXiv:gr-qc/9312023} \BibitemShut {NoStop}%
\bibitem [{\citenamefont {Elgood}\ \emph {et~al.}(2020)\citenamefont {Elgood}, \citenamefont {Meessen},\ and\ \citenamefont {Ort\'\i{}n}}]{Elgood:2020svt}%
  \BibitemOpen
  \bibfield  {author} {\bibinfo {author} {\bibfnamefont {Z.}~\bibnamefont {Elgood}}, \bibinfo {author} {\bibfnamefont {P.}~\bibnamefont {Meessen}},\ and\ \bibinfo {author} {\bibfnamefont {T.}~\bibnamefont {Ort\'\i{}n}},\ }\bibfield  {title} {\bibinfo {title} {{The first law of black hole mechanics in the Einstein-Maxwell theory revisited}},\ }\href {https://doi.org/10.1007/JHEP09(2020)026} {\bibfield  {journal} {\bibinfo  {journal} {JHEP}\ }\textbf {\bibinfo {volume} {09}},\ \bibinfo {pages} {026}},\ \Eprint {https://arxiv.org/abs/2006.02792} {arXiv:2006.02792 [hep-th]} \BibitemShut {NoStop}%
\bibitem [{\citenamefont {Elgood}\ \emph {et~al.}(2021{\natexlab{a}})\citenamefont {Elgood}, \citenamefont {Mitsios}, \citenamefont {Ort\'\i{}n},\ and\ \citenamefont {Pere\~n\'\i{}guez}}]{Elgood:2020mdx}%
  \BibitemOpen
  \bibfield  {author} {\bibinfo {author} {\bibfnamefont {Z.}~\bibnamefont {Elgood}}, \bibinfo {author} {\bibfnamefont {D.}~\bibnamefont {Mitsios}}, \bibinfo {author} {\bibfnamefont {T.}~\bibnamefont {Ort\'\i{}n}},\ and\ \bibinfo {author} {\bibfnamefont {D.}~\bibnamefont {Pere\~n\'\i{}guez}},\ }\bibfield  {title} {\bibinfo {title} {{The first law of heterotic stringy black hole mechanics at zeroth order in \ensuremath{\alpha}'}},\ }\href {https://doi.org/10.1007/JHEP07(2021)007} {\bibfield  {journal} {\bibinfo  {journal} {JHEP}\ }\textbf {\bibinfo {volume} {07}},\ \bibinfo {pages} {007}},\ \Eprint {https://arxiv.org/abs/2012.13323} {arXiv:2012.13323 [hep-th]} \BibitemShut {NoStop}%
\bibitem [{\citenamefont {Elgood}\ \emph {et~al.}(2021{\natexlab{b}})\citenamefont {Elgood}, \citenamefont {Ort\'\i{}n},\ and\ \citenamefont {Pere\~n\'\i{}guez}}]{Elgood:2020nls}%
  \BibitemOpen
  \bibfield  {author} {\bibinfo {author} {\bibfnamefont {Z.}~\bibnamefont {Elgood}}, \bibinfo {author} {\bibfnamefont {T.}~\bibnamefont {Ort\'\i{}n}},\ and\ \bibinfo {author} {\bibfnamefont {D.}~\bibnamefont {Pere\~n\'\i{}guez}},\ }\bibfield  {title} {\bibinfo {title} {{The first law and Wald entropy formula of heterotic stringy black holes at first order in $\alpha'$}},\ }\href {https://doi.org/10.1007/JHEP05(2021)110} {\bibfield  {journal} {\bibinfo  {journal} {JHEP}\ }\textbf {\bibinfo {volume} {05}},\ \bibinfo {pages} {110}},\ \Eprint {https://arxiv.org/abs/2012.14892} {arXiv:2012.14892 [hep-th]} \BibitemShut {NoStop}%
\bibitem [{\citenamefont {Reall}\ and\ \citenamefont {Santos}(2019)}]{Reall:2019sah}%
  \BibitemOpen
  \bibfield  {author} {\bibinfo {author} {\bibfnamefont {H.~S.}\ \bibnamefont {Reall}}\ and\ \bibinfo {author} {\bibfnamefont {J.~E.}\ \bibnamefont {Santos}},\ }\bibfield  {title} {\bibinfo {title} {{Higher derivative corrections to Kerr black hole thermodynamics}},\ }\href {https://doi.org/10.1007/JHEP04(2019)021} {\bibfield  {journal} {\bibinfo  {journal} {JHEP}\ }\textbf {\bibinfo {volume} {04}},\ \bibinfo {pages} {021}},\ \Eprint {https://arxiv.org/abs/1901.11535} {arXiv:1901.11535 [hep-th]} \BibitemShut {NoStop}%
\bibitem [{\citenamefont {Ma}\ \emph {et~al.}(2023)\citenamefont {Ma}, \citenamefont {Pang},\ and\ \citenamefont {L\"u}}]{Ma:2022gtm}%
  \BibitemOpen
  \bibfield  {author} {\bibinfo {author} {\bibfnamefont {L.}~\bibnamefont {Ma}}, \bibinfo {author} {\bibfnamefont {Y.}~\bibnamefont {Pang}},\ and\ \bibinfo {author} {\bibfnamefont {H.}~\bibnamefont {L\"u}},\ }\bibfield  {title} {\bibinfo {title} {{Negative corrections to black hole entropy from string theory}},\ }\href {https://doi.org/10.1007/s11433-023-2257-6} {\bibfield  {journal} {\bibinfo  {journal} {Sci. China Phys. Mech. Astron.}\ }\textbf {\bibinfo {volume} {66}},\ \bibinfo {pages} {121011} (\bibinfo {year} {2023})},\ \Eprint {https://arxiv.org/abs/2212.03262} {arXiv:2212.03262 [hep-th]} \BibitemShut {NoStop}%
\bibitem [{\citenamefont {Baron}\ and\ \citenamefont {Marques}(2021)}]{Baron:2020xel}%
  \BibitemOpen
  \bibfield  {author} {\bibinfo {author} {\bibfnamefont {W.}~\bibnamefont {Baron}}\ and\ \bibinfo {author} {\bibfnamefont {D.}~\bibnamefont {Marques}},\ }\bibfield  {title} {\bibinfo {title} {{The generalized Bergshoeff-de Roo identification. Part II}},\ }\href {https://doi.org/10.1007/JHEP01(2021)171} {\bibfield  {journal} {\bibinfo  {journal} {JHEP}\ }\textbf {\bibinfo {volume} {01}},\ \bibinfo {pages} {171}},\ \Eprint {https://arxiv.org/abs/2009.07291} {arXiv:2009.07291 [hep-th]} \BibitemShut {NoStop}%
\bibitem [{\citenamefont {Hronek}\ and\ \citenamefont {Wulff}(2021{\natexlab{b}})}]{Hronek:2021nqk}%
  \BibitemOpen
  \bibfield  {author} {\bibinfo {author} {\bibfnamefont {S.}~\bibnamefont {Hronek}}\ and\ \bibinfo {author} {\bibfnamefont {L.}~\bibnamefont {Wulff}},\ }\bibfield  {title} {\bibinfo {title} {{String theory at order \ensuremath{\alpha}'$^{2}$ and the generalized Bergshoeff-de Roo identification}},\ }\href {https://doi.org/10.1007/JHEP11(2021)186} {\bibfield  {journal} {\bibinfo  {journal} {JHEP}\ }\textbf {\bibinfo {volume} {11}},\ \bibinfo {pages} {186}},\ \Eprint {https://arxiv.org/abs/2109.12200} {arXiv:2109.12200 [hep-th]} \BibitemShut {NoStop}%
\bibitem [{\citenamefont {Hronek}\ \emph {et~al.}(2022)\citenamefont {Hronek}, \citenamefont {Wulff},\ and\ \citenamefont {Zacarias}}]{Hronek:2022dyr}%
  \BibitemOpen
  \bibfield  {author} {\bibinfo {author} {\bibfnamefont {S.}~\bibnamefont {Hronek}}, \bibinfo {author} {\bibfnamefont {L.}~\bibnamefont {Wulff}},\ and\ \bibinfo {author} {\bibfnamefont {S.}~\bibnamefont {Zacarias}},\ }\bibfield  {title} {\bibinfo {title} {{The \ensuremath{\alpha}'$^{2}$ correction from double field theory}},\ }\href {https://doi.org/10.1007/JHEP11(2022)090} {\bibfield  {journal} {\bibinfo  {journal} {JHEP}\ }\textbf {\bibinfo {volume} {11}},\ \bibinfo {pages} {090}},\ \Eprint {https://arxiv.org/abs/2206.10640} {arXiv:2206.10640 [hep-th]} \BibitemShut {NoStop}%
\bibitem [{\citenamefont {Chen}\ and\ \citenamefont {Stein}(2018)}]{Chen:2018jed}%
  \BibitemOpen
  \bibfield  {author} {\bibinfo {author} {\bibfnamefont {B.}~\bibnamefont {Chen}}\ and\ \bibinfo {author} {\bibfnamefont {L.~C.}\ \bibnamefont {Stein}},\ }\bibfield  {title} {\bibinfo {title} {{Deformation of extremal black holes from stringy interactions}},\ }\href {https://doi.org/10.1103/PhysRevD.97.084012} {\bibfield  {journal} {\bibinfo  {journal} {Phys. Rev. D}\ }\textbf {\bibinfo {volume} {97}},\ \bibinfo {pages} {084012} (\bibinfo {year} {2018})},\ \Eprint {https://arxiv.org/abs/1802.02159} {arXiv:1802.02159 [gr-qc]} \BibitemShut {NoStop}%
\bibitem [{\citenamefont {Cano}\ and\ \citenamefont {David}(2023)}]{Cano:2023dyg}%
  \BibitemOpen
  \bibfield  {author} {\bibinfo {author} {\bibfnamefont {P.~A.}\ \bibnamefont {Cano}}\ and\ \bibinfo {author} {\bibfnamefont {M.}~\bibnamefont {David}},\ }\bibfield  {title} {\bibinfo {title} {{The extremal Kerr entropy in higher-derivative gravities}},\ }\href {https://doi.org/10.1007/JHEP05(2023)219} {\bibfield  {journal} {\bibinfo  {journal} {JHEP}\ }\textbf {\bibinfo {volume} {05}},\ \bibinfo {pages} {219}},\ \Eprint {https://arxiv.org/abs/2303.13286} {arXiv:2303.13286 [hep-th]} \BibitemShut {NoStop}%
\bibitem [{\citenamefont {Cassani}\ \emph {et~al.}(2023)\citenamefont {Cassani}, \citenamefont {Ruip\'erez},\ and\ \citenamefont {Turetta}}]{Cassani:2023vsa}%
  \BibitemOpen
  \bibfield  {author} {\bibinfo {author} {\bibfnamefont {D.}~\bibnamefont {Cassani}}, \bibinfo {author} {\bibfnamefont {A.}~\bibnamefont {Ruip\'erez}},\ and\ \bibinfo {author} {\bibfnamefont {E.}~\bibnamefont {Turetta}},\ }\bibfield  {title} {\bibinfo {title} {{Boundary terms and conserved charges in higher-derivative gauged supergravity}},\ }\href {https://doi.org/10.1007/JHEP06(2023)203} {\bibfield  {journal} {\bibinfo  {journal} {JHEP}\ }\textbf {\bibinfo {volume} {06}},\ \bibinfo {pages} {203}},\ \Eprint {https://arxiv.org/abs/2304.06101} {arXiv:2304.06101 [hep-th]} \BibitemShut {NoStop}%
\bibitem [{\citenamefont {Cano}\ and\ \citenamefont {David}(2024)}]{Cano:2024tcr}%
  \BibitemOpen
  \bibfield  {author} {\bibinfo {author} {\bibfnamefont {P.~A.}\ \bibnamefont {Cano}}\ and\ \bibinfo {author} {\bibfnamefont {M.}~\bibnamefont {David}},\ }\bibfield  {title} {\bibinfo {title} {{Near-horizon geometries and black hole thermodynamics in higher-derivative AdS$_{5}$ supergravity}},\ }\href {https://doi.org/10.1007/JHEP03(2024)036} {\bibfield  {journal} {\bibinfo  {journal} {JHEP}\ }\textbf {\bibinfo {volume} {03}},\ \bibinfo {pages} {036}},\ \Eprint {https://arxiv.org/abs/2402.02215} {arXiv:2402.02215 [hep-th]} \BibitemShut {NoStop}%
\bibitem [{\citenamefont {Maharana}\ and\ \citenamefont {Schwarz}(1993)}]{Maharana:1992my}%
  \BibitemOpen
  \bibfield  {author} {\bibinfo {author} {\bibfnamefont {J.}~\bibnamefont {Maharana}}\ and\ \bibinfo {author} {\bibfnamefont {J.~H.}\ \bibnamefont {Schwarz}},\ }\bibfield  {title} {\bibinfo {title} {{Noncompact symmetries in string theory}},\ }\href {https://doi.org/10.1016/0550-3213(93)90387-5} {\bibfield  {journal} {\bibinfo  {journal} {Nucl. Phys. B}\ }\textbf {\bibinfo {volume} {390}},\ \bibinfo {pages} {3} (\bibinfo {year} {1993})},\ \Eprint {https://arxiv.org/abs/hep-th/9207016} {arXiv:hep-th/9207016} \BibitemShut {NoStop}%
\bibitem [{\citenamefont {Sen}(1994)}]{Sen:1994fa}%
  \BibitemOpen
  \bibfield  {author} {\bibinfo {author} {\bibfnamefont {A.}~\bibnamefont {Sen}},\ }\bibfield  {title} {\bibinfo {title} {{Strong - weak coupling duality in four-dimensional string theory}},\ }\href {https://doi.org/10.1142/S0217751X94001497} {\bibfield  {journal} {\bibinfo  {journal} {Int. J. Mod. Phys. A}\ }\textbf {\bibinfo {volume} {9}},\ \bibinfo {pages} {3707} (\bibinfo {year} {1994})},\ \Eprint {https://arxiv.org/abs/hep-th/9402002} {arXiv:hep-th/9402002} \BibitemShut {NoStop}%
\end{thebibliography}%
\let\addcontentsline\oldaddcontentsline

\end{document}